\documentclass[journal,12pt,onecolumn,letterpaper]{IEEEtran}
\usepackage{arxiv}
\usepackage[utf8]{inputenc} 
\usepackage[T1]{fontenc}    
\usepackage{hyperref}       
\usepackage{url}            
\usepackage{booktabs}       
\usepackage{amsfonts}       
\usepackage{nicefrac}       
\usepackage{microtype}      
\usepackage{lipsum}		
\usepackage{graphicx}
\usepackage{doi}
\usepackage{siunitx}
\usepackage{booktabs}
\usepackage{graphicx}
\usepackage{mathtools}

\title{Consensus Algorithms of Distributed Ledger Technology - A Comprehensive Analysis}

\author{\href{https://orcid.org/0009-0009-8297-7627}{\includegraphics[scale=0.06]{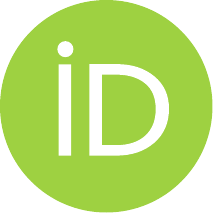}\hspace{1mm}Ahmad J.~ Alkhodair} \\
	Computer Engineering Department \\
	University of Tabuk\\
	Tabuk, Saudi Arabia \\
	\texttt{aalkhodair@ut.edu.sa} \\
	\And
	\href{https://orcid.org/0000-0003-2959-6541}{\includegraphics[scale=0.06]{orcid.pdf}\hspace{1mm}Saraju P.~Mohanty} \\
	Department of Computer Science and Engineering \\
	University of North Texas\\
	Denton, TX 76203, USA \\
	\texttt{saraju.mohanty@unt.edu} \\
	\And
	\href{https://orcid.org/0000-0002-1616-7628}{\includegraphics[scale=0.06]{orcid.pdf}\hspace{1mm}Elias Kougianos} \\
	Department of Electrical Engineering \\
	University of North Texas\\
	Denton, TX 76203, USA \\
	USA \\
	\texttt{elias.kougianos@unt.edu} \\
}

\renewcommand{\shorttitle}{\textit{arXiv} Consensus Algorithms Comparative Analysis}

\hypersetup{
pdftitle={A template for the arxiv style},
pdfsubject={q-bio.NC, q-bio.QM},
pdfauthor={David S.~Hippocampus, Elias D.~Striatum},
pdfkeywords={First keyword, Second keyword, More},
}

\begin{document}
\maketitle

\begin{abstract}
The most essential component of every Distributed Ledger Technology (DLT) is the Consensus Algorithm (CA), which enables users to reach a consensus in a decentralized and distributed manner. Numerous CA exist, but their viability for particular applications varies, making their trade-offs a crucial factor to consider when implementing DLT in a specific field. This article provided a comprehensive analysis of the various consensus algorithms used in distributed ledger technologies (DLT) and blockchain networks. We cover an extensive array of thirty consensus algorithms. Eleven attributes including hardware requirements, pre-trust level, tolerance level, and more, were used to generate a series of comparison tables evaluating these consensus algorithms.  In addition, we discuss DLT classifications, the categories of certain consensus algorithms, and provide examples of authentication-focused and data-storage-focused DLTs. In addition, we analyze the pros and cons of particular consensus algorithms, such as Nominated Proof of Stake (NPoS), Bonded Proof of Stake (BPoS), and Avalanche. In conclusion, we discuss the applicability of these consensus algorithms to various Cyber Physical System (CPS) use cases, including supply chain management, intelligent transportation systems, and smart healthcare.
\end{abstract}


\setcounter{section}{0} 
\section{Introduction}
\label{Introduction}

Distributed Ledger Technologies (DLTs) have disrupted business paradigms and changed how companies handle data and transactions. DLTs are predicted to reach \$2,082.8 million \cite{fortunebusinessinsights2020} of transactions. The Quorum blockchain platform streamlines cross-border payments and reduces expenses. DLTs provide secure and efficient medical record sharing, improving patient care and privacy. Walmart and other retail giants use blockchain technology to improve supply chain traceability and product quality. As demand for secure, transparent, and efficient solutions rises, researchers, developers, and industry professionals must understand DLTs' success mechanisms \cite{Walmart}.

DLTs are a novel method for administering and exchanging data across a network of participants without a central authority. They provide a secure, transparent, and tamper-resistant method for recording and sharing information, which has the potential to improve trust, productivity, and collaboration across a variety of applications and industries \cite{ASurveyonBlockchainTechnologyEvolutionArchitectureandSecurity}.

The most prominent and extensively adopted form of DLT is the blockchain. It is a decentralized digital ledger that uses cryptographic techniques and consensus algorithms to maintain an immutable chain of blocks, or records. Each block contains a set of transactions, and once a block is added to the chain it becomes practically impossible to alter the information it contains without the agreement of all network participants  \cite{ComparativeResearchOnBlockchainConsensusAlgorithms}.

Key concepts related to distributed ledger technology and blockchain include the use of cryptographic hashing and consensus mechanism guarantees that the data stored on the blockchain is secure and unchangeable. Once a block is added to the chain, it is extremely difficult to alter its content, which helps preserve the ledger's integrity and credibility \cite{BlockchainConsensusAlgorithmsASurvey}.

The blockchain eliminates the requirements for a centralized authority to govern and maintain the ledger. Instead, the responsibility for maintaining and updating it is distributed across the network nodes. This decentralized approach improves the system's security, transparency, and resistance to censorship \cite{DistributedLedgerTechnologyDLTTheBeginningofaTechnologicalRevolutionforBlockchain}.

\begin{figure}[htbp]
\centering
\includegraphics[width=16.5cm]{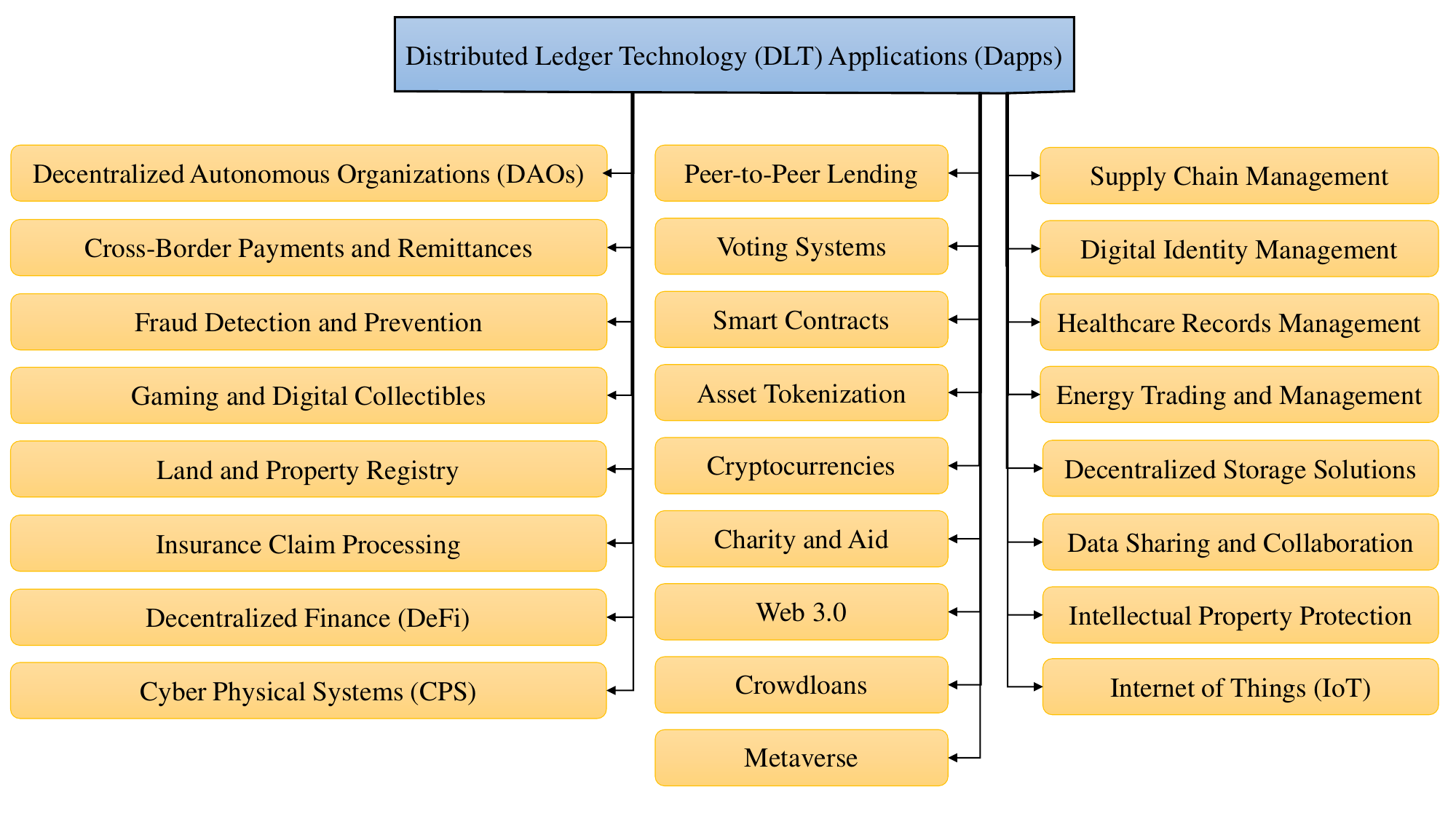}
\caption{DLT Applications.}
\label{FIG:DLTApplications}
\end{figure}

All network participants can view the transactions recorded on the blockchain, fostering transparency and allowing anyone to audit and verify the data. This feature is especially useful in applications where data integrity and trust are essential, such as financial transactions and supply chain management \cite{AJourneyfromCommitProcessinginDistributedDatabasestoConsensusinBlockchain}.
Multiple applications of DLTs that use different consensus mechanisms are shown in Fig. \ref{FIG:DLTApplications} \cite{8386948}. The blockchain, or DLTs in general have been applied to almost all aspects of human life including healthcare, agriculture, pharmaceutical, transportation, and supply chain \cite{9288682, DBLP:journals/sensors/VangipuramMKR22, DBLP:journals/corr/abs-2201-04754, https://doi.org/10.1049/ntw2.12041}.

For example, cryptocurrencies are digital or virtual currencies that use encryption to make sure that transactions are safe and that control is not centralized. Supply Chain Management (SCM) is the process of making things and services easier to track, more transparent, and more efficient. Voting systems must be safe, clear, and unchangeable. Smart contracts should automatically put the terms of the contract into effect when certain conditions are met. Decentralized Finance (DeFi) means financial services that are made on decentralized platforms, which eliminate and make financial products easier to get. Digital identity management is the process of making sure that digital identities are safe and can be checked. This improves privacy and speeds up the authentication process. The Internet of Things (IoT) makes it possible for connected devices to share data in a safe and clear way, which improves trust and coordination. Intellectual property protection improves the way copyrights, patents, and trademarks are managed by keeping records that cannot be changed. Healthcare records management is the safe and efficient storage, sharing, and control of sensitive patient data by all healthcare providers. Energy trading and management decentralized systems involves making, using, and trading energy and support renewable energy and efficient energy markets. Cross-border payments and remittances make foreign transactions faster, cheaper, and safer by cutting out middlemen and limiting fees.

Asset tokenization is the process of turning real or digital assets into digital tokens and putting them in a blockchain. This makes it easier to share ownership and trade.
Decentralized Autonomous Organizations (DAOs) are organizations without a central authority or hierarchy that are run by pre-set rules and consensus processes. Data sharing and collaboration data between people, groups, or devices must be done in a way that keeps the data private and secure. Processing insurance claims requires streamlining and automating the way they are handled, limiting fraud, and making the customer experience better. Land and property registry requires keeping clear, safe, and unchangeable records of who owns land and property, which builds trust and reduces disagreements. Decentralized storage solutions are networks of secure, distributed data storage that make data more secure, reliable, and easy to reach. Peer-to-peer lending takes place when people trade and borrow money directly from each other, without any middlemen. This gives people better rates and easier access to credit. Gaming and digital collectibles involves making, selling, and proving ownership of unique digital assets like non-fungible tokens (NFTs). Finally, fraud discovery and prevention demands better security and openness in transactions and data management, which lowers the risk of fraud and other bad things happening. The applications that suit DLT usage increase every day due to the wide variety of ledgers and consensus algorithms.  

Consensus algorithms are an essential component of DLT systems, as they play a crucial role in sustaining the distributed ledger's integrity, security, and stability. These algorithms provide a method for network participants to concur on the ledger's validity and state, ensuring that all nodes have a consistent and accurate view of the data. Transaction validation, network synchronization, conflict resolution, and overall system security are the responsibility of consensus mechanisms. A DLT system can be tailored to satisfy specific goals and requirements, such as scalability, energy efficiency, and decentralization, by selecting the appropriate consensus algorithm \cite{BlockchainConsensusProtocolsStateoftheartandFutureDirections}.

\subsection{Paper Organization}

\begin{figure}[htp]
	\centering
	\includegraphics[width=16.5cm]{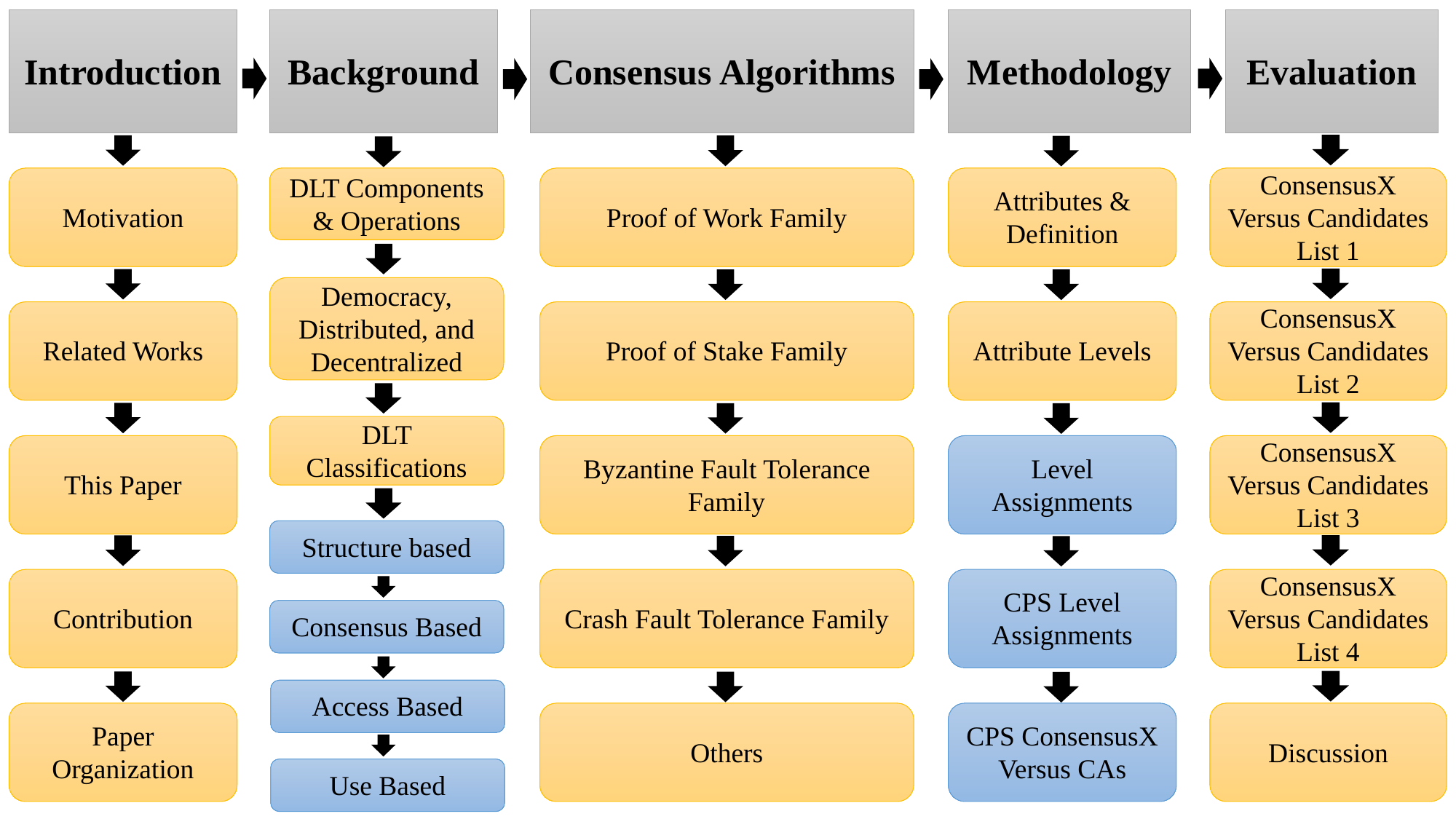}
	\caption{Paper Organization.}
\label{FIG:PaperOrganization}
\end{figure}

The paper is organized as shown in Fig. \ref{FIG:PaperOrganization}. The paper organization starts with an introduction in Sec. \ref{Introduction} that presents the background, motivation, and purpose of the study, as well as an overview of the DLT and its applications. Finally, it presents the significance of consensus algorithms. Sec. \ref{sec:ContributionsoftheCurrentPaper} presents firstly the crucial role of consensus algorithms and then moves to the motivation of the paper. Sec. \ref{sec:Related Work} discusses related works and their contributions. Sec. \ref{sec:Background} discusses DLT classifications and their consensus mechanisms. Sec. \ref{sec:GeneralOverviewforConsensusAlgorithms} explores multiple consensus algorithms and their pros and cons. Next, Sec. \ref{sec:Methodology} presents the method used to evaluate the consensus algorithms. Sec. \ref{sec:Evaluation} examines the evaluation and their results. Sec. \ref{sec:Conclusion} concludes the work presented in this paper. 

\section{Contributions of the Current Paper}
\label{sec:ContributionsoftheCurrentPaper}

In the realm of DLTs, consensus algorithms play a crucial role in ensuring the security, integrity, and consistency of the shared ledger across network participants. A multitude of consensus algorithms have been proposed, each with distinct features and trade-offs. When evaluating these algorithms, it is essential to consider their suitability for specific applications. For instance, Proof of Work (PoW), employed by Bitcoin and, initially, Ethereum, offers robust security but suffers from high energy consumption and limited scalability. In contrast, Proof of Stake (PoS), adopted by Ethereum 2.0 and Cardano, addresses these limitations while maintaining adequate security, making it more suitable for energy-conscious and scalable applications. Delegated Proof of Stake (DPoS), as seen in EOS and Tron, further enhance scalability, but at the cost of centralization. Practical Byzantine Fault Tolerance (PBFT) and its variants, used in permissioned DLT systems like Hyperledger Fabric, offer high transaction throughput and low latency, ideal for enterprise applications. Ultimately, selecting the appropriate consensus algorithm hinges on the specific requirements of the target application, with considerations such as security, scalability, decentralization, and energy efficiency driving the decision-making process.

\subsection{Motivation}

The motivation for studying consensus algorithms in the context of DLTs stems from the wide range of challenges and requirements that different DLT systems encounter. These diverse systems necessitate the development and implementation of various consensus algorithms, each with their unique properties, trade-offs, and performance characteristics. As DLTs are being adopted across numerous industries, it is crucial to have a comprehensive understanding of the underlying consensus mechanisms that govern their operation.

One of the primary challenges faced by DLT systems is scalability, which is as the number of participants and transactions in a DLT system grows, and ensures that the system is able to handle the increased load while maintaining performance and security. Different consensus algorithms address scalability in various ways, affecting the overall efficiency and throughput of the system.

Another challenge is ensuring the security and integrity of the distributed ledger. It is a critical requirement for any DLT system. Consensus algorithms play a vital role in maintaining security by preventing attacks, such as double-spending or Sybil attacks, and ensuring that only valid transactions are added to the ledger.

Decentralization is a key characteristic of many DLT systems, which removes the need for a central authority and distributes control among network participants. Consensus algorithms must strike a balance between decentralization and other factors, such as efficiency and security, to ensure the optimal functioning of the system.

Moreover, energy efficiency is vital. Some consensus mechanisms, such as PoW, consume significant amounts of energy, which has environmental and economic implications. As a result, there is a growing interest in exploring alternative consensus algorithms that are more energy-efficient while maintaining the desired levels of security and performance.

Lastly, latency, which in some applications such as financial transactions or real-time data sharing is crucial in minimizing the amount of time spent in transaction confirmation and data propagation. Consensus algorithms play a significant role in determining the speed at which transactions are validated and added to the ledger.

For researchers, developers, and practitioners working with DLTs, understanding the various consensus algorithms is essential to make informed decisions about which mechanism is best suited for their specific use case. Categorizing and comparing these algorithms based on their properties, trade-offs, and performance allows for a more in-depth analysis and a better understanding of the advantages and limitations of each approach. This knowledge can ultimately lead to the development of more efficient, secure, and scalable DLT systems that cater to the diverse needs of different industries and applications.


\subsection{Contributions}

The main goals and objectives of this article are to provide a comprehensive review of different consensus algorithms by:

\begin{itemize}
	\item Presenting an in-depth analysis of various consensus mechanisms used in DLTs, highlighting their features, advantages, and limitations. This will help readers gain a better understanding of the diverse consensus algorithms and their applicability to different use cases and requirements. 
	\item Developing a classification scheme for consensus algorithms: Propose a systematic classification of consensus algorithms based on their underlying properties, such as proof-based, voting-based, weight-based, randomized, and Directed Acyclic Graph (DAG)-based mechanisms. This classification will facilitate the comparison and analysis of different consensus mechanisms, making it easier for researchers and practitioners to identify suitable algorithms for their specific needs. 
	\item Investigating different DLT architectures: Examine various DLT architectures, focusing on how data is stored and linked within these systems. This investigation will provide insights into the design choices and trade-offs associated with different DLT architectures, helping readers make informed decisions when selecting or developing DLT solutions. 
	\item Analyzing the relationship between consensus algorithms and DLT architectures: Explore the interplay between consensus mechanisms and DLT architectures, identifying how the choice of consensus algorithm can impact the overall performance, security, and scalability of a DLT system. This analysis will contribute to the understanding of how consensus algorithms and DLT architectures can be optimized to meet the diverse requirements of different industries and applications. 
	\item Contributing to the existing body of knowledge: By offering a comprehensive review, classification, and analysis of consensus algorithms and DLT architectures, this article aims to contribute to the ongoing research and development efforts in the field of DLTs. The insights provided in this article will not only benefit researchers and developers working with DLTs but will also inform decision-makers, policymakers, and other stakeholders interested in the potential applications and implications of these technologies. 
\end{itemize} 
By achieving these goals and objectives, this article will add significant value to the existing literature on consensus algorithms and DLTs, facilitating a deeper understanding of these technologies and their potential impact on various industries and applications.

\section{Related Prior Works}
\label{sec:Related Work}

We have reviewed prior works contributing to various type of surveys and overview works on blockchain and DLT. Table \ref{TBL:RelatedWorks} compares the contribution of each related work to the current paper.

The work in \cite{bano2017consensusSurveys} provides an extensive overview of various consensus algorithms used in blockchain technology, focusing on their strengths and weaknesses. The authors analyze each algorithm's trade-offs in terms of decentralization, security, and performance, ultimately highlighting the ongoing challenges in achieving a balance between these factors. 

The work \cite{zhang2019necessitySurveys} investigates the Bitcoin Unlimited mining protocol, emphasizing the importance of a prescribed block validity consensus. Through their analysis, the authors reveal potential issues with Bitcoin Unlimited, including risks of blockchain forks, which can compromise the security and stability of the network. 

Authors in \cite{vukolic2015questSurveys}  examine the scalability challenges faced by blockchain technology, comparing Proof-of-Work (PoW) and Byzantine Fault Tolerance (BFT) replication mechanisms. The author highlights the trade-offs between decentralization, security, and performance, suggesting that a combination of PoW and BFT might offer a more scalable and secure solution for blockchain systems. 

The work in \cite{xin2018researchSurveys} provides a comparative study of various blockchain consensus algorithms, focusing on their strengths, weaknesses, and applicability in different scenarios. The authors highlight the importance of selecting the appropriate consensus algorithm based on specific use cases and system requirements, emphasizing the trade-offs among security, performance, and decentralization. 

A comprehensive overview of blockchain technology in their NIST report, focusing on the underlying concepts, consensus mechanisms, and various use cases has been presented in \cite{yaga2018blockchainSurveys}. The authors discuss the challenges and limitations of current blockchain implementations, offering guidance for organizations interested in adopting or developing blockchain-based systems.

The work \cite{wang2019surveySurveys} provides a systematic survey of various consensus algorithms employed in blockchain technology. The authors discuss the strengths and weaknesses of each algorithm, highlighting their suitability for different application scenarios. The survey offers valuable insights for researchers and practitioners exploring consensus mechanisms for their blockchain projects. 

The work in \cite{nguyen2019surveySurveys} presents an extensive survey on consensus algorithms utilized in blockchain systems, discussing their properties and functions. The authors compare the algorithms based on performance, security, and decentralization, offering a valuable resource for understanding the trade-offs and selection criteria for different blockchain applications. 

The work \cite{merkle2019comparativeSurveys} conducts a comparative analysis of blockchain consensus algorithms, evaluating their performance, security, and decentralization properties. The authors emphasize the importance of selecting the right consensus mechanism for specific use cases, and their study serves as a helpful guide for practitioners and researchers in the field of blockchain technology. 

A comprehensive survey of consensus mechanisms in blockchain technology, specifically focusing on applications in IoT networks has been provided in \cite{abbasi2021surveySurveys}. The authors analyze the properties and requirements of each consensus algorithm, offering valuable insights into their suitability for various IoT use cases and challenges. 

The work in \cite{garg2021blockchainSurveys} provides a thorough review of blockchain consensus algorithms, discussing their properties, advantages, and disadvantages. The authors examine various consensus mechanisms in the context of diverse application scenarios, offering a valuable resource for researchers and practitioners seeking to choose the right algorithm for their blockchain projects. 

The authors in \cite{pustokhina2020consensusSurveys} perform a comparative analysis and classification of consensus algorithms in blockchain technology. The authors evaluate each mechanism based on key characteristics, providing a systematic overview that helps readers understand their strengths and weaknesses, as well as their suitability for different application domains. 

The work in \cite{soleymani2020comprehensiveSurveys} conducts a comprehensive study of blockchain consensus algorithms, examining their features, security aspects, and performance. The authors delve into various consensus mechanisms, offering a detailed comparison that aids researchers and practitioners in understanding the trade-offs and selecting appropriate algorithms for their blockchain projects. 

The work \cite{ASurveyofBlockchainConsensusAlgorithmsDeng} provides a survey of different types of consensus algorithms used in blockchain networks and discusses their strengths and weaknesses. It also describes how consensus algorithms are integrated with other technologies, such as BFT (Byzantine Fault Tolerance), credit mechanisms, and artificial intelligence algorithms. The paper provides a detailed analysis of three types of consensus algorithms: those based on certain attribute value proof of peers, those based on peer voting mechanism, and Paxos class consensus algorithm. Additionally, the paper conducts comparative research according to Munde ll's impossible triangle theory and gives the development direction of consensus algorithm. Overall, the paper provides a useful reference for researchers to conduct in-depth research on blockchain consensus algorithms. 

In \cite{ASurveyonConsensusAlgorithmsinBlockchainBasedonPostQuantumCryptosystemsJose}, a survey and analysis of different consensus algorithms used in Blockchain based on Post Quantum Cryptosystems has been presented. It explains that in a decentralized network, trust among the nodes is important and consensus algorithms are used to ensure that every node agrees before any data is added to the network. The paper also discusses how quantum computing can impact the security of Blockchain and proposes solutions to make Blockchain systems quantum safe. Overall, the paper provides a comprehensive overview of consensus algorithms in Blockchain and their relationship with post-quantum cryptography. 

In \cite{ASurveyonEfficientConsensusMechanismforElectricityInformationAcquisitionSystemJu}, consensus mechanisms for electricity information acquisition systems has been discussed. It analyzes the problems faced by the current power system data sharing and proposes three consensus algorithms that are suitable for solving these problems. The paper compares and improves these algorithms to make them more efficient and energy-saving. The proposed solutions have the potential to improve data sharing in the power system, increase efficiency, and reduce energy consumption. Overall, this paper provides valuable insights into the challenges faced by the current power system data sharing and proposes efficient solutions to overcome them. 

The work in \cite{ASurveyonConsensusMechanismsforBlockchainTechnologyGu} presents a comprehensive survey of the latest consensus mechanisms used in blockchain systems, along with their advantages and disadvantages. It systematically studies the state-of-the-art consensus mechanisms, and looks into their pros and cons, respectively. The two most widely used consensus mechanisms are Proof of Work (PoW) and Proof of Stake (PoS), but there are also other consensus algorithms that derive from either method such as Delegated Proof of Stake (DPoS). The paper compares these algorithms from multiple perspectives including security, fault tolerance, validation speed, energy consumption, and level of decentralization. It also takes a closer look at the different problems each algorithm has to deal with. 

The work in \cite{ASurveyonConsensusAlgorithmsinBlockchainbasedApplicationsArchitectureTaxonomyandOperationalIssuesIslam} contains a comprehensive survey of consensus algorithms in blockchain-based applications. It presents a taxonomy of consensus algorithms, including proof-based and voting-based algorithms, and provides a comparative discussion of their performances, efficiencies, and uses in blockchains. The paper also analyzes the application domains of consensus algorithms in terms of development tools, uses, and environments. Additionally, it highlights the challenges in blockchain applications regarding functional and non-functional issues. Overall, this paper provides valuable insights for stakeholders interested in understanding the trends and issues related to consensus algorithms in blockchain technology. 

In \cite{SurveyontheapplicationofblockchaininDigitalRightsProtectionGuichun}, the application of blockchain in digital rights protection, with a focus on the improvements made by researchers to the consensus algorithm has been presented. The consensus algorithm is the core of blockchain technology, but it cannot be directly applied to digital rights protection due to its drawbacks. Therefore, researchers have optimized and improved the consensus algorithm to meet the needs of digital rights protection. The paper reviews these improvements and analyzes their impact on digital rights protection. 

In \cite{ASurveyonDecentralizedConsensusMechanismsforCyberPhysicalSystemsBodkhe}, a survey on Decentralized Consensus Mechanisms for Cyber Physical Systems has been presented. The authors discuss the challenges of deploying decentralized consensus protocols to ensure fairness, trust, and transparency in operations. They also present an evaluation model of consensus approaches based on defined metrics and categorize them through parameters such as incentive, performance, data model, energy-efficiency, and exposure likelihood.  

A comprehensive review of consensus algorithms in blockchain technology has been presented in \cite{ThreeDimensionalTradeoffsforConsensusAlgorithmsAReviewLi}. It proposes a multi-dimensional tradeoff model to guide the construction of consensus algorithms, taking into account the technical constraints that limit the large-scale application of blockchain, namely scalability, security, and decentralization. The paper compares and analyzes various classical consensus algorithms and focuses on their design principles under the multi-dimensional tradeoff model. It also describes the performance indicators of different consensus algorithms and provides different solutions for blockchain in different dimensions. Overall, this paper offers operational advice to developers and users on how to design optimal consensus algorithms for their specific needs. 

The work in \cite{ASurveyPaperOnConsensusAlgorithmOfMobileHealthcareInBlockchainNetworkRwibasira} surveys the consensus algorithms used in mobile healthcare blockchain networks. It discusses the importance of consensus algorithms in blockchain distributed ledgers and provides guidance on selecting the most suitable consensus algorithm for a given network. The paper also presents a theoretical approach to improving existing blockchain structures in mobile healthcare settings. However, it does not provide specific details on the proposed approach or its potential limitations.


\begin{table}[htbp]
	\caption{Related Works. \label{TBL:RelatedWorks}}
	\centering
	\begin{tabular}{|p{0.30\linewidth}|p{0.65\linewidth}|}
\hline
		\textbf{Work}                  & \textbf{Description}	 \\
		\hline
Bano et al. [2017] \cite{bano2017consensusSurveys} & Assessment of algorithm's decentralization, security, and performance trade-offs, stressing the continuous difficulties in balancing these variables.\\
		\hline
		
		Xin et al. [2018] \cite{xin2018researchSurveys}   & The authors emphasize security, performance, and decentralization trade-offs when choosing a consensus method based on use cases and system needs.\\
		\hline
		
		Yaga et al. [2018] \cite{yaga2018blockchainSurveys}  &  A discussion on the challenges and limitations of current blockchain implementations, offering guidance for organizations interested in adopting or developing blockchain-based systems.\\
		\hline
		
		Wang and Wang [2019] \cite{wang2019surveySurveys}  &  The authors discuss the strengths and weaknesses of each algorithm, highlighting their suitability for different application scenarios.\\
		\hline
		
		Nguyen et al. [2019] \cite{nguyen2019surveySurveys}  &  The authors compare the algorithms based on performance, security, and decentralization, offering a valuable resource for understanding the trade-offs and selection criteria for different blockchain applications.\\
		\hline
		
		Merkle and Niebling [2019] \cite{merkle2019comparativeSurveys}   &  Compare blockchain consensus algorithms for performance, security, and decentralization. Their work helps blockchain practitioners and researchers choose the optimal consensus mechanism for certain use cases.\\
		\hline
		
		Abbasi et al. [2021] \cite{abbasi2021surveySurveys}  &  Give a complete survey of blockchain consensus algorithms for IoT networks. The authors investigate each consensus algorithm's features and needs, providing useful insights into their applicability for specific IoT use cases and issues.\\
		\hline
		
		Garg and Choudhary [2021] \cite{garg2021blockchainSurveys}  &  The authors examine various consensus mechanisms in the context of diverse application scenarios, offering a valuable resource for researchers and practitioners seeking to choose the right algorithm for their blockchain projects.\\
		\hline
		
		Pustokhina et al. [2020] \cite{pustokhina2020consensusSurveys}  &  The authors analyze each mechanism based on important criteria, giving readers a comprehensive overview of their strengths and shortcomings and applicability for different application fields.\\
		\hline
		
		Soleymani and Anjomshoa [2020] \cite{soleymani2020comprehensiveSurveys}   &  The authors delve into various consensus mechanisms, offering a detailed comparison that aids researchers and practitioners in understanding the trade-offs and selecting appropriate algorithms for their blockchain projects.\\
		\hline
		
		Deng et al. [2022] \cite{ASurveyofBlockchainConsensusAlgorithmsDeng}  &   The study analyzes peer voting, attribute value proof, and Paxos class consensus algorithms.\\
		\hline
		
		Jose and V [2022] \cite{ASurveyonConsensusAlgorithmsinBlockchainBasedonPostQuantumCryptosystemsJose}   &  Analyzes Post Quantum Cryptosystem-based Blockchain consensus techniques. \\
		\hline
		
		Ju et al. [2021] \cite{ASurveyonEfficientConsensusMechanismforElectricityInformationAcquisitionSystemJu}  & The authors investigate power system data sharing issues and provides three consensus algorithms to address them. The paper compares and optimizes several techniques for conserving energy. \\
		\hline
		
		Gu et al. [2021] \cite{ASurveyonConsensusMechanismsforBlockchainTechnologyGu}  &  The study compares various algorithms on security, fault tolerance, validation speed, energy consumption, and decentralization and examines each algorithm's issues.\\
		\hline
		
		Islam et al. [2023] \cite{ASurveyonConsensusAlgorithmsinBlockchainbasedApplicationsArchitectureTaxonomyandOperationalIssuesIslam}  & The authors compares proof-based and voting-based consensus algorithms for performance, efficiency, and blockchain applications and explores consensus algorithm development tools, uses, and contexts.\\
		\hline
		
		Guichun et al. [2020] \cite{SurveyontheapplicationofblockchaininDigitalRightsProtectionGuichun}   &  Examines blockchain's use in digital rights protection and researchers' consensus algorithm upgrades. The study examines improvements and their effects on digital rights protection.\\
		\midrule
		
Bodkhe et al. [2020] \cite{ASurveyonDecentralizedConsensusMechanismsforCyberPhysicalSystemsBodkhe}  &  Decentralized Consensus Mechanisms for CPS for fairness, trust, and transparency in operations present obstacles and  evaluate consensus approaches using motivation, performance, data model, energy-efficiency, and likelihood.\\
		\hline
		
		Li et al. [2022] \cite{ThreeDimensionalTradeoffsforConsensusAlgorithmsAReviewLi}   & Authors use a multi-dimensional tradeoff model to guide consensus algorithm construction, considering technical restrictions like scalability, security, and decentralization that limit large-scale blockchain deployment.\\
		\hline
		
		Rwibasira and Suchithra [2020] \cite{ASurveyPaperOnConsensusAlgorithmOfMobileHealthcareInBlockchainNetworkRwibasira}  & Authors examine blockchain consensus methods and how to choose the best one for a network. The paper also proposes upgrading mobile healthcare blockchain systems theoretically. \\
		\hline
		
\textbf{Current Work} & Explores 30 consensus algorithms types and their ledger structure, evaluate them based on 11 attributes, and test their suitability to CPS applications.\\
		\hline
	\end{tabular}
\end{table}

\section{Background: Distributed Ledger Technology (DLT)}
\label{sec:Background}

One of the most revolutionary developments in recent years has been the introduction of blockchain and Distributed Ledger Technologies (DLTs), which have altered many facets of our life. These innovative methods of data storage and transaction processing may cause widespread change across a variety of sectors, from banking and supply chain management to government and healthcare. Yet, it is essential to note that blockchain technology and DLT are not the same thing. 

The blockchain is a sort of distributed ledger technology (DLT) that uses a distributed network of computers to record transactions in a way that is both public and immutable. Blocks of transactions are linked using cryptographic hashes to create a linear and immutable chain. Because of how this system is set up, it is impossible to make illegal changes or spend money twice. Bitcoin, a digital currency that facilitates peer-to-peer transactions without a central authority, is the most well-known application of blockchain technology.

In contrast, distributed ledger technology is a catch-all phrase for any digital system in which numerous nodes work together to keep a single, authoritative record of all transactions. While blockchains do fall under the umbrella term of distributed ledger technology (DLT), not all DLTs are blockchains. There are DLTs that don't rely on a sequential chaining of blocks but instead employ different data structures and consensus procedures. The primary feature of DLTs is their decentralized nature, which improves security, fosters transparency, and minimizes the need for a centralized authority.

Blockchain and other forms of distributed ledger technologies rely heavily on consensus mechanisms. To keep the ledger honest and consistent, these algorithms make sure that every node in the network has the same information.

Table \ref{TBL:DLTBC} outlines the differences between Distributed Ledger Technology (DLT) and Blockchain technology.

\begin{table}[htbp]
	\caption{DLT Versus BC.}
\label{TBL:DLTBC}
	\centering
	\begin{tabular}{|p{0.15\linewidth}|p{0.40\linewidth}|p{0.35\linewidth}|}
		\hline
			\textbf{Attribute}          & \textbf{Distributed Ledger Technology (DLT)}	                         & \textbf{Blockchain Technology}\\
			\hline
			Data Structure              & Various structures, such as directed acyclic graphs (DAGs)             & Linear chain of blocks linked with hashes\\
			
			\hline
			Consensus Mechanisms        & Multiple options, e.g., PBFT, DPoS, Federated Consensus                & Commonly PoW or PoS \\
			
			\hline
			Degree of Decentralization  & Varies; can include partially decentralized models                     &  Typically highly decentralized   \\
			
			\hline
			Scalability \& Performance   &  Potentially higher throughput and performance                        & Can face scalability challenges \\
			
			\hline
			Privacy \& Confidentiality  &  More flexible; can support permissioned or private channels           & Transparent and publicly accessible ledgers      \\
			
			\hline
			Use Cases \& Applications   &  Broad range of industries, e.g., supply chain, healthcare, governance & Often associated with cryptocurrencies \& DeFi     \\
			
			\hline
	\end{tabular}
\end{table}

\subsection{Distributed Ledger Technology Versus Blockchain Technology}
\label{sec:DistributedLedgerTechnologyVersusBlockchainTechnology}

DLT and the blockchain are related. Both systems use a distributed network of nodes to manage a shared digital ledger, but they differ significantly. The main differences between blockchain and distributed ledger technologies are as follows:

\begin{enumerate}
	\item Data structures distinguish distributed ledger and blockchain technologies. Blockchain uses cryptographic hashed blocks added consecutively to a chain to record and verify financial transactions. This approach makes transactions public and immutable. DLTs can leverage data structures like DAGs instead of just chaining blocks.
	
	\item Distributed ledger technology and blockchain require consensus methods, although their implementations vary. Blockchain consensus methods include PoW and PoS. DLTs can use Practical Byzantine Fault Tolerance (PBFT), Delegated Proof of Stake (DPoS), and Federated Consensus.
	
	\item DLT and blockchain are decentralized, although to different degrees. In hybrid decentralized/centralized distributed ledger technology, only a fraction of nodes are trusted to validate transactions and update the ledger. In a blockchain network, any node can contribute to the validation process.
	
	\item DLTs' data architecture and consensus processes may outperform blockchains in scalability and performance. DLTs based on DAGs or other data structures can run transactions concurrently to boost transaction throughput. Blockchain systems' linear block structure and resource-intensive consensus mechanisms like PoW make scaling and optimization difficult.
	
	\item DLTs may protect user data better than public blockchains. Permissioned DLTs with private channels keep transaction details private by limiting access to the parties involved. Public blockchains provide transaction records.
	
	\item DLT and blockchain have the potential to disrupt numerous industries, but their distinctions make them more suited to certain use cases. Blockchain technology is often used for cryptocurrencies and DeFi (Decentralized Finance). DLTs can be customized for supply chain management, healthcare, government, and governance because to their flexible data formats and consensus methods.
	
	\item DLT and blockchain both use distributed networks to store shared digital ledgers, but they differ in data structures, consensus mechanisms, degree of decentralization, scalability, privacy, and application space. Understanding these distinctions is crucial for using the best technology in every context.
\end{enumerate}

\subsection{DLT Components}

DLT is an umbrella term that encompasses multiple technologies that enable secure, transparent, and decentralized record-keeping across a network of participants. To comprehend DLT from a top-down perspective, it is necessary to deconstruct and build upon its fundamental components. A step-by-step operational analysis of DLT, beginning with its smallest component and progressing to its most obvious parts follows:

\textbf{Cryptography}: The use of cryptographic algorithms to assure data security, integrity, and authentication is the foundation of DLT. Hashing, digital signatures, and public-private key cryptography are crucial for protecting data and facilitating communication between network participants without requiring trust.

\textbf{Transactions:} Transactions are the fundamental data element in a distributed ledger. These operations include value transfers, record updates, and the execution of smart contracts. The creators of transactions digitally sign them and distribute them to the network for validation and processing.

\textbf{Consensus Mechanisms:} DLT systems rely on consensus algorithms to maintain a secure and consistent ledger, permitting network participants to concur on the validity of transactions. Proof of Work (PoW), Proof of Stake (PoS), and Byzantine Fault Tolerance (BFT) are popular consensus mechanisms. Each mechanism has tradeoffs in areas such as security, efficiency, and resource usage.

\textbf{Ledger Structure:} The ledger in a DLT is a data structure that records transactions and maintains an exhaustive, verifiable record of all network activities. The ledger structure can take numerous forms, such as a linear blockchain such as Bitcoin, and Ethereum or a Directed Acyclic Graph (DAG) such as IOTA, depending on the specific DLT implementation.

\textbf{Nodes \& Network Architecture:} In a DLT, nodes (computers) validate transactions, store the ledger, and communicate with one another in order to maintain the ledger. Full nodes (which store the entire ledger) and lightweight nodes (which store only a portion of the ledger) can have various roles and responsibilities. The architecture of the network is distributed and decentralized, with no singular point of failure or control.

\textbf{DLT Applications:} DLT has numerous applications across a wide range of industries and use cases, including finance (cryptocurrencies, remittance, asset tokenization), supply chain management (provenance tracking, inventory control), healthcare (secure data sharing, patient records), and identity management (digital identity, access control).

A top-down approach for understanding DLT entails analyzing its fundamental components, ranging from cryptographic methods and transactions to consensus mechanisms, ledger structures, network architecture, and prospective applications. This all-encompassing perspective clarifies the diverse layers and components that converge to form a secure, transparent, and decentralized system for record-keeping and data management.

\begin{figure}[htp]
	\centering
	\includegraphics[width=16.5cm]{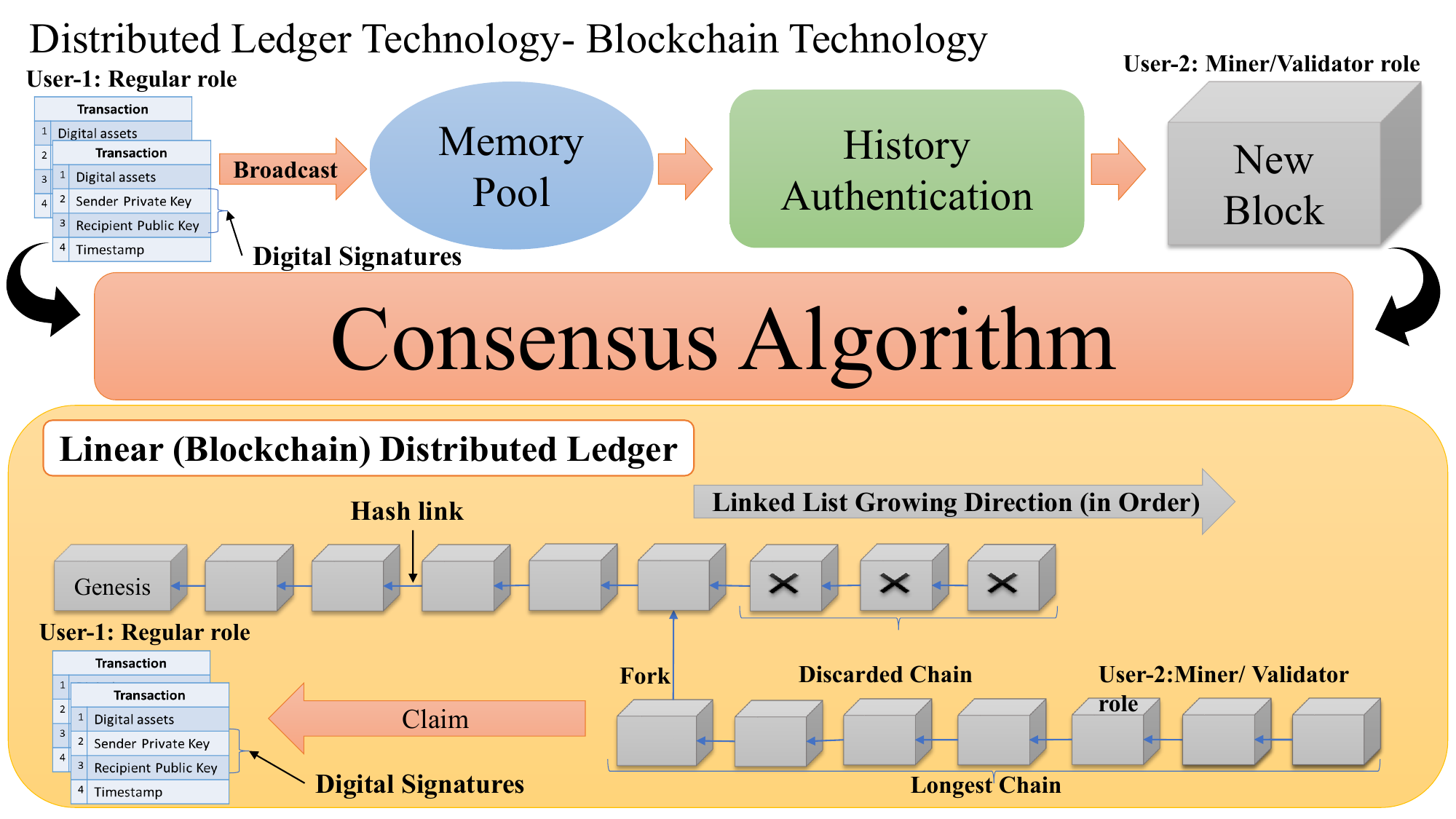}
	\caption{DLT-Blockchain Depiction.}
\label{FIG:DLTBlockchain}
\end{figure}

\begin{figure}[htp]
	\centering
	\includegraphics[width=16.5cm]{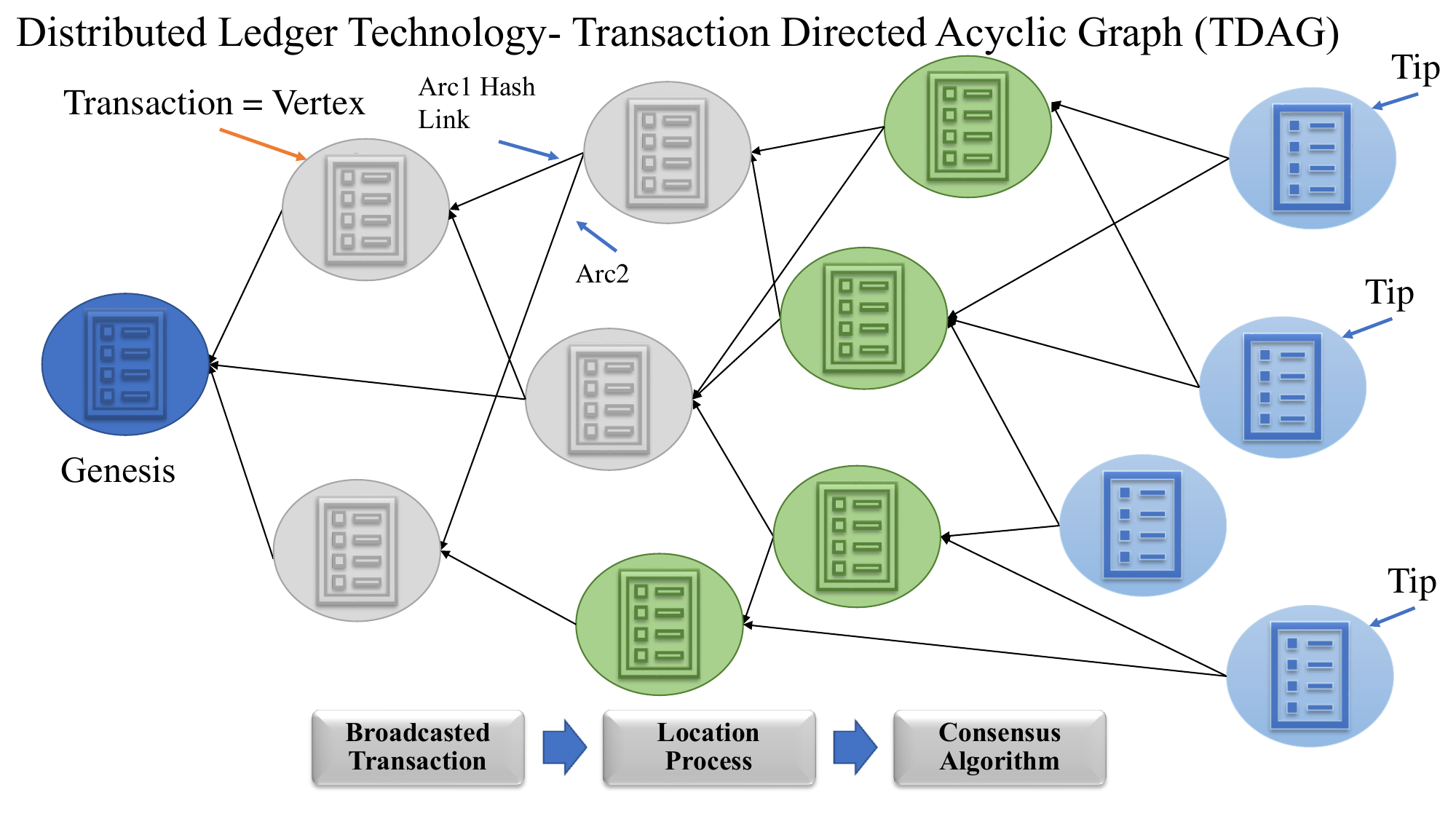}
	\caption{DLT-Transaction Directed Acyclic Graph (TDAG) Depiction.}
\label{FIG:DLTOthers1}
\end{figure}

\begin{figure}[htp]
	\centering
	\includegraphics[width=16.5cm]{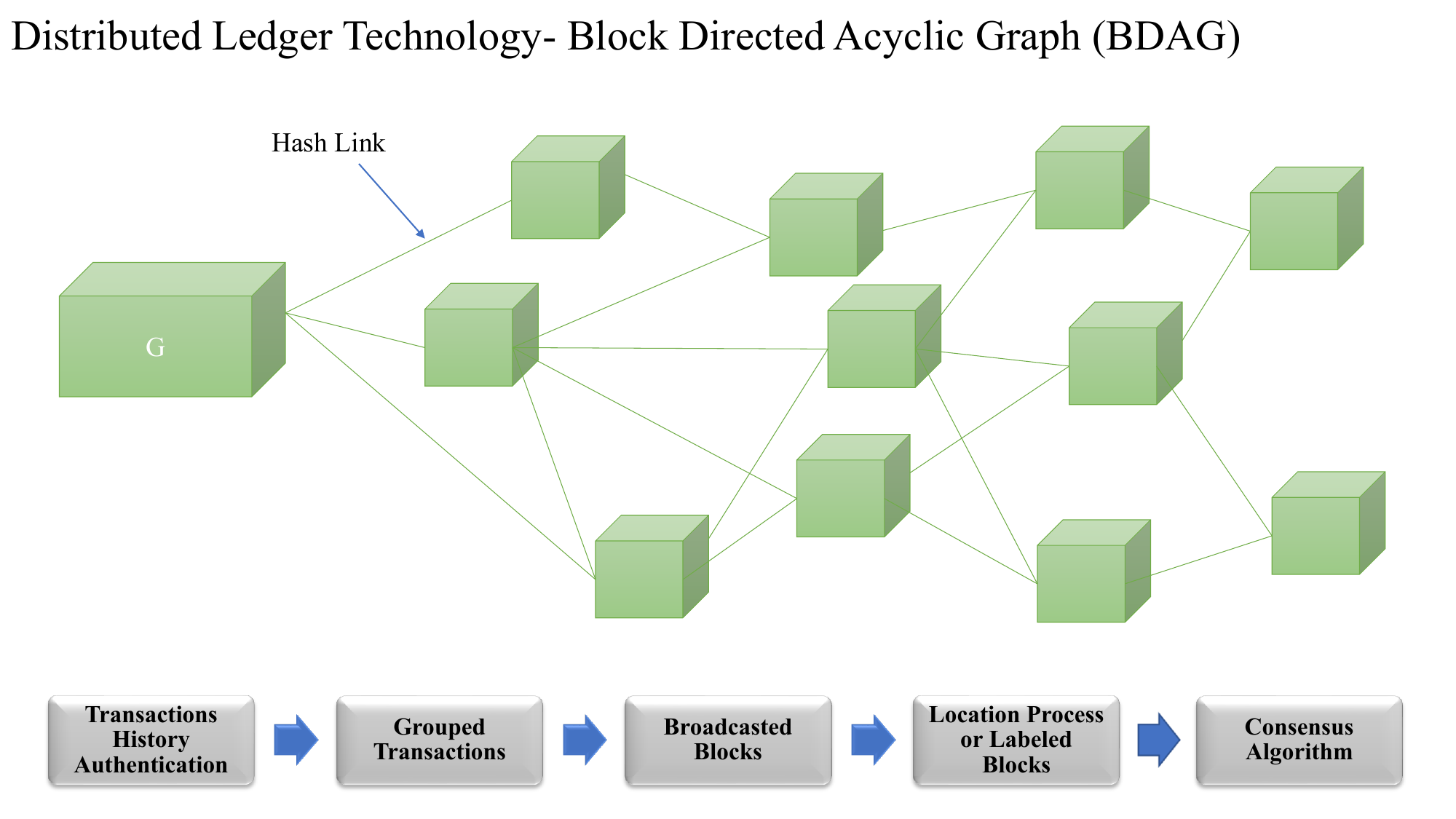}
	\caption{DLT-Block Directed Acyclic Graph (BDAG) Depiction.}
\label{FIG:DLTOthers2}
\end{figure}

\begin{figure}[htp]
	\centering
	\includegraphics[width=16.5cm]{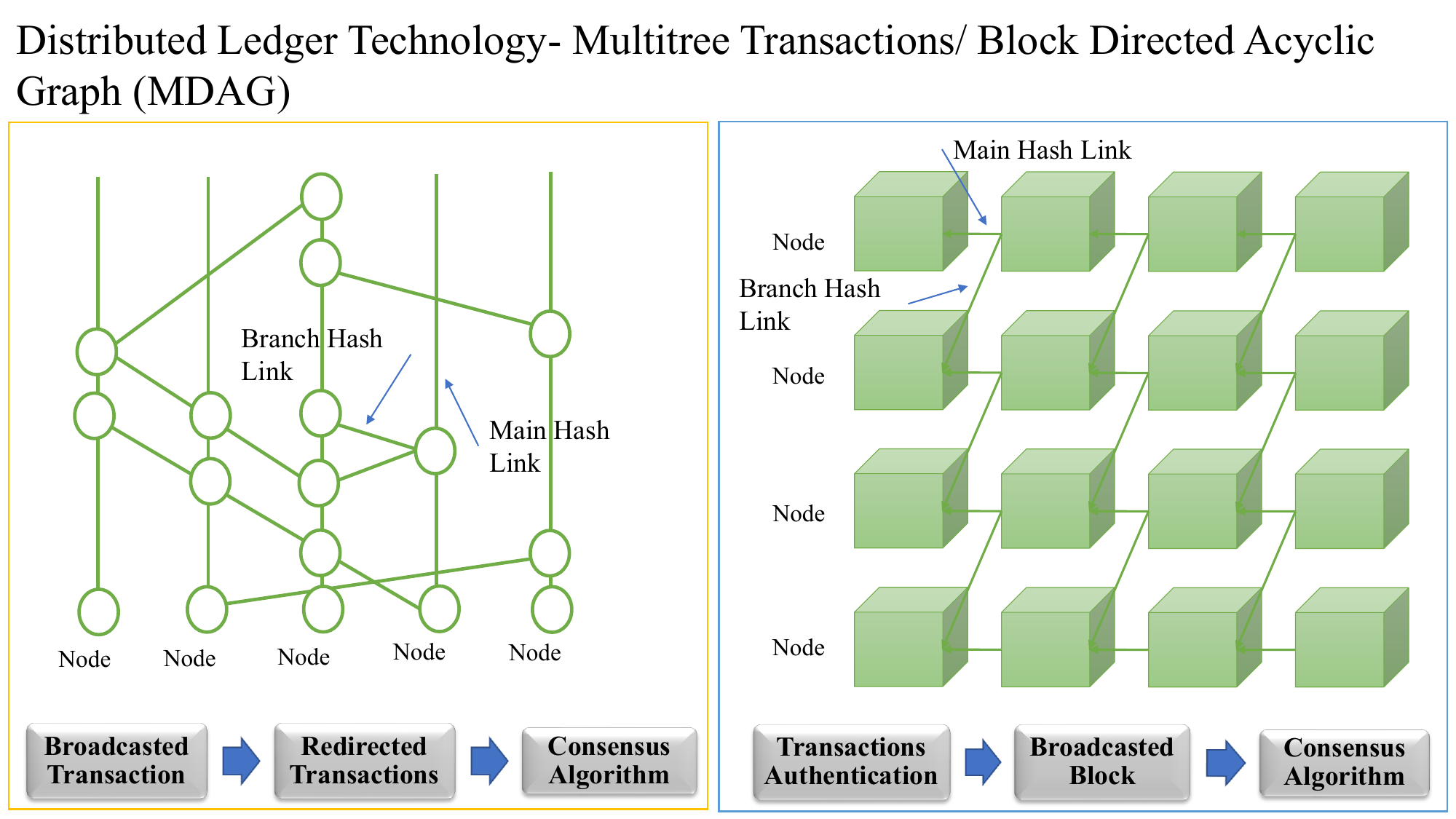}
	\caption{DLT-Multitree Directed Acyclic Graph (MDAG) Depiction Depiction.}
\label{FIG:DLTOthers3}
\end{figure}

Figures \ref{FIG:DLTBlockchain}, \ref{FIG:DLTOthers1}, \ref{FIG:DLTOthers2}, and \ref{FIG:DLTOthers3} illustrate BC and DLT components at a general high level description.  

\subsection{Consensus Operations}

\textbf{Mining:} Mining is typically associated with Proof of Work (PoW) consensus algorithms such as Bitcoin's and IOTA's networks. To add a new block of transactions to the blockchain, it is required to solve sophisticated cryptographic puzzles. The first miner to successfully solve these riddles is rewarded with newly created tokens (such as Bitcoins) and transaction fees. Mining contributes to the network's security and decentralization by making it computationally costly to initiate attacks or manipulate the blockchain. Similarly, To add a new transaction to DAG, it is required to solve a lightweight puzzles in IOTA.  

\textbf{Validation:} Validation is the procedure of confirming the accuracy of blocks and transactions in a DLT network. Validators verify that transactions conform to the network's regulations, including verifying digital signatures, ensuring inputs and outputs are consistent, and preventing double-spending. Once validators have determined that a transaction is legitimate, it is added to the blockchain via a new block. Selecting validators and deciding which transactions to include in a block are performed differently by distinct consensus algorithms.

\textbf{Voting:} Some consensus algorithms, such as Delegated Proof of Stake (DPoS) and Practical Byzantine Fault Tolerance (PBFT), use voting to reach consensus on the network's state. In these algorithms, a group of validators or delegates are selected, and they deliberate on the legitimacy of transactions and the order in which they should be added to the blockchain. The network reaches consensus when the majority of validators (or a predefined threshold) concur on the same state. Voting helps preserve the network's decentralized nature, provides fault tolerance, and ensures that no singular entity controls the network.

\textbf{Authentication:} Authentication in DLT is the procedure of confirming the identity of network participants. This is accomplished using cryptographic techniques, such as public-key cryptography, in which each participant possesses a pair of keys—a private key and a public key. The private key is utilized to sign transactions, whereas the public key is utilized to verify the signature. This enables the network to confirm that transactions are legitimate and authorized by the rightful proprietor of the digital assets or accounts involved. In certain DLT systems, such as permissioned blockchains or private DLTs, authentication also entails verifying that a participant is authorized to access the network, typically through the use of additional identity management systems or access control mechanisms.

\subsection{Understanding Democracy, Distributed, and Decentralized Concepts in Blockchain and Distributed Ledger Technology}
In the context of blockchain technology or distributed ledger technology (DLT), the terms democracy, distributed, and decentralized are often used to describe various aspects of the technology's design and governance. Here's an explanation of each term and how they relate to blockchain and DLT:

\textbf{Democracy:} This term refers to the decision-making process within a blockchain network, which can be either centralized or decentralized. In a democratic blockchain, decisions are made collectively by the network's participants, often through voting mechanisms or consensus protocols. This ensures that no single entity has control over the network and its development, fostering a fair and transparent system. Examples of democratic governance models in blockchain include Decentralized Autonomous Organizations (DAOs) and on-chain governance.

\textbf{Distributed:} A distributed system refers to a network that is spread across multiple computers or nodes, which work together to maintain and update the shared ledger. In a distributed ledger, each node maintains a copy of the entire ledger, which is constantly updated and synchronized with the rest of the network. This distribution of responsibility provides increased resilience against failures or attacks, as there is no single point of failure. Both blockchain and other forms of DLT are distributed systems by design.

\textbf{Decentralized:} Decentralization is a key characteristic of blockchain and DLT, which eliminates the need for a central authority to control and maintain the ledger. In a decentralized system, control is dispersed among the network participants, with no single entity having the power to make unilateral decisions. Decentralization enhances the security, transparency, and censorship-resistance of the system. It also helps to prevent single points of failure and reduces the risk of fraud or manipulation.

The terms democracy, distributed, and decentralized are all important aspects of blockchain technology and DLT. They describe the decision-making processes, network architecture, and control mechanisms that underpin these innovative systems, ensuring their security, transparency, and reliability.

\begin{figure}[htp]
	\centering
	\includegraphics[width=16.5cm]{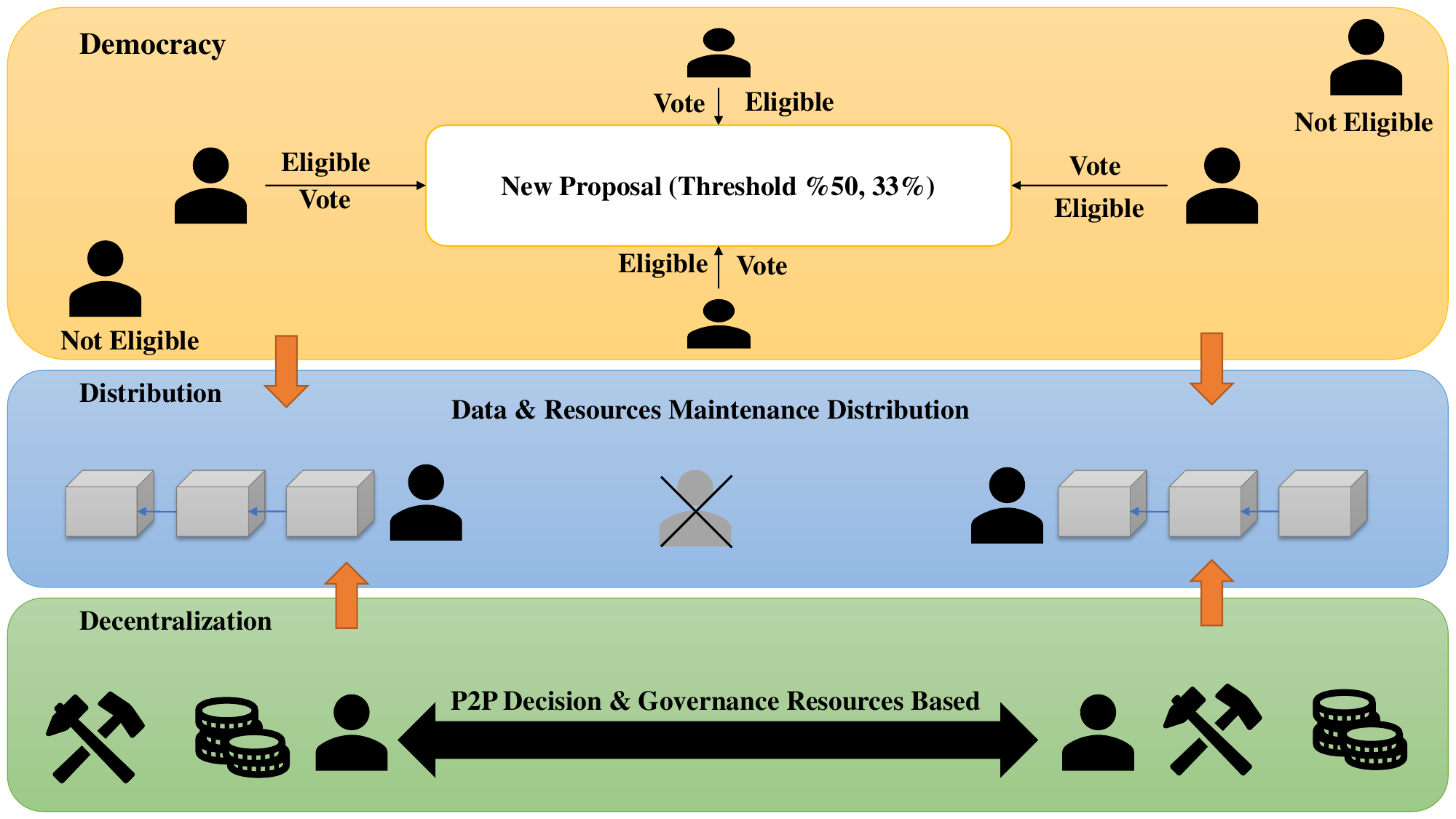}
	\caption{Democracy, Distribution, Decentralization Concepts.}
\label{FIG:DemocracyDistributionDecentralization}
\end{figure}

Figure \ref{FIG:DemocracyDistributionDecentralization} depicts the concepts of Democracy, Distribution, and Decentralization in the context of DLT. 

\subsection{DLT Classifications}
DLT can be classified into various types based on their structure, consensus mechanisms, access control, and use cases. Here is a general classification scheme:

\subsection{Structure-based Classification}

\begin{figure}[htp]
	\centering
	\includegraphics[width=16.5cm]{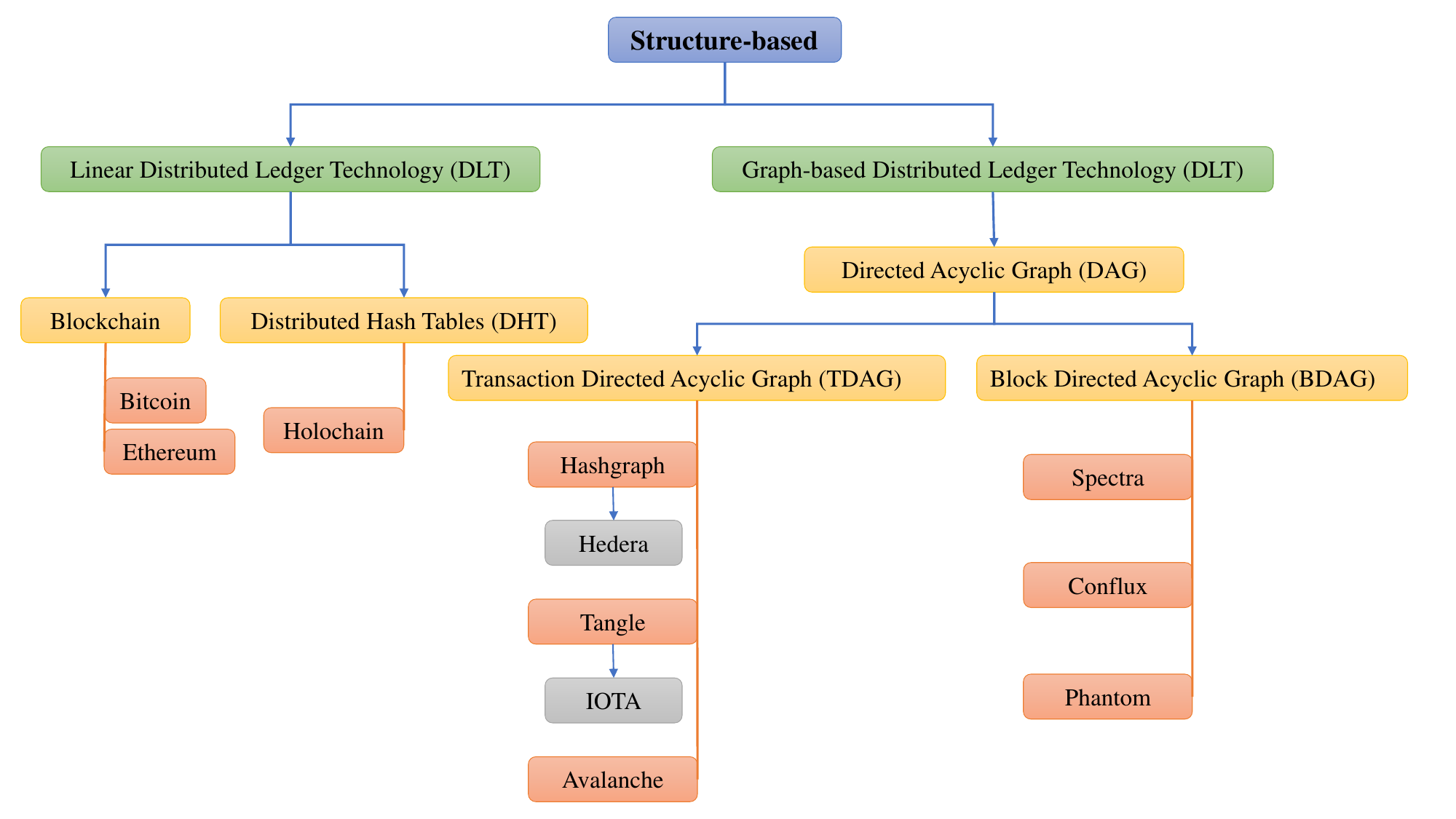}
	\caption{Structure-Based Classification.}
\label{FIG:DLTClassificationStructureBased}
\end{figure}   

\subsubsection{Linear DLT} Ledgers with a linear data structure, such as Blockchains (BCs), where data is organized into blocks and linked sequentially. Examples include Bitcoin and Ethereum. Distributed Hash Tables (DHTs), where data is organized into transactions and linked sequentially. An example is Holochain (Gossip Protocol).

\subsubsection{Graph-based DLT} Ledgers with a graph data structure, such as Directed Acyclic Graphs (DAGs), where data is organized into nodes or vertices and linked in a non-linear, non-circular manner. Examples include IOTA's Tangle and Hedera Hashgraph.

Figure \ref{FIG:DLTClassificationStructureBased} presents the most common structures used in DLT. 

\subsection{Consensus Mechanism-based Classification}

\begin{figure}[htp]
	\centering
	\includegraphics[width=16.5cm]{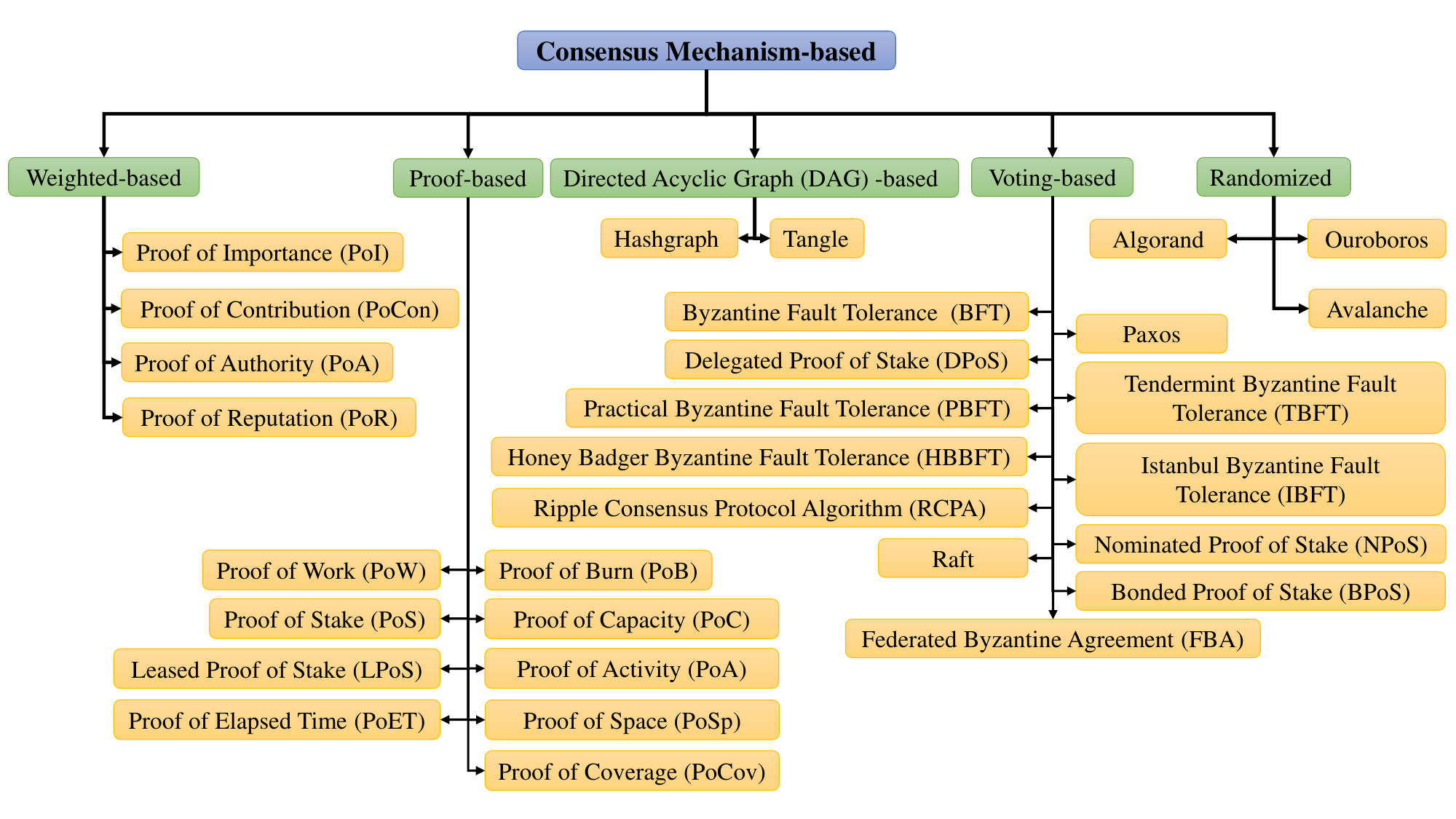}
	\caption{Consensus-Based Classification.}
\label{FIG:DLTClassificationConsensusBased}
\end{figure}   

\subsubsection{Proof-based}

Proof-based consensus algorithms are a category of consensus mechanisms that rely on participants providing some form of proof or evidence to gain the right to create new blocks or validate transactions. The objective of proof-based algorithms is to prevent malicious activities, such as double-spending or Sybil attacks, and ensure the security and reliability of the network. Proof-based algorithms require participants to demonstrate they have made a commitment or invested resources to be eligible to perform specific tasks in the system.

Such schemes include Proof of Work (PoW), Proof of Stake (PoS), Leased Proof of Stake (LPoS), Proof of Elapsed Time (PoET), Proof of Burn (PoB), Proof of Capacity (PoC), Proof of Activity (PoA), Proof of Space (PoSpace), and Proof of Coverage (PoC).

\subsubsection{Voting-based}

Voting-based consensus algorithms are a category of consensus mechanisms in which the participants in a distributed system reach an agreement through a process that involves exchanging messages, proposing blocks, and voting on the validity of those blocks. The objective of voting-based algorithms is to achieve consensus in a decentralized manner, despite the possibility of faults, delays, or malicious actions from some participants. In voting-based consensus algorithms, nodes typically follow a set of predefined rules to propose, validate, and vote on transactions or blocks. The proposed blocks or transactions are considered valid and added to the blockchain or distributed ledger when a sufficient number of votes (usually a supermajority) have been collected, indicating that the network agrees on the state of the system.

Examples are Delegated Proof of Stake (DPoS), Byzantine Fault Tolerance (BFT), Practical Byzantine Fault Tolerance (PBFT), Federated Byzantine Agreement (FBA), Ripple Consensus Protocol Algorithm (RCPA), Raft, Paxos, Honey Badger BFT, Tendermint BFT, Istanbul BFT, Nominated Proof of Stake (NPoS), and Bonded Proof of Stake (BPoS). 

\subsubsection{Weight-based}

Weighted-based consensus algorithms belong to a category of consensus mechanisms in which the influence or decision-making power of the participants in a distributed system is determined by a specific attributes or weights associated with them. This weight can be derived from various factors, such as the amount of cryptocurrency held, the reputation of the participant, or their contribution to the network. The objective of weight-based algorithms is to ensure that the consensus process is fair and resistant to manipulation, while also incentivizing desirable behavior among participants. In weight-based consensus algorithms, the likelihood of a participant being selected to create a new block, validate transactions, or have a higher voting power is proportional to their weight in the system. This mechanism helps to prevent attacks or manipulation by making it more difficult for malicious participants to control the network.

Such weight-based schemes include Proof of Authority (PoA), Proof of Importance (PoI), Proof of Reputation (PoR), and Proof of Contribution (PoC).

\subsubsection{Randomized}

Randomized consensus algorithms are a category of consensus mechanisms in which the selection of participants responsible for creating new blocks or validating transactions is based on a random process. These algorithms aim to provide a fair and unbiased mechanism for participant selection, ensuring the security and decentralization of the network. Randomized consensus algorithms often incorporate cryptographic techniques, such as verifiable random functions (VRFs) or cryptographic sortition, to generate random numbers or sequences that determine the selection of block proposers or validators. The randomization process helps to prevent manipulation and collusion among participants, as the chances of being selected are unpredictable and not directly influenced by factors such as computational power, stake, or reputation.

Examples include Randomized PoS (Ouroboros), Randomized Pure PoS (PPoS) (Algorand), and Avalanche.

\subsubsection{DAG-based}

DAG-based (Directed Acyclic Graph) consensus algorithms are a category of consensus mechanisms that use a graph-based data structure instead of a linear chain of blocks, as seen in traditional blockchain systems. In a DAG-based system, transactions or data are represented as vertices in the graph, and the edges represent the relationship between them. This data structure allows for multiple transactions to be linked and processed concurrently, enabling higher scalability and faster transaction throughput. DAG-based consensus algorithms aim to overcome some of the limitations of traditional blockchain systems, such as the need for global consensus on a single chain and the associated inefficiencies that arise from this requirement. In DAG-based systems, local consensus can be achieved, and the overall network can maintain its integrity without requiring every node to agree on the state of the entire ledger.

Tangle (PoW), and Hashgraph (BFT) (Gossip about Gossip Protocol) are common examples.

Figure \ref{FIG:DLTClassificationConsensusBased} presents the most common consensus algorithms used in DLT. 

\subsection{Access Control-based Classification}

\textbf{Public DLT:} Open, permissionless networks where anyone can join, participate, and validate transactions. Examples include Bitcoin and Ethereum.
\textbf{Private DLT:} Restricted, permissioned networks where access is granted only to authorized participants, often used for enterprise solutions. Examples include Hyperledger Fabric and R3 Corda.
\textbf{Consortium DLT:} Semi-private networks where a group of organizations jointly governs and controls the network. Examples include Quorum and B3i \cite{blockchainAccess1} \cite{blockchainAccess} \cite{blockchainAccess}.

\begin{figure}[htp]
	\centering
	\includegraphics[width=16.5cm,trim=0cm 10cm 0cm 0cm]{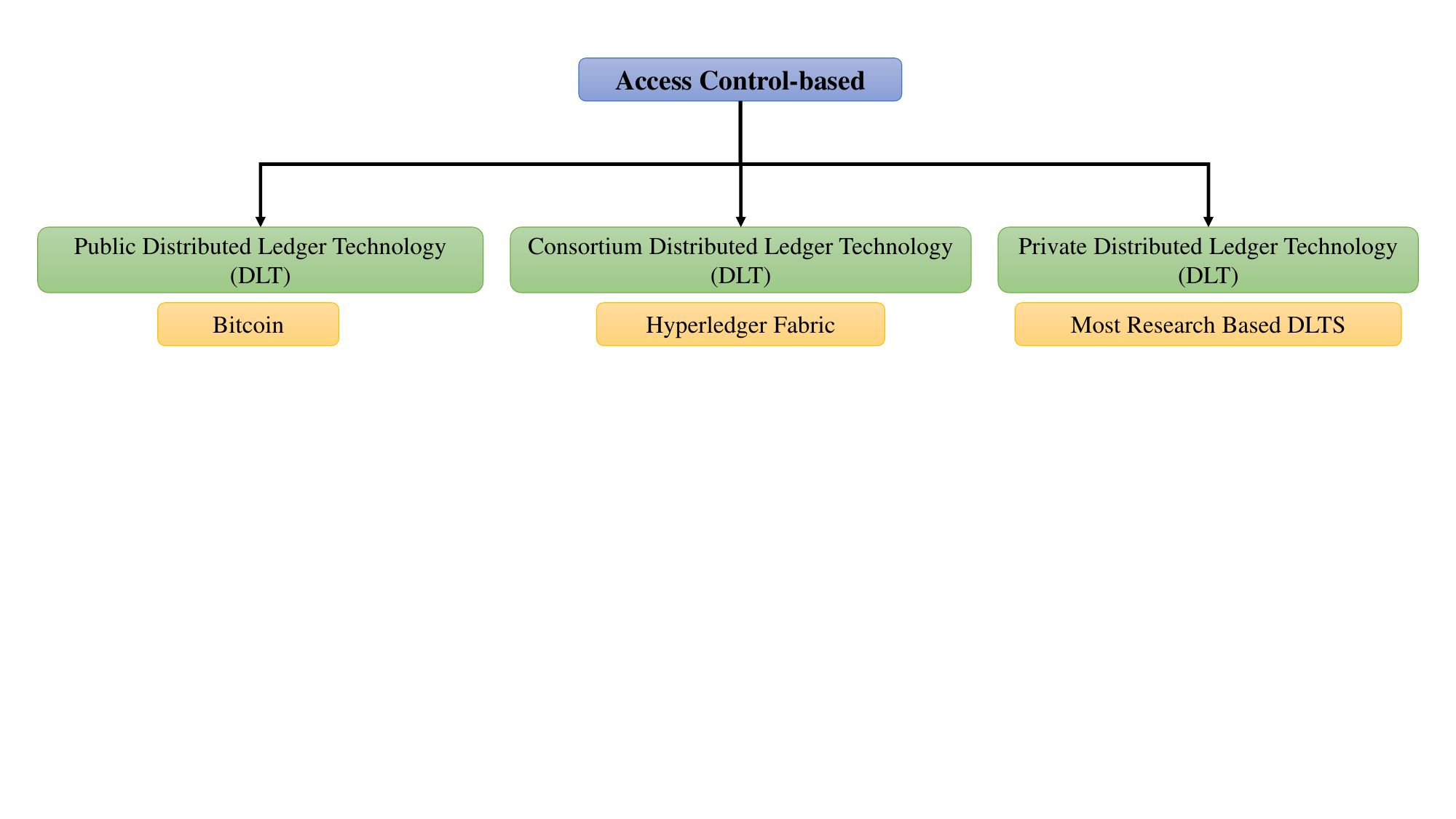}
	\caption{Access-Based Classification.}
	\label{FIG:DLTClassificationsAccessBased}
\end{figure}

Figure \ref{FIG:DLTClassificationsAccessBased} list the types of access control in DLT. 

\subsection{Use Case-based Classification}

\textbf{Cryptocurrency:} DLTs primarily focused on facilitating cryptocurrency transactions. Examples include Bitcoin, Litecoin, and Monero.

\textbf{Smart Contract:} DLTs designed to enable the deployment and execution of smart contracts. Examples include Ethereum, Cardano, and Tezos.

\textbf{Data Storage:} DLTs designed for decentralized data storage and sharing. Examples include Filecoin, Storj, and Sia.

\textbf{Identity and Authentication:} DLTs built for managing digital identities and authentication. Examples include Sovrin and Civic. Presentes in \cite{PoAh1} and \cite{PoAh2} a novel consensus based in private network operates and distributed on predefined trusted edge nodes.

\textbf{Supply Chain Management:} DLTs tailored for improving supply chain transparency and traceability. Examples include VeChain and Waltonchain. In \cite{PharmaChain}, as an example of Decentralized Supply Chain Management (DSCM) in research, the authors used a particular consensus algorithm (PoA) which is known for its suitability to SCM. 

\textbf{Smart Healthcare:} A decentralized, distributed, intelligent healthcare system eliminates the sole point of failure and third-party control of healthcare data. This means that users have greater control over their medical data and can rest easy knowing that it is secure and confidential. As n  research example  \cite{FortifiedChain2} addresses the problem of single points of failure (SPoF) in healthcare systems by proposing a decentralized distributed smart healthcare system that eliminates the SPoF and third-party control over healthcare data.

These classifications provide a high-level overview of different DLT types and can help in understanding the various aspects of these technologies. Some DLTs may fall under multiple categories, depending on their specific features and use cases.

\begin{figure}[htp]
	\centering
	\includegraphics[width=16.5cm,trim=0cm 8cm 0cm 0cm]{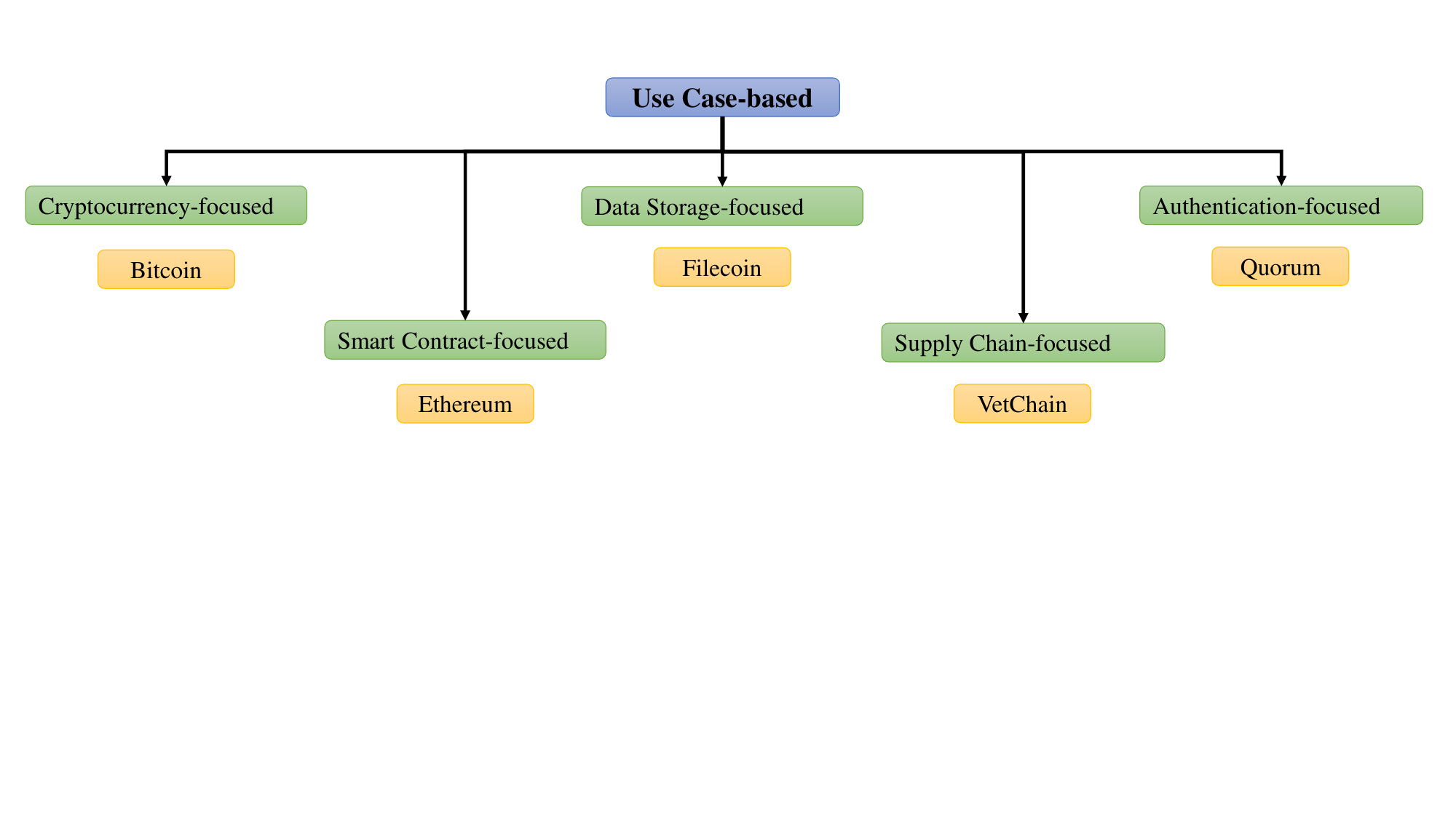}
	\caption{Use-Based Classification.}
\label{FIG:DLTClassificationsUseBased}
\end{figure}

Figure \ref{FIG:DLTClassificationsUseBased} presents use case classification based on established networks. 

\begin{figure}[htp]
	\includegraphics[width=16.5cm]{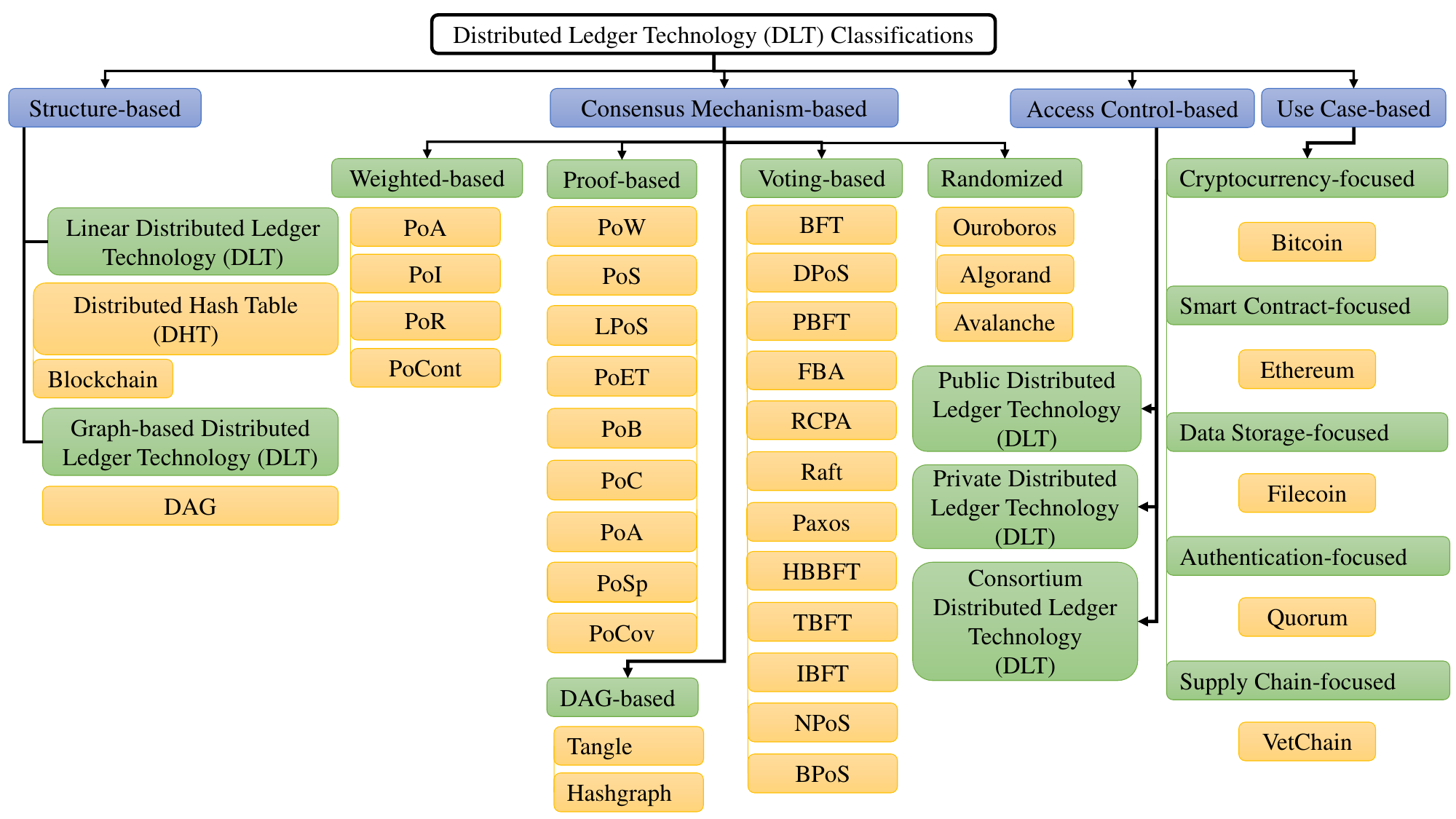}
	\caption{DLT Classifications.}
\label{FIG:DLTClassificatios}
\end{figure}

Figure.\ref{FIG:DLTClassificatios} illustrates a high level classification of DLT in general.  

\section{General Overview for Consensus Algorithms}
\label{sec:GeneralOverviewforConsensusAlgorithms}

There are many consensus algorithms in use, with new ones being developed as the field of distributed ledger technologies evolves. While it is challenging to list all consensus algorithms in the market, some of the most common and notable ones are listed in Table \ref{TBL:ConsensusAlgorithms} based on their primary and secondary categories. 

\begin{table}[htbp]
	\caption{Classification of Consensus Algorithms\label{TBL:ConsensusAlgorithms}}
	\centering
	\begin{tabular}{|p{0.15\linewidth}|p{0.30\linewidth}|p{0.25\linewidth}|p{0.20\linewidth}|}
		\hline
			\textbf{Algorithm}                          & \textbf{Primary Category} & \textbf{Secondary Category}  & \textbf{Examples} \\
			\hline
			Proof of Work (PoW)                         & Proof-based               &      NA                           & Bitcoin \\
			Proof of Stake (PoS)                        & Proof-based               &      NA                           & Ethereum 2.0  \\
			Delegated Proof of Stake (DPoS)             & Voting-based              & Proof-based                       & EOS, Lisk, Tron \\
			Leased Proof of Stake (LPoS)                & Proof-based               & Voting-based                      & Waves \\
			Nominated Proof of Stake (NPoS)             & Voting-based              &      NA                           & Polkadot \\
			Bonded Proof of Stake (BPoS)                & Voting-based              &      NA                           & Waves \\
			Proof of Authority (PoA)                    & Weighted-based            &      NA                           & VeChain\\
			Proof of Contribution (PoCon)               & Weighted-based            & Proof-based (PoW\&PoS elements)   & \\
			Byzantine Fault Tolerance (BFT)             & Voting-based              &       NA                          &  \\
			Practical BFT (PBFT)                        & Voting-based              &       NA                          & Hyperledger Fabric \\
			Federated Byzantine Agreement (FBA)         & Voting-based              &       NA                          & Stellar  \\
			Ripple Consensus Protocol Algorithm (RCPA)  & Voting-based              &       NA                          & XRP Ledger \\
			Raft Consensus Algorithm                    & Voting-based              &       NA                          & \\
			Paxos                                       & Voting-based              &       NA                          & \\
			Honey Badger BFT                            & Voting-based              &       NA                          & \\
			Avalanche Consensus                         & Voting-based              & DAG-Based (PoW elements)          & Avalanche platform \\
			Proof of Elapsed Time (PoET)                & Proof-based               &       NA                          & Intel's Sawtooth \\
			Proof of Burn (PoB)                         & Proof-based               & Voting-based                      & Slimcoin \\
			Proof of Importance (PoI)                   & Weighted-based            &       NA                          & NEM \\
			Proof of Capacity (PoC)                     & Proof-based               &       NA                          & Burstcoin \\
			Proof of Activity (PoA)                     & Proof-based               &       NA                          & Decred, Espers \\
			Proof of Space (PoSpace)                    & Proof-based               &       NA                          & Chia Network \\
			Tendermint BFT                              & Voting-based              & Proof-based (PoS elements)        & Cosmos Network \\
			Istanbul BFT                                & Voting-based              &       NA                          & Quorum \\
			Proof of Reputation (PoR)                   & Weighted-based            &       NA                          & Idena  \\
			Proof of Coverage (PoC)                     & Proof-based               &       NA                          & Helium Network \\
			Algorand                                    & Randomized                & Proof-Based (Pure PoS)            & Algorand platform\\
			Ouroboros                                   & Randomized                & Proof-based (PoS elements)        & Cardano \\
			Tangle                                      & DAG-based                 & Proof-based (Lightweight PoW)     & IOTA\\
			Hashgraph                                   & DAG-based                 & Voting-based (Original BFT)       & Hedera \\
			\hline
	\end{tabular}
\end{table}

\subsection{Proof of Work (PoW)}

\begin{figure}[htp]
	\includegraphics[width=16.5cm,trim=0cm 12cm 0cm 0cm]{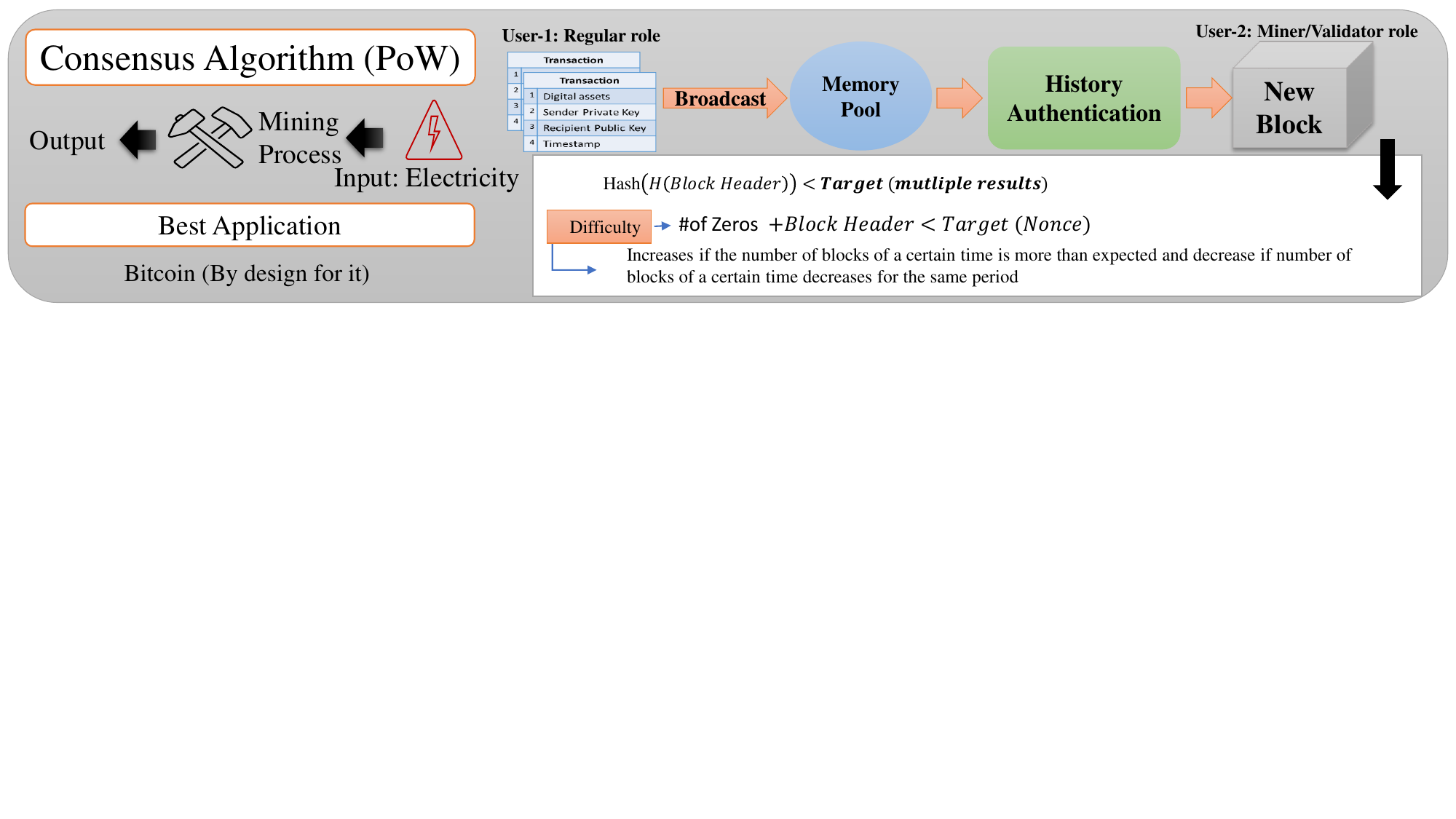}
	\caption{PoW Operations \label{FIG:PoW}.}
\end{figure}

Proof of Work (PoW), is a consensus technique utilized by numerous blockchain networks, including Bitcoin and Ethereum 1.0. The PoW algorithm enables nodes, known as miners, to compete in solving challenging cryptographic problems as shown in figure \ref{FIG:PoW} in order to validate transactions, generate new blocks, and secure the blockchain network \cite{PoWEvaluationofPerformanceandSecurityofProofofWorkandProofofStakeusingBlockchain}. The primary goal of PoW is to discourage malicious conduct by performing energy- and resource-consuming computationally intensive activities \cite{PoWATemplateforAlternativeProofofWorkforCryptocurrencies}.

The PoW consensus algorithm operates as follows:

\begin{enumerate}
	\item Transaction Pool: Transactions are broadcast by users to the network. These transactions are then accumulated in a transaction pool, awaiting inclusion by miners in a subsequent block \cite{PoWOntheSecurityandPerformanceofProofofWorkBlockchains}.
	\item Block Candidate: A miner selects a group of unconfirmed transactions from the transaction pool, prioritizing those with the highest transaction fees. The miner then generates a block candidate, which consists of a block header containing essential data, such as the preceding block's hash, the Merkle tree root of the selected transactions, a timestamp, and a nonce \cite{PoWAquicklookatCryptocurrencyMiningProofofWork}.
	\item Hash Function: Miners create a hash value for the block header using a cryptographic hash function, such as SHA-256 in the case of Bitcoin. A hash function accepts an input of arbitrary length and returns a fixed-length, seemingly random output (hash). Hash functions possess three fundamental properties: determinism, resistance to preimages, and collision resistance \cite{PoWHashcash}.
	\item Proof of Work Puzzle: The PoW puzzle requires miners to identify a nonce, a number appended to the block header, such that the block header's resulting hash is less than or equal to a target value. The target value is determined by the network's current difficulty level, which regularly changes to maintain a constant block generation pace (e.g., approximately every 10 minutes for Bitcoin).
	\item Mining: Miners continuously modify the nonce and recompute the hash of the block header until they find a valid solution that satisfies the desired condition. This procedure needs a considerable amount of energy and resources and is computationally intensive. The first miner to discover a correct solution may add the new block to the network.
	\item Block Validation and Propagation: After discovering a valid solution, miners propagate the new block to the network. Other network nodes validate the block to ensure that the PoW solution is correct and that the transactions are genuine. If the block is genuine, it is added to the local copy of the blockchain, and nodes begin processing the next block using the hash of the newly added block as the reference in the new block header.
	\item Block Reward: The miner who successfully adds a new block to the blockchain is rewarded with newly minted cryptocurrency (e.g., Bitcoin) and the total transaction fees from the included transactions. This payment acts as an incentive for miners to participate in the Proof-of-Work protocol and safeguard the network \cite{PoWSAnalysisonBlockchainConsensusMechanismBasedonProofofWorkandProofofStake}.
\end{enumerate}

The most successful and widely recognized application of Proof of Work (PoW) consensus algorithm is Bitcoin because it has been designed for it. Rules embedded in the system such as longest chain and 10 minutes average per a block is to keep it sustainable and reliable until all 21 million coins are mined. The mechanism advantages and disadvantages are listed in table \ref{tab:pow}.

\begin{table}[htbp]
	\centering
	\caption{Advantages and Disadvantages of PoW.}
\label{tab:pow}
	\begin{tabular}{|p{0.45\linewidth}|p{0.45\linewidth}|}
		\hline
		\textbf{Advantages} & \textbf{Disadvantages} \\ \hline
		Security                        & 1. Energy consumption             \\ \hline
		Decentralization                & 2. Centralization risk            \\ \hline
		Sybil resistance                & 3. Hardware requirements          \\ \hline
		Proven track record             & 4. Inefficient resource allocation\\ \hline
		Block appending timing          & 5. Limited throughput             \\ \hline
	\end{tabular}
\end{table}

\subsection{Lightweight PoW}

Tangle consensus is unique to IOTA's Directed Acyclic Graph (DAG) data structure. This data structure lacks blocks and miners, unlike blockchains. Instead, transactions are interconnected in a web-like form. The consensus is a lightweight PoW that requires a participant to validate two previous transactions in order to add a new one \cite{tangle}. The mechanism advantages and disadvantages are listed in table \ref{tab:tangle}.

\begin{table}[htbp]
	\centering
	\caption{Advantages and Disadvantages of Tangle.}
\label{tab:tangle}
	\begin{tabular}{|p{0.45\linewidth}|p{0.45\linewidth}|}
		\hline
		\textbf{Advantages} & \textbf{Disadvantages} \\ \hline
		
		Scalability & Network maturity \\ \hline
		No transaction fees & Coordinator reliance \\ \hline
		Decentralization & Double-spending resistance concerns \\ \hline
		Energy efficiency & Lower adoption \\ \hline
		Adaptability & Complexity \\ \hline
	\end{tabular}
\end{table}

\subsection{Proof of Stake (PoS) Family}

Proof of Stake (PoS): A more energy-efficient alternative to PoW, PoS is designed to address some of the limitations of PoW, such as energy consumption and centralization risks. PoS relies on a participant's stake in the network, usually measured by the amount of cryptocurrency they hold or have locked up, to determine their chances of creating a new block and validating transactions. Many blockchain implementations, including Ethereum, have added smart contracts to the Bitcoin model. These smart contracts are self-executing and may be programmed to do a wide range of tasks, opening the door to new possibilities for automation and simplification of business operations \cite{EvaluationofPerformanceandSecurityofProofofWorkandProofofStakeusingBlockchain}.
The Proof of Stake consensus algorithm, depicted in figure \ref{FIG:PoS}, operates as follows:

\begin{figure}[htp]
	\includegraphics[width=16.5cm,trim=0cm 12cm 0cm 0cm]{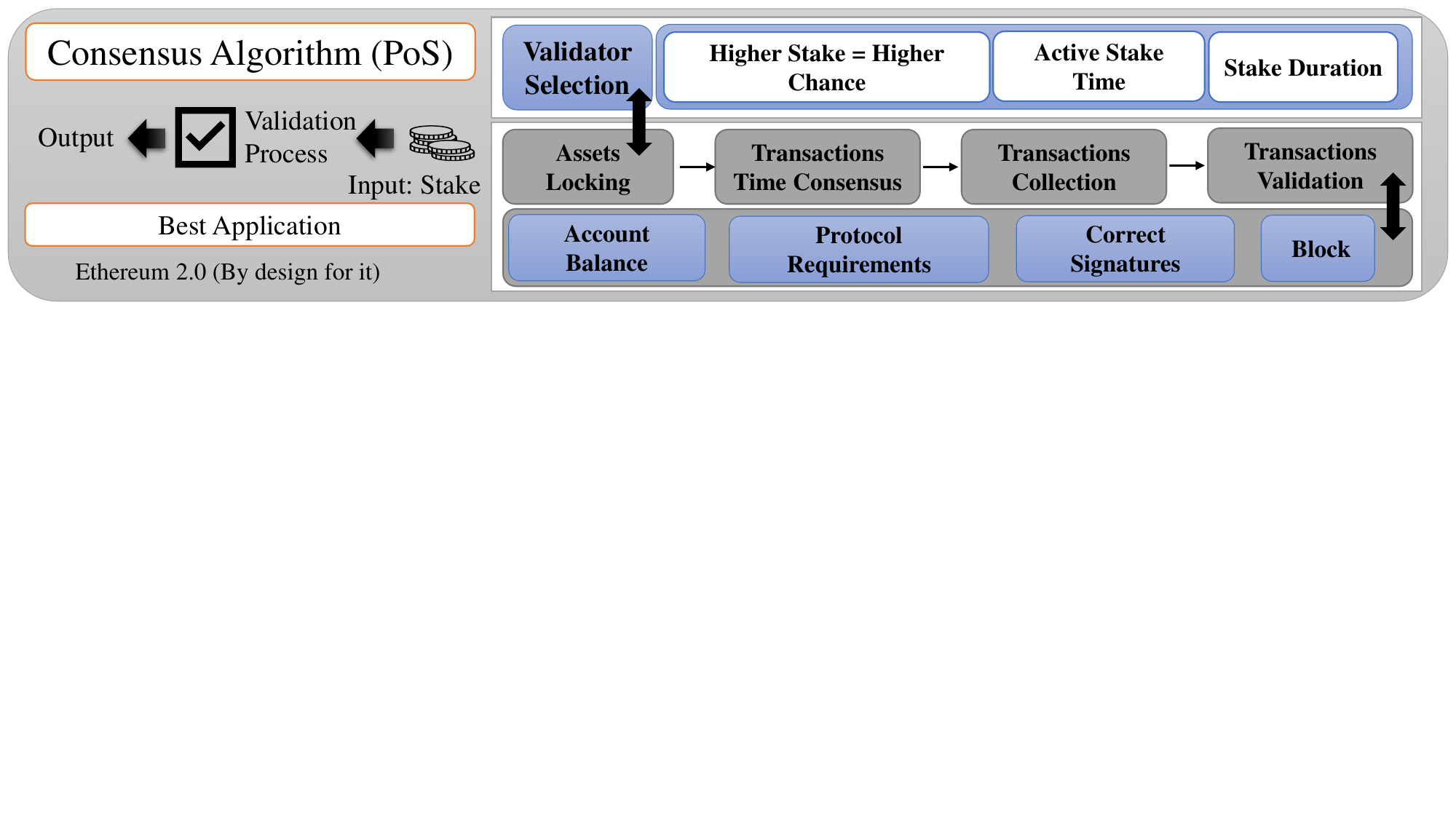}
	\caption{PoS Operations.}
\label{FIG:PoS}
\end{figure}   

\begin{enumerate}
	\item Validators and Staking: Unlike PoW, where miners compete to solve cryptographic puzzles, PoS networks rely on validators to create new blocks and validate transactions. Validators are chosen based on the amount of cryptocurrency they hold and are willing to ``stake'' or lock up as collateral. Staking typically involves a validator depositing their cryptocurrency in a special wallet or smart contract, which is then locked for a predetermined period \cite{EvaluationofPerformanceandSecurityofProofofWorkandProofofStakeusingBlockchain}.
	\item Selection Algorithm: Validators are chosen to create new blocks and validate transactions through a selection algorithm, which varies among PoS implementations. Common selection methods include randomization, coin age-based selection, and delegated proof of stake (DPoS), where token holders vote for validators \cite{ComparativeAnalysisoftheBlockchainConsensusAlgorithmBetweenProofofStakeandDelegatedProofofStake}.
	\item Block Creation and Validation: The selected validator creates a new block containing a set of unconfirmed transactions and broadcasts it to the network. Other validators in the network check the block's validity, ensuring that the transactions are valid and the creating validator had the right to generate the block. If the block is valid, it is added to the blockchain, and the process continues with the selection of the next validator.
	\item Block Rewards and Transaction Fees: In PoS networks, validators are usually rewarded with transaction fees from the transactions included in the block, rather than receiving newly minted cryptocurrency as in PoW systems. Some PoS implementations, however, may still offer block rewards in addition to transaction fees.
	\item Slashing and Penalties: PoS networks often incorporate mechanisms to penalize validators who act maliciously or fail to validate transactions properly. This process, known as slashing, can result in the loss of a portion or all of the staked cryptocurrency, which serves as a strong deterrent against malicious behavior.
	\item Security and Attacks: PoS networks are designed to be secure against attacks, such as 51\% attacks and long-range attacks. The cost of acquiring a controlling stake in a PoS network is typically much higher than the potential rewards, making it economically irrational for an attacker to attempt to manipulate the network.
	\item Energy Efficiency: PoS networks are generally more energy-efficient than PoW networks, as they do not require participants to expend large amounts of computational resources and energy to solve cryptographic puzzles. This makes PoS networks more environmentally friendly and cost-effective. 
\end{enumerate}

Various PoS implementations exist in the market, with Ethereum 2.0 (currently transitioning from PoW to PoS), Cardano, and Tezos being some notable examples. Each implementation may have its unique features and variations of the PoS algorithm, but the core principles remain the same. The mechanism advantages and disadvantages are listed in Table \ref{tab:pos}.

\begin{table}[htbp]
	\centering
	\caption{Advantages and Disadvantages of PoS.}
\label{tab:pos}
	\begin{tabular}{|p{0.45\linewidth}|p{0.45\linewidth}|}
		\hline
		\textbf{Advantages} & \textbf{Disadvantages} \\ \hline
		1. Energy Efficiency                & 1. Initial Distribution            \\ \hline
		2. Decentralization                 & 2. "Rich Get Richer" Effect        \\ \hline
		3. Security                         & 3. Lack of Proven Track Record     \\ \hline
		4. Inflation Control                & 4. Complexity                      \\ \hline
		5. Lower Entry Barrier              & 5. Centralization Risk             \\ \hline
	\end{tabular}
\end{table}

\subsubsection{Delegated Proof of Stake (DPoS)}

Delegated Proof of Stake (DPoS) is a consensus algorithm that builds upon the concepts of Proof of Stake (PoS), aiming to provide a more efficient, scalable, and democratic method of achieving consensus in a blockchain network. DPoS introduces a voting mechanism for selecting validators and focuses on a smaller number of elected delegates, which can result in faster transaction processing and increased accountability. DPoS was introduced by Daniel Larimer in 2014 as part of the BitShares blockchain platform \cite{DPoSandroulaki2018hyperledger}. The mechanism advantages and disadvantages are listed in Table \ref{tab:dpos}.

\begin{table}[htbp]
	\centering
	\caption{Advantages and Disadvantages of DPoS.}
\label{tab:dpos}
	\begin{tabular}{|p{0.45\linewidth}|p{0.45\linewidth}|}
		\hline
		\multicolumn{1}{|c|}{\textbf{Advantages}} & \multicolumn{1}{c|}{\textbf{Disadvantages}} \\ \hline
		1. Scalability and Efficiency       & 1. Centralization Risk            \\ \hline
		2. Energy Efficiency                & 2. Voting Apathy and Voter Influence  \\ \hline
		3. Democracy and Accountability     & 3. Bribery and Corruption         \\ \hline
		4. Incentive to Act Honestly        & 4. Cartel Formation               \\ \hline
		5. Lower Entry Barrier              & 5. Complexity                     \\ \hline
	\end{tabular}
\end{table}

\subsubsection{Leased Proof of Stake (LPoS)}

Leased Proof of Stake (LPoS) is a consensus mechanism that extends the concept of Proof of Stake (PoS) to allow users to lease their stakes to other network participants. This enables small stakeholders to participate in the staking process and earn rewards, even if they don't possess enough tokens to become validators themselves. LPoS was first implemented by the Waves Platform \cite{BlockchainAidedEdgeComputingMarketSmartContractandConsensusMechanisms}. The mechanism advantages and disadvantages are listed in Table \ref{tab:lpos}.

\begin{table}[htbp]
	\centering
	\caption{Advantages and Disadvantages of LPoS.}
\label{tab:lpos}
	\begin{tabular}{|p{0.45\linewidth}|p{0.45\linewidth}|}
		\hline
		\multicolumn{1}{|c|}{\textbf{Advantages}} & \multicolumn{1}{c|}{\textbf{Disadvantages}} \\ \hline
		1. Inclusivity                      & 1. Complexity                      \\ \hline
		2. Scalability and Efficiency       & 2. Trust in Validators             \\ \hline
		3. Security                         & 3. Centralization Risk             \\ \hline
		4. Reward Distribution              & 4. Reduced Staking Control         \\ \hline
		5. Enhanced Decentralization        & 5. Liquidity Risk                   \\ \hline
	\end{tabular}
\end{table}

\subsubsection{Pure Proof-of-Stake (Algorand)}

(PPoS) Pure Proof of Stake. This technique is different from traditional Proof of Stake (PoS) methods in that it ensures a high level of decentralization, security, and speed by choosing validators for block proposal and agreement in a random and secret way \cite{algorand}. The mechanism advantages and disadvantages are listed in Table \ref{tab:ppos}.

\begin{table}[htbp]
	\centering
	\caption{Advantages and Disadvantages of PPoS.}
\label{tab:ppos}
	\begin{tabular}{|p{0.45\linewidth}|p{0.45\linewidth}|}
		\hline
		\textbf{Advantages} & \textbf{Disadvantages} \\ \hline
		Scalability & Wealth Concentration \\ \hline
		Decentralization & Initial Distribution \\ \hline
		Energy Efficiency & Network Maturity \\ \hline
		Security & Potential for Centralization \\ \hline
		Inclusivity & Barrier to Entry \\ \hline
	\end{tabular}
\end{table}

\subsubsection{Ouroboros}

Ouroboros is a Proof of Stake (PoS) consensus mechanism. The protocol is made to keep decentralization while providing high levels of security, scalability, and energy economy. The procedure picks a committee of stakeholders at random for each slot, and the committee members decide together which block goes next in the chain. A random seed, which is changed at the end of each epoch, is used to choose the committee members \cite{ouroboros2017whitepaper}. The mechanism advantages and disadvantages are listed in Table \ref{tab:ouroboros}.

\begin{table}[htbp]
	\centering
	\caption{Advantages and Disadvantages of the Ouroboros Consensus Mechanism.}
\label{tab:ouroboros}
	\begin{tabular}{|p{0.45\linewidth}|p{0.45\linewidth}|}
		\hline
		\textbf{Advantages} & \textbf{Disadvantages} \\ \hline
		Energy Efficiency & Complexity \\ \hline
		Security & Potential centralization risk \\ \hline
		Decentralization & Initial bootstrapping \\ \hline
		Scalability & Network connectivity \\ \hline
		Incentivization & Adaptation \\ \hline
	\end{tabular}
\end{table}

\subsubsection{Nominated Proof of Stake (NPoS)}

Nominated Proof of Stake (NPoS) is a variant of the Proof of Stake (PoS) consensus algorithm, with some unique features that differentiate it from other PoS-based mechanisms. It was introduced by the Polkadot network, a scalable and interoperable blockchain platform. NPoS is designed to provide a secure, scalable, and efficient consensus mechanism for blockchain networks. It is particularly suitable for platforms like Polkadot, which focus on interoperability and cross-chain communication \cite{polkadot2016whitepaper}. The mechanism advantages and disadvantages are listed in Table \ref{tab:npos}.

\begin{table}[htbp]
	\centering
	\caption{Advantages and Disadvantages of the NPoS Consensus Mechanism.}
\label{tab:npos}
	\begin{tabular}{|p{0.45\linewidth}|p{0.45\linewidth}|}
		\hline
		\textbf{Advantages} & \textbf{Disadvantages} \\ \hline
		Decentralization & Complexity \\ \hline
		Security & Potential centralization of power \\ \hline
		Scalability & Slashing risks \\ \hline
		Inclusive participation & Inflation \\ \hline
		Incentive alignment & Barrier to entry \\ \hline
	\end{tabular}
\end{table}

\subsubsection{Bonded Proof of Stake (BPoS)}

Bonded Proof of Stake (BPoS) is a variation of the Proof of Stake (PoS) consensus algorithm, in which validators are required to lock up or bond a certain amount of their tokens as collateral. This bond acts as a security deposit, incentivizing validators to act honestly and discouraging malicious behavior. If a validator is found to be acting maliciously, their bonded tokens can be confiscated or slashed as a penalty. By requiring validators to have a stake in the network, BPoS aims to ensure the network's security and trustworthiness \cite{BPoS}. The mechanism advantages and disadvantages are listed in Table \ref{tab:bpos}.

\begin{table}[htbp]
	\centering
	\caption{Advantages and Disadvantages of the BPoS Consensus Mechanism.}
\label{tab:bpos}
	\begin{tabular}{|p{0.45\linewidth}|p{0.45\linewidth}|}
		\hline
		\textbf{Advantages} & \textbf{Disadvantages} \\ \hline
		Enhanced Security & Centralization Risk \\ \hline
		Energy Efficiency & Barrier to Entry \\ \hline
		Incentivized Participation & Slashing Penalties \\ \hline
		Sybil Attack Resistance & Limited Token Circulation \\ \hline
	\end{tabular}
\end{table}

Table \ref{TBL:PoSFamily} compares all Proof of Stake (PoS) variants in term of their operation and highlights the differences.

\begin{table}[htbp]
\caption{Comparison of PoS Family In term of Operations \cite{ComparativeAnalysisoftheBlockchainConsensusAlgorithmBetweenProofofStakeandDelegatedProofofStake}.}
\label{TBL:PoSFamily}
	\centering
	\begin{tabular}{|p{0.10\linewidth}|p{0.18\linewidth}|p{0.18\linewidth}|p{0.20\linewidth}|p{0.18\linewidth}|}
		\hline
		\textbf{Differences} & Participation and Inclusivity & Validator Selection & Reward Distribution & 	Network Decentralization \\
		\hline
        \textbf{PoS} \cite{DevelopmentinConsensusProtocolsFromPoWtoPoStoDPoS} & Limited to major stakeholders & Based on stake and selection algorithm & Based on stake & Larger validator set\\
		\hline
        \textbf{DPoS} \cite{Comparativeanalysisofblockchainconsensusalgorithms} & Token holders can vote for validators & Voting system to elect delegates & Elected delegates receive rewards & Smaller number of elected delegates\\
		\hline
        \textbf{LPoS} \cite{BlockchainAidedEdgeComputingMarketSmartContractandConsensusMechanisms} & Users can lease their stakes to validators or staking pools &  Similar to PoS, but includes leased stakes& Validators receive rewards based on total stake and distribute among leasing users & Similar to PoS, larger validator set\\
		\hline
        \textbf{PPoS} & All token holders can participate & Cryptographic sortition & Proportional to stake & High decentralization\\
		\hline
		\textbf{Ouroboros} & All token holders can participate & Based on stake and randomization & Proportional to stake & High decentralization\\
		\hline 
		\textbf{NPoS} & Nominated validators & Nominated by token holders & Proportional to stake & Decentralization based on nominations \\
		\hline
		\textbf{BPoS} & Validators post bonds & Validators with bonded stakes & Proportional to bonded stake & Decentralization based on bonded stakes\\
		\hline
	\end{tabular}
\end{table}


\subsection{Byzantine Fault Tolerance (BFT) Family}

The original Byzantine Fault Tolerance (BFT) algorithm was introduced by Castro and Liskov in their 1999 paper ``Practical Byzantine Fault Tolerance''. The algorithm was designed to provide a robust and secure consensus mechanism for distributed systems in the presence of malicious or faulty nodes \cite{OBFTzheng2017overview}. 
The original BFT algorithm operates in three phases: request, pre-prepare, and commit. The steps of each phase are as follows \cite{OBFTcastro1999practical, OBFTvukolic2015quest}:

\begin{enumerate}
	\item Request Phase: A client sends a request to the primary node, which is responsible for proposing a set of transactions. The request includes a sequence number and a unique client ID.
	
	\item Pre-Prepare Phase: The primary node receives the request from the client and assigns a sequence number. It then creates a message that includes the sequence number, the proposed transaction set, and a digest of the previous message. The primary node sends this message to the backup nodes for validation.
	
	\item Prepare Phase: Upon receiving the message from the primary node, the backup nodes validate the message by checking the sequence number, the digest, and the proposed transaction set. If the message is valid, the backup nodes send a ``prepare'' message to all other nodes in the network, indicating their agreement with the proposed transaction set.
	
	\item Commit Phase: Once the backup nodes have reached a two-thirds majority agreement, they send a ``commit'' message to all other nodes in the network, indicating that the proposed transaction set has been approved. At this point, the transaction set is added to the ledger and can be queried by other nodes in the network.
\end{enumerate}

One of the key features of the original BFT algorithm is that it requires a two-thirds majority of the backup nodes to agree on a proposed transaction set before it can be approved. This ensures that the network remains secure even if up to one-third of the nodes are faulty or malicious. Another important aspect of the original BFT algorithm is fault detection and recovery. BFT algorithms include mechanisms for detecting Byzantine faults, such as timeouts or replica divergence detection. If a fault is detected, the system can take corrective action, such as removing the faulty node or resynchronizing the state of the system.

The original BFT algorithm is a robust and secure consensus mechanism that enables distributed systems to reach agreement on a set of transactions, even in the presence of malicious or faulty nodes. Its technical operations include client request, pre-prepare, prepare, and commit phases, as well as fault detection and recovery mechanisms \cite{OBFTmao2018state} \cite{OBFTvukolic2015quest}. Its advantages and disadvantages are listed in Table \ref{tab:bft}.

\begin{table}[htbp]
	\centering
\caption{Advantages \& Disadvantages of BFT Algorithm \cite{LeopardTowardsHighThroughputPreservingBFTforLargescaleSystems}.}
\label{tab:bft}
	\begin{tabular}{|p{0.45\linewidth}|p{0.45\linewidth}|}
		\hline
		\textbf{Advantages} \cite{OBFTzheng2017overview}& \textbf{Disadvantages} \cite{OBFTzheng2017overview}.\\ \hline
		Robustness & Complexity \\ \hline
		Security & Resource-intensive \\ \hline
		Speed & Limited scalability \\ \hline
		Fault tolerance & Single point of failure (primary node vulnerability) \\ \hline
	\end{tabular}
\end{table}

\subsubsection{Practical Byzantine Fault Tolerance (PBFT)}

\begin{figure}[htp]
	\includegraphics[width=16.0cm]{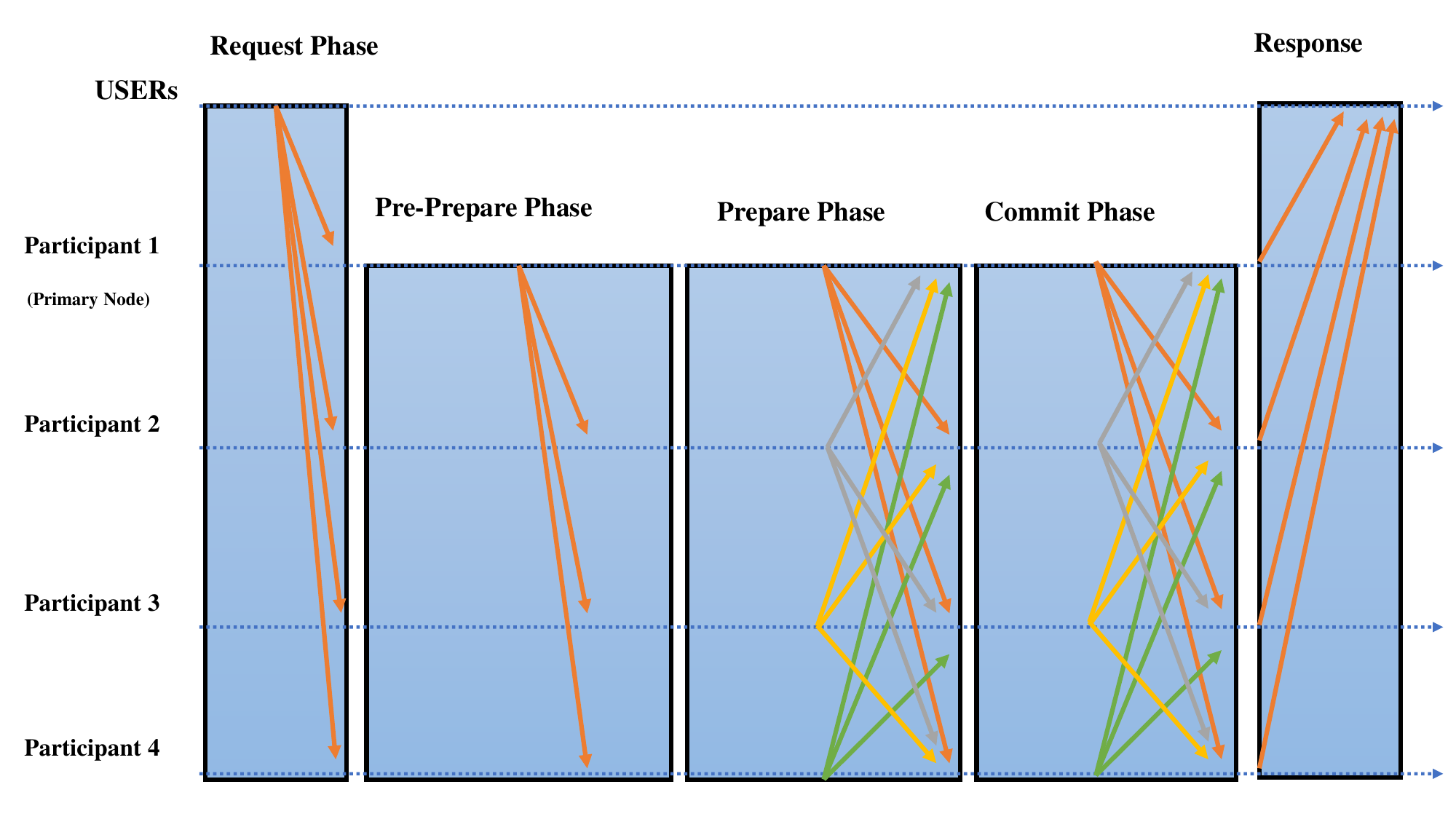}
	\caption{PBFT Operations.}
\label{FIG:PBFT}
\end{figure}

Practical Byzantine Fault Tolerance (PBFT) is a consensus algorithm that extends the original Byzantine Fault Tolerance (BFT) algorithm to improve scalability and reduce communication overhead. PBFT is designed to be used in permissioned blockchain networks, where the network participants are known and trusted \cite{practicalBFT}. The advantages and disadvantages are listed in Table \ref{tab:pbft}. Figure \ref{FIG:PBFT} illustrates the operation of PBFT as a major variant of BFT and the most common one among the family.

\begin{table}[htbp]
	\centering
	\caption{Advantages \& Disadvantages of PBFT Algorithm.}
	\label{tab:pbft}
	\begin{tabular}{|p{0.45\linewidth}|p{0.45\linewidth}|}
		\hline
		\textbf{Advantages} & \textbf{Disadvantages} \\ \hline
		High Security and Reliability & Implementation Complexity \\ \hline
		Fast Processing & Resource Intensive \\ \hline
		Scalability & Limited Decentralization \\ \hline
		Well-suited for Permissioned Networks & Centralization \\ \hline
	\end{tabular}
\end{table}

\subsubsection{Federated Byzantine Agreement (FBA)}

Federated Byzantine Agreement (FBA) is variation of BFT. It involves a network of nodes that form quorum slices to reach consensus. Stellar uses a version of FBA called the Stellar Consensus Protocol (SCP). FBA is a consensus mechanism that operates differently from traditional Byzantine Fault Tolerance (BFT) algorithms. Instead of requiring every node to communicate with every other node, FBA allows nodes to choose their quorum slices, leading to a more decentralized and efficient consensus process \cite{FBA}. The advantages and disadvantages are listed in Table \ref{tab:fba}.

\begin{table}[htbp]
	\centering
	\caption{Advantages and Disadvantages of FBA Algorithm.}
\label{tab:fba}
	\begin{tabular}{|p{0.45\textwidth}|p{0.45\textwidth}|}
		\hline
		\textbf{Advantages} & \textbf{Disadvantages} \\
		\hline
		Scalability: Offers better scalability due to reduced communication overhead. & Complexity: Can be complex to implement and understand. \\
		\hline
		Decentralization: Promotes decentralization by allowing nodes to choose their own quorum slices. & Potential Network Splits: Poorly chosen quorum slices can lead to network splits. \\
		\hline
		Asynchronous Operation: Resilient to network latency and functions efficiently in an asynchronous environment. & Slower Finality: May take longer to reach consensus depending on the network configuration and quorum slices. \\
		\hline
		Flexible Trust: Allows nodes to determine their own trust relationships for a more robust network. & Vulnerability to Sybil Attacks: More vulnerable if malicious nodes can create enough fake nodes to influence quorum slices. \\
		\hline
	\end{tabular}
\end{table}

\subsubsection{Istanbul BFT}

Istanbul Byzantine Fault Tolerance (IBFT) is a consensus algorithm designed for blockchain networks with permissioned access. Practical Byzantine Fault Tolerance (PBFT) is a variant that provides high security, fault tolerance, and rapid transaction processing. Nodes in IBFT use a leader-based methodology to propose and validate blocks. As long as fewer than one-third of the nodes are flawed or malicious, the algorithm assures the system's functionality. IBFT is ideally adapted for enterprise-level applications in which network participants are known and trusted \cite{IBFT} \cite{IBFT1}. The advantages and disadvantages are listed in Table \ref{tab:ibft}.

\begin{table}[htbp]
	\centering
	\caption{Advantages and Disadvantages of IBFT Algorithm.}
\label{tab:ibft}
	\begin{tabular}{|p{0.45\linewidth}|p{0.45\linewidth}|}
		\hline
		\textbf{Advantages} & \textbf{Disadvantages} \\
		\hline
		High Security and Reliability & Implementation Complexity \\
		\hline
		Fast Transaction Processing & Resource Intensive \\
		\hline
		Scalability & Limited Decentralization \\
		\hline
		Energy Efficiency & Centralization Risk \\
		\hline
		Finality & \\
		\hline
	\end{tabular}
\end{table}

\subsubsection{Honey Badger BFT}

Honey Badger Byzantine Fault Tolerance (HBBFT) is an asynchronous consensus technique that tolerates Byzantine errors without compromising efficiency, durability, or security. For distributed systems with unreliable networks, HBBFT can handle arbitrary message delays.

Using threshold cryptography, HBBFT nodes generate batches of encrypted transactions. Each node contributes transactions and decrypts the bulk without exposing individual transactions. Decryption submits the batch to the blockchain.

Honey Badger BFT's strengths are its ability to withstand up to one-third of nodes being defective or malicious, its tolerance to network latency, and its excellent security guarantees. It is used to build distributed systems and blockchain networks like Nym \cite{honeyBFT}. The advantages and disadvantages are listed in Table \ref{tab:hbbft}.

\begin{table}[htbp]
	\centering
	\caption{Advantages and Disadvantages of HBBFT Algorithm.}
\label{tab:hbbft}
	\begin{tabular}{|p{0.45\textwidth}|p{0.45\textwidth}|}
		\hline
		\textbf{Advantages} & \textbf{Disadvantages} \\
		\hline
		Asynchronous communication & Complexity \\
		\hline
		Robustness & Resource-intensive \\
		\hline
		Security & Limited decentralization \\
		\hline
		Adaptability & Communication overhead \\
		\hline
		High throughput & Less mature \\
		\hline
	\end{tabular}
\end{table}

\subsubsection{Ripple Consensus Protocol Algorithm (RCPA)}

Ripple Consensus Protocol Algorithm (RCPA) is the consensus mechanism used by the Ripple network (XRP Ledger) to validate transactions and maintain its distributed ledger. Ripple is a blockchain-based platform designed primarily for cross-border payments and real-time gross settlement. RCPA is a unique consensus mechanism that differs from Proof of Work (PoW), Proof of Stake (PoS), and other commonly known consensus algorithms. RCPA is a Byzantine fault-tolerant (BFT) consensus algorithm that uses a voting process among a group of validator nodes to agree on the state of the ledger \cite{ripple}. The advantages and disadvantages are listed in Table \ref{tab:rcpa}.

\begin{table}[htbp]
	\centering
	\caption{Advantages \& Disadvantages of RCPA Algorithm.}
\label{tab:rcpa}
	\begin{tabular}{|p{0.45\textwidth}|p{0.45\textwidth}|}
		\hline
		\textbf{Advantages} & \textbf{Disadvantages} \\
		\hline
		Fast transaction confirmations & Centralization \\
		\hline
		Energy efficiency & Trust in validators \\
		\hline
		Scalability & Limited decentralization benefits \\
		\hline
		Resistance to certain attacks & Control over validators \\
		\hline
		Lower transaction fees & \\
		\hline
	\end{tabular}
\end{table}

\subsection{Ignite (Tendermint)}

Tendermint is a Byzantine Fault-Tolerant (BFT) consensus algorithm and a blockchain engine that lets developers build decentralized applications in a safe and efficient manner. Tendermint uses a BFT mechanism called Tendermint Core that can be used in both private and public blockchain networks \cite{tendermint}. The advantages and disadvantages are listed in Table \ref{tab:ignite}.

\begin{table}[htbp]
	\centering
	\caption{Advantages and Disadvantages of Tendermint Algorithm.}
\label{tab:ignite}
	\begin{tabular}{|p{0.45\textwidth}|p{0.45\textwidth}|}
		\hline
		\textbf{Advantages} & \textbf{Disadvantages} \\ \hline
		Fast finality & Centralization concerns \\ \hline
		Scalability & Vulnerability to long-range attacks \\ \hline
		Energy efficiency & Validator collusion \\ \hline
		Byzantine Fault Tolerance & Initial validator setup \\ \hline
		Easier upgrades and governance & Staking requirements \\ \hline
	\end{tabular}
\end{table}

\subsection{Hashgraph}

Hashgraph is a distributed ledger technology (DLT) that uses a consensus method called ``Swirlds Consensus Algorithm.'' It is an alternative to standard blockchain networks that promises to be faster, safer, and more fair. Hashgraph uses a data format called a Directed Acyclic Graph (DAG), in which transactions are arranged in a graph instead of in a chain of blocks. Swirlds Consensus Algorithm is an Asynchronous Byzantine Fault Tolerance (aBFT) \cite{hashgraph}. The advantages and disadvantages are listed in Table \ref{tab:Hashgraph}.

\begin{table}[htbp]
	\centering
	\caption{Advantages and Disadvantages of Hashgraph.}
\label{tab:Hashgraph}
	\begin{tabular}{|p{0.45\linewidth}|p{0.45\linewidth}|}
		\hline
		\textbf{Advantages} & \textbf{Disadvantages} \\
		\hline
		High throughput & Centralization (patented technology) \\
		\hline
		Fairness & Licensing model \\
		\hline
		Asynchronous Byzantine Fault Tolerance (aBFT) & Complexity \\
		\hline
		No mining (energy-efficient) & Limited adoption \\
		\hline
	\end{tabular}
\end{table}

\subsection{Avalanche}
Avalanche is a consensus system that aims to provide high throughput, low latency, and strong security. It is made for different kinds of decentralized applications and works best on open networks that don't require permission. The Avalanche consensus is a new way to reach a decision. It is a leaderless, metastable, Byzantine Fault Tolerant protocol \cite{Avalanch}. The advantages and disadvantages are listed in Table \ref{tab:Avalanche}.

\begin{table}[htbp]
	\caption{Advantages \& Disadvantages of Avalanche.}
\label{tab:Avalanche}
	\centering
	\begin{tabular}{|p{0.45\linewidth}|p{0.45\linewidth}|}
    \hline
	\textbf{Advantages}    & \textbf{Disadvantages} \\ \hline 
    High Scalability       & Relatively New Technology \\ \hline
    Decentralization       & Unclear Long-term Stability \\ \hline
    Quick Finality         & Security Dependence on Network Size \\ \hline
    Energy Efficiency      & Complexity in Protocol Design \\ \hline
    Adaptive Consensus Mechanism & Adoption Challenges \\ \hline

	\end{tabular}
\end{table}

Table \ref{TBL:BFTFamily} compares BFT variants based on their operation and network complexity. 

\begin{table}[htbp]
	\caption{Comparison of BFT Family In Term of Operations.}
\label{TBL:BFTFamily}
	\centering
	\begin{tabular}{|p{0.10\linewidth}|p{0.15\linewidth}|p{0.15\linewidth}|p{0.15\linewidth}|p{0.15\linewidth}|p{0.10\linewidth}|}
		\hline
		\textbf{Differences} & Voting         & Network        & Users           & Complexity      & Threshold       \\
		\hline
		\textbf{BFT}    & Rounds of voting   & Sync. network  & All nodes       & O(n$^2$)        & Up to 1/3       \\
		\hline
		\textbf{PBFT}    & Voting phases  & Sync. network  & All nodes       & O(n$^2$)        & Up to 1/3       \\
		\hline
		\textbf{BFA}    & Quorum slices and intersecting quorums & Partially sync. network & Quorum slices of nodes & O(n)            & Configurable    \\
		\hline
		\textbf{IBFT}    & Round-robin leader selection & Sync. network  & All nodes       & O(n$^2$)        & Up to 1/3       \\
		\hline
		\textbf{HBBFT}    & Parallel proposals and reconstruction & Async. network & All nodes       & O(n$^2$)        & Up to 1/3       \\
		\hline
		\textbf{RCPA}    & Trusted validators propose and vote on transactions & Sync. or partially sync. network & Trusted validators & O(n)            & Configurable    \\
		\hline
		\textbf{Ignite}    & Propose and vote on blocks & Partially sync. network & Validators with voting power & O(n$^2$)        & Up to 1/3       \\
		\hline
		\textbf{Hashg.}   & Gossip-based voting & Async. network & All nodes       & O(n*log(n))     & Up to 1/3       \\
		\hline
		\textbf{Aval.}   & Chained consensus with multiple subnets & Partially sync. network & Validators with voting power & O(n)            & Up to 1/2       \\
		\hline
	\end{tabular}
\end{table}

\subsection{Crash Fault Tolerance (CFT) Family}

Crash Fault Tolerance (CFT) is a group of consensus algorithms that can handle faults in distributed systems that are not Byzantine. Byzantine Fault Tolerance (BFT), on the other hand, deals with random faults, such as malicious behavior. CFT focuses on handling crashes, failures, and nodes that stop responding. The CFT family is made up of consensus algorithms that keep the system running and consistent even if some nodes crash or stop working \cite{cft}.

Algorithm operations in CFT involve the following steps:
\begin{enumerate}
	
	\item Communication: Nodes in the network communicate with each other by exchanging messages. They rely on reliable and timely message delivery to coordinate their actions and reach consensus.
	
	\item Agreement: Nodes must agree on a specific value or state proposed by a designated leader or a group of nodes. This can be achieved through voting, where nodes cast their votes for a proposal, or through other coordination mechanisms, like rounds of message exchanges.
	
	\item Failure detection: Nodes monitor each other's behavior to detect failures. If a node doesn't receive a message within a specified timeframe or fails to respond to queries, it may be assumed to be unresponsive or crashed.
	
	\item Recovery: Once a failure is detected, the remaining nodes work together to recover from the crash fault. This may involve electing a new leader, redistributing the workload, or reconfiguring the system to continue operation without the faulty node.
	
	\item Consistency: Despite the presence of crash faults, the non-faulty nodes must ensure that the system maintains consistency. This means that all nodes must eventually agree on a consistent state or value, even if some nodes have crashed or become unresponsive.

\end{enumerate}

CFT consensus algorithms are designed to handle non-malicious failures in distributed systems. However, they are not equipped to deal with Byzantine faults, where nodes can behave arbitrarily or maliciously. For systems that need to withstand Byzantine faults, Byzantine Fault Tolerance (BFT) algorithms are used instead. Table \ref{tab:cft} lists advantages and disadvantages of the algorithm.

\begin{table}[htbp]
	\centering
	\caption{Advantages and Disadvantages of CFT Algorithm.}
\label{tab:cft}
	\begin{tabular}{|p{0.45\linewidth}|p{0.45\linewidth}|}
		\hline
		\textbf{Advantages} & \textbf{Disadvantages} \\
		\hline
		Simplicity & Limited fault tolerance \\
		\hline
		Efficiency & Trust assumptions \\
		\hline
		Robustness & Vulnerability to attacks \\
		\hline
		Faster consensus & Dependence on timely communication \\
		\hline
	\end{tabular}
\end{table}

\subsubsection{Paxos}

A variant of CFT and an algorithm for getting all the nodes in a distributed system to agree on a single value or operation, even if some of the nodes are broken. It is especially useful for fault-tolerant distributed systems where nodes can fail or become unreachable \cite{Paxos1} \cite{Paxos2}. Table \ref{tab:paxos} lists advantages and disadvantages of the algorithm.

\begin{table}[htbp]
	\centering
	\caption{Advantages and Disadvantages of Paxos Algorithm.}
\label{tab:paxos}
	\begin{tabular}{|p{0.45\linewidth}|p{0.45\linewidth}|}
		\hline
		\textbf{Advantages} & \textbf{Disadvantages} \\
		\hline
		Fault Tolerance & Complexity \\
		\hline
		Safety & Liveness Issues \\
		\hline
		Asynchronous Communication & Communication Overhead \\
		\hline
		Adaptable & Dependency on a Leader \\
		\hline
	\end{tabular}
\end{table}

\subsubsection{RAFT} 

RAFT is a distributed consensus algorithm that is made to be easy to understand and use while ensuring fault tolerance, safety, and liveness. It can be used instead of the Paxos algorithm, which is more complicated \cite{RAFT} \cite{RaftConsensusAlgorithmforPrivateBlockchains}. Table \ref{tab:raft} lists advantages and disadvantages of the algorithm.

\begin{table}[htbp]
	\centering
	\caption{Advantages and Disadvantages of RAFT Algorithm.}
\label{tab:raft}
	\begin{tabular}{|p{0.45\linewidth}|p{0.45\linewidth}|}
		\hline
		\textbf{Advantages} & \textbf{Disadvantages} \\
		\hline
		Understandability & Performance \\
		\hline
		Fault Tolerance & Scalability \\
		\hline
		Strong Consistency & Partial Network Partitions\\
		\hline
		Leader Election & Complexity \\
		\hline
		Dynamic Membership Changes & \\
		\hline
	\end{tabular}
\end{table}

%

In Table \ref{TBL:CFTFamily} the differences are represented in term of operations between CFT variants. 
\begin{table}[htbp]
	\caption{Comparison of CFT Family In Term of Operations.}
\label{TBL:CFTFamily}
	\centering
	\begin{tabular}{|p{0.10\linewidth}|p{0.15\linewidth}|p{0.15\linewidth}|p{0.10\linewidth}|p{0.15\linewidth}|p{0.15\linewidth}|}
		\hline
		\textbf{Differences} & \textbf{Goal} & \textbf{Model} & \textbf{Roles} & \textbf{Protocol} & \textbf{Recovery} \\
		\hline
		CFT & Ensure system reliability despite crash failures & Assumes crash failures only & No specific roles & Varies based on specific implementation & Relies on replication and backups \\
		\hline
		Paxos & Ensure consistency and agreement in distributed systems & Assumes crash failures only & Proposers, Acceptors, Learners & Three phases: Prepare, Propose, Learn & Learners apply committed decisions \\
		\hline
		RAFT & Ensure consistency and agreement in distributed systems & Assumes crash failures only & Leader, Followers, Candidates & Three phases: Leader Election, Log Replication, Safety & Nodes store and apply committed entries \\
		\hline
	\end{tabular}
\end{table}


\subsection{Proof of Elapsed Time (PoET)}

Proof of Elapsed Time (PoET) is a consensus algorithm initially developed by Intel for permissioned blockchain systems. It is designed to provide a fair, energy-efficient, and highly scalable alternative to Proof of Work (PoW) and other consensus mechanisms. PoET leverages Intel's Software Guard Extensions (SGX) technology, a set of hardware extensions providing secure enclaves to protect sensitive data and code execution from external access or tampering \cite{PoET1}.

How the Proof of Elapsed Time Algorithm operates:
\begin{enumerate}
	\item Trusted Execution Environment (TEE): PoET relies on a Trusted Execution Environment (TEE) provided by Intel SGX, which enables the execution of trusted code in a secure enclave, ensuring the integrity and confidentiality of the code and data.
	
	\item Wait Time Lottery: PoET operates as a ``wait time lottery,'' where participating nodes (validators) generate a random wait time and go to sleep for that duration. The validator with the shortest wait time wakes up first, creates a new block, and broadcasts it to the network. Other validators verify the block and add it to the blockchain if it is valid. The process then repeats for the next block.
	
	\item Wait Time Calculation: Each validator generates its random wait time within the secure enclave using the TEE. This wait time calculation uses the previous block's data, the validator's local information, and a target wait time set by the network. The resulting wait time is cryptographically signed by the enclave to ensure its authenticity and prevent tampering.
	
	\item Winning Validator: When a validator's wait time elapses, it wakes up and creates a new block proposal. The block proposal includes the validator's wait time, the secure enclave's signature, and the transactions to be included in the block. The validator then broadcasts its block proposal to the network.
	
	\item Block Verification: Other validators in the network verify the block proposal by checking the following: The block proposal's wait time is less than or equal to the target wait time. The secure enclave's signature is valid, confirming that the wait time was generated within a TEE. The transactions in the block are valid. If the block proposal passes the verification, validators add the block to their local copy of the blockchain.
\end{enumerate}

The primary advantages of PoET are its energy efficiency, as it does not require the computationally intensive mining process of PoW, and its fairness, as each participating node has an equal chance of creating a block. However, PoET's reliance on Intel's SGX technology raises concerns about centralization, as it requires specific hardware and trust in Intel as the technology provider. Additionally, potential vulnerabilities in SGX could compromise the security and integrity of the PoET consensus mechanism \cite{PoET2}. Table \ref{tab:poet} lists advantages and disadvantages of the algorithm.

\begin{table}[htbp]
	\centering
	\caption{Advantages and Disadvantages of PoET.}
\label{tab:poet}
	\begin{tabular}{|p{0.45\linewidth}|p{0.45\linewidth}|}
		\hline
		\textbf{Advantages} & \textbf{Disadvantages} \\
		\hline
		Energy efficiency & Dependence on Intel SGX\\
		\hline
		Fairness & Trust in Intel \\
		\hline
		Scalability & SGX Vulnerabilities\\
		\hline
		Security & Permissioned networks\\
		\hline
	\end{tabular}
\end{table}

\subsection{Proof of Burn (PoB)}

Proof of Burn (PoB) is an alternative consensus algorithm that aims to address some of the drawbacks of Proof of Work (PoW), such as energy inefficiency and the centralization of mining power. PoB requires participants to ``burn'' or permanently destroy a certain amount of cryptocurrency to obtain the right to create new blocks and earn rewards. This process simulates the consumption of resources, like PoW, but without the significant energy expenditure associated with mining \cite{PoBslimcoin}.

How the Proof of Burn Algorithm operates:
\begin{enumerate}
	\item Coin Burning: In PoB, participants (validators or miners) destroy a certain amount of cryptocurrency, usually by sending it to an unspendable address (known as an ``eater address''). This process is called ``burning'' the coins, as they are removed from circulation and become irretrievable.
	
	\item Burned Coins as Proof: The act of burning coins serves as a proof of the participant's commitment to the network. The more coins a participant burns, the higher their chances of being selected to create a new block. This simulates the resource consumption of PoW but without the energy-intensive mining process.
	
	\item Block Creation: The protocol periodically selects a participant to create a new block based on the amount of cryptocurrency they have burned. This selection can be done using various algorithms, such as a weighted random approach where the probability of being chosen is proportional to the amount of cryptocurrency burned.
	
	\item Block Rewards: The selected participant creates a new block and receives a block reward, typically in the form of newly minted cryptocurrency. This incentivizes participants to burn coins, as they have the potential to earn more coins in the long run by creating new blocks.
	
	\item Burn-and-Mint Equilibrium: Over time, an equilibrium is reached between the amount of cryptocurrency burned and the amount minted through block rewards. This equilibrium helps regulate the rate of coin burning and ensures that the overall cryptocurrency supply remains stable.
\end{enumerate}

Proof of Burn is an alternative consensus mechanism that addresses some of the challenges associated with Proof of Work, such as energy inefficiency and centralization. By requiring participants to burn cryptocurrency to gain the right to create new blocks, PoB offers a more energy-efficient approach to achieving consensus while still simulating resource consumption. However, the permanent destruction of cryptocurrency and potential wealth concentration issues are significant concerns that may limit its adoption in certain applications \cite{PoBslimcoin}. Table \ref{tab:pob} lists advantages and disadvantages of the algorithm.

\begin{table}[htbp]
	\centering
	\caption{Advantages and Disadvantages of PoB \cite{PoBbartoletti2017empirical, PoBcounterparty, proofofburn}.}
\label{tab:pob}
	\begin{tabular}{|p{0.45\linewidth}|p{0.45\linewidth}|}
		\hline
		\textbf{Advantages} & \textbf{Disadvantages} \\
		\hline
		Energy Efficiency & Coin Destruction \\
		\hline
		Reduced Centralization & Wealth Concentration \\
		\hline
		ASIC Resistance & Limited Adoption \\
		\hline
		Security & Barrier to Entry \\
		\hline
	\end{tabular}
\end{table}

\subsection{Proof of Coverage (PoC)}

Proof of Coverage (PoC) is a consensus algorithm used predominantly in wireless networks, such as the Helium blockchain, to validate network participant coverage. The algorithm incentivizes participants to deploy wireless coverage devices known as hotspots that provide connectivity for Internet of Things devices. These locations are rewarded for demonstrating their coverage area and service quality \cite{ProofofCoverage}.
The Proof of Capacity Algorithm operates as follows:
\begin{enumerate}
	
	\item Challenge Creation: The network or hotspot generates a challenge consisting of encrypted data that must be transmitted and received by hotspots.
	
	\item Beacon Transmission: The challenged hotspot transmits a beacon comprising the encrypted data as evidence that it has transmitted the signal. As evidence, the transaction is recorded on the blockchain.
	
	\item Witnessing: Neighboring locations within the coverage area receive and act as witnesses for the beacon. They affirm receipt of the beacon and submit their validation to the blockchain to validate the transmission.
	
	\item Reward Distribution: Hotspots that participate in the challenge-response process by transmitting beacons and witnessing transmissions are rewarded with the native cryptocurrency (e.g., Helium tokens, HNT).
\end{enumerate}

Its applicability is limited to specific use cases and relies on the assumption that participants act honestly in providing accurate information about their coverage area and service quality. Table \ref{tab:pocoverage} lists advantages and disadvantages of the algorithm.

\begin{table}[htbp]
	\centering
	\caption{Advantages and Disadvantages of PoC.}
\label{tab:pocoverage}
	\begin{tabular}{|p{0.45\linewidth}|p{0.45\linewidth}|}
		\hline
		\textbf{Advantages} & \textbf{Disadvantages} \\
		\hline
		Incentivized Network Expansion & Limited Applicability \\
		\hline
		Energy Efficiency & Trust in Participants \\
		\hline
		Decentralization & Geographical Limitations \\
		\hline
		Security & Sybil Attacks \\
		\hline
		Cost-Effectiveness & \\
		\hline
	\end{tabular}
\end{table}

\subsection{Proof of Capacity (PoC)}

Proof of Capacity (PoC) is a consensus algorithm used in some blockchain networks as an alternative to the energy-intensive Proof of Work (PoW) mechanism. Also known as Proof of Space, PoC relies on the available storage capacity of network participants to achieve consensus and maintain the network's security \cite{PoCdziembowski2015proofs, proofspace}.
The Proof of Capacity Algorithm operates as follows:
\begin{enumerate}
	\item In PoC, participants, called ``farmers,'' allocate a portion of their storage capacity to store a large dataset called ``plot'' files. These plot files contain unique solutions to cryptographic problems, similar to nonces in the PoW mechanism. When a new block needs to be added to the blockchain, the network initiates a competition among the farmers. The farmers scan their plot files to find the best solution for the cryptographic problem presented in the new block.
	
	\item The winning farmer is the one who finds the solution with the shortest deadline, which is a specific value associated with the solution. The deadline represents the time it takes for the farmer to mine the block. The shorter the deadline, the faster the farmer can mine the block. The winning farmer then has the right to add the block to the blockchain and receive the mining reward.
	
	\item To participate in PoC, farmers must perform a one-time process called ``plotting.'' During plotting, the farmer generates the plot files by computing all possible solutions for cryptographic problems and storing them on their allocated storage. The plotting process is computationally intensive but only needs to be done once, after which the farmer can participate in the PoC network with minimal resource usage.
	
	\item Energy Efficiency: PoC is more energy-efficient than PoW, as it requires less computational power for mining. Instead of constantly solving complex problems using a large amount of processing power, PoC requires only storage space and low-power processing to read from the storage.
	
	\item Fairness and Decentralization: PoC provides a more level playing field for participants, as it does not require specialized mining hardware like PoW. This can result in increased decentralization, as more participants can afford to join the network, reducing the influence of large mining pools.
	
	\item Lower Barrier to Entry: As PoC relies on storage capacity rather than computational power, the barrier to entry is lower for participants. Many individuals already own storage devices with unused capacity, making it easier for them to join a PoC-based network.
	
	\item Hardware Longevity: PoC-based mining is less demanding on hardware compared to PoW mining, which can lead to longer-lasting hardware and reduced electronic waste \cite{PoCpark2018chained}. 
\end{enumerate}

Table \ref{tab:pocapacity} lists advantages and disadvantages of the algorithm. 

\begin{table}[htbp]
	\centering
\caption{Advantages and Disadvantages of PoC \cite{PoCdziembowski2015proofs, PoCpark2018chained, PoCshoker2017sustainable}.}
\label{tab:pocapacity}
	\begin{tabular}{|p{0.45\linewidth}|p{0.45\linewidth}|}
		\hline
		\textbf{Advantages} & \textbf{Disadvantages} \\
		\hline
		Energy Efficiency & Storage Centralization \\
		\hline
		Fairness and Decentralization & Data Storage Inefficiency \\
		\hline
		Lower Barrier to Entry & Vulnerability to Sybil Attacks \\
		\hline
		Hardware Longevity & Slower Adoption \\
		\hline
	\end{tabular}
\end{table}

\subsection{Proof of Importance (PoI)}

Proof of Importance (PoI) is a consensus mechanism used in certain blockchain networks, including NEM, to determine which participants have the authority to establish new blocks and validate transactions. PoI is designed to recompense active network participants, encourage more equitable reward distribution, and discourage coin hoarding \cite{PoI1}.

The PoI algorithm operation is as follows:
\begin{enumerate}
	\item Account balance evaluation: The PoI algorithm considers each participant's account balance when determining their significance. A participant's importance score is greater if they have a greater interest in the network.
	
	\item Transaction Activity evaluation: The PoI algorithm investigates each participant's transaction history, with more active participants receiving higher importance scores. This rewards users who initiate transactions and actively contribute to the network.
	
	\item Network Topology Consideration:The PoI algorithm takes into account the significance of other network participants with whom the user has conducted business. This encourages users to conduct business with other influential and active users, thereby enhancing the network.
	
	\item Importance: Calculation of importance score The PoI algorithm incorporates the aforementioned factors to determine an importance score for each participant. This score represents the participant's contribution to the network and is used to determine the participant's likelihood of being selected to create new blocks and validate transactions.
	
	\item Block validation: Participants with higher importance scores are more likely to be selected to create new blocks and validate transactions. This selection process promotes active network participation and a more decentralized, equitable distribution of rewards. 
\end{enumerate}

Table \ref{tab:poi} lists advantages and disadvantages of the algorithm.

\begin{table}[htbp]
	\centering
	\caption{Advantages and Disadvantages of PoI. }
\label{tab:poi}
	\begin{tabular}{|p{0.45\linewidth}|p{0.45\linewidth}|}
		\hline
		\textbf{Advantages}          & \textbf{Disadvantages}            \\ \hline
		Incentivizes active nodes    & Complexity                        \\ \hline
		Fairness                     & Potential for manipulation        \\ \hline
		Decentralization             & Limited adoption                  \\ \hline
		Encourages network growth    & Initial importance assignment     \\ \hline
	\end{tabular}
\end{table}

\subsection{Proof of Reputation (PoR)}
Proof of Reputation (PoR) is a consensus algorithm used in some blockchain networks to establish trust and determine who can produce new blocks and validate transactions. By rewarding participants with positive reputations, Proof of Reputation seeks to encourage network activity, security, and equity \cite{ProofOfReputation}.

The PoR algorithm operation is the following:
\begin{enumerate}
	
	\item Reputation assessment: The PoR algorithm evaluates each participant's reputation based on their past actions, compliance with network norms, and contribution to the network as a whole. This may include transaction history, up-time, and other metrics defined by the particular implementation.
	
	\item Calculation of reputation scores: The algorithm computes a reputation score for each participant that reflects their credibility and prominence within the network. The precise method for calculating reputation scores may differ between implementations, but typically involves combining multiple factors into a holistic evaluation.
	
	\item Reputation: Frequently, PoR systems employ a weighting mechanism that assigns greater weights to individuals with higher reputation scores. This weight affects the likelihood that a participant will be selected to generate new blocks and validate transactions.
	
	\item Block validation: Participants with higher reputation scores (and, therefore, higher weights) have a greater chance of being selected to create new blocks and validate transactions. This selection procedure encourages network participants to maintain a positive reputation and promotes a more decentralized, secure, and equitable distribution of rewards.
\end{enumerate}

Taking these pros and cons into account, PoR can be a good way to reach a consensus in apps that value trust, security, and decentralization. But in some situations, its complexity, ability to be manipulated, and the fact that image is hard to define may be problems \cite{ProofOfReputation}. Table \ref{tab:por} lists advantages and disadvantages of the algorithm.
\begin{table}[htbp]
	\centering
	\caption{Advantages and Disadvantages of PoR.}
\label{tab:por}
	\begin{tabular}{|p{0.45\linewidth}|p{0.45\linewidth}|}
		\hline
		\textbf{Advantages}          & \textbf{Disadvantages}            \\ \hline
		Incentivizes good behavior   & Complexity                        \\ \hline
		Decentralization             & Potential for manipulation        \\ \hline
		Security                     & Subjectivity                      \\ \hline
		Adaptability                 & Initial reputation assignment    \\ \hline
	\end{tabular}
\end{table}

\subsection{Proof of Contribution (PoC)}

Blockchain networks use Proof of Contribution (PoC) to reward network contributors. PoC rewards network validators or nodes that provide useful resources or services. Storage, processing power, and data transport are examples. PoC supports network engagement and decentralization by compensating contributors \cite{Proofofcont}.

Operations of the PoC algorithm:
\begin{enumerate}
	\item Contribution tracking: The algorithm keeps track of what each person has contributed. Contributions can be things like computing power, storage space, or network bandwidth, or they can be acts like validating transactions, taking part in governance, or providing other services.
	
	\item Contribution scoring: The algorithm gives each person a score based on what they have contributed. This number shows how much the member has contributed to the network as a whole. It can be calculated using a weighted formula that looks at various types of contributions and how important they are to the network as a whole.
	
	\item Selection: Choosing who gets rewards or tasks: The algorithm uses the input scores to choose who gets rewards or tasks in the network. For example, people with higher scores may be more likely to be chosen as validators or to get a share of newly created tokens.
	
	\item Reward distribution: Once the players have been chosen, the algorithm divides the rewards based on how well they contributed. This can happen through the creation of new tokens or through transaction fees.
	
	\item Updates and Adjustment: Updates and changes are always being made. The program is constantly updating the contribution scores and changing the way rewards are given based on how people are contributing. This makes sure that the network stays busy and gives participants a reason to stay involved.
\end{enumerate}

Table \ref{tab:pocont} lists advantages and disadvantages of the algorithm.

\begin{table}[h]
	\centering
	\caption{Advantages and Disadvantages of Proof of Cont.}
\label{tab:pocont}
	\begin{tabular}{|p{0.45\linewidth}|p{0.45\linewidth}|}
		\hline
		\textbf{Advantages} & \textbf{Disadvantages} \\
		\hline
		Incentivizes active participation & Complexity \\
		\hline
		Decentralization & Potential for gaming the system \\
		\hline
		Fair distribution of rewards & Difficulty in quantifying contributions \\
		\hline
		Encourages diverse contributions & Scalability concerns \\
		\hline
		Dynamic and adaptable & Centralization risks \\
		\hline
	\end{tabular}
\end{table}

Due to the implementation similarity of PoR, PoI, and PoCont, Table \ref{TAB:Weight} compares them based on their operations. 

\begin{table}[htbp]
	\caption{Comparison of PoR, PoI, and PoC In Term of Operations.}
\label{TAB:Weight}
	\centering
	\begin{tabular}{|p{0.10\linewidth}|p{0.10\linewidth}|p{0.15\linewidth}|p{0.15\linewidth}|p{0.15\linewidth}|p{0.15\linewidth}|}
		\hline
		\textbf{Differences} & \textbf{Basis for consensus} & \textbf{Calculation method} & \textbf{Incentives} & \textbf{Sybil attack resistance} & \textbf{Target use case} \\
		\hline
		\textbf{PoR} & Based on the reputation of participants & Reputation determined by participant's history of positive contributions and penalties for negative behavior & Higher reputation scores receive more significant rewards & Resistant due to the difficulty of creating multiple fake identities with high reputation scores & Platforms that require trust and positive behavior, such as social networks and content curation platforms \\
		\hline
		\textbf{PoI} & Based on the importance of participants & Importance calculated using factors like account balance, transaction activity, and interacting node importance & Higher importance scores receive more significant rewards & Resistant due to the difficulty of creating multiple fake identities with high importance scores & Networks that aim to reward active and influential participants, such as NEM's blockchain \\
		\hline
		\textbf{PoC} & Based on the contributions of participants & Contribution determined by evaluating the value or resources provided by the participant & Participants with more contributions receive more significant rewards & Resistant due to the difficulty of creating multiple fake identities with high contribution scores & Networks that aim to reward participants based on their contributions, such as resource sharing or computational power \\
		\hline
	\end{tabular}
\end{table}

\subsection{Proof of Authority (PoA)}

Proof of Authority (PoA) is a consensus algorithm used in permissioned blockchain networks, where a set of pre-approved, trusted nodes (known as authorities or validators) are responsible for validating transactions and creating new blocks. PoA emphasizes the identity and reputation of the validators as a means to ensure network security, rather than relying on computational power or stake as in Proof of Work (PoW) or Proof of Stake (PoS) systems \cite{PoAvechain2018whitepaper}.

How the Proof of Authority consensus algorithm operates:
\begin{enumerate}
	\item Authority Selection: In a PoA-based blockchain, a limited number of validators are chosen based on their identity and reputation. These validators are usually known entities, such as organizations or individuals, with a high level of trustworthiness. The selection process can vary depending on the implementation, but it typically involves manual vetting, a governance mechanism, or a combination of both.
	\item Block Creation: Validators take turns in a round-robin fashion to create new blocks. When a validator's turn comes, it collects and validates the pending transactions, then creates a new block that includes the valid transactions, a reference to the previous block's hash, a timestamp, and the validator's digital signature. The digital signature serves as the validator's seal of approval, confirming that the block complies with the network's rules.
	\item Block Validation: Once a new block is created, the other validators in the network verify the block by checking the transactions and ensuring that the digital signature is valid. If the majority of validators approve the block, it is added to the blockchain, and the transactions within it are considered confirmed.
	\item Incentives and Penalties: Validators are typically rewarded for their work in validating and creating blocks. The rewards can be in the form of transaction fees or native cryptocurrency. On the other hand, if a validator behaves maliciously or fails to follow the network's rules, it may face penalties, such as being removed from the validator set or forfeiting a security deposit.
	\item Governance: PoA networks often include governance mechanisms to manage the validator set, handle disputes, and make decisions regarding network upgrades or changes. Governance can involve voting by the validators themselves or other participants in the network, depending on the implementation \cite{PoAvechain2018whitepaper}.
\end{enumerate}

Table \ref{tab:poa} lists advantages and disadvantages of the algorithm.

\begin{table}[htbp]
	\centering
	\caption{Advantages and Disadvantages of PoA.}
\label{tab:poa}
	\begin{tabular}{|p{0.45\linewidth}|p{0.45\linewidth}|}
		\hline
		\textbf{Advantages} & \textbf{Disadvantages} \\ \hline
		Energy efficiency   & Centralization                                 \\ \hline
		Scalability         & Trust in authorities                           \\ \hline
		Security            & Limited decentralization benefits              \\ \hline
		Predictable block times  & Vulnerability to regulatory pressure       \\ \hline
		Reduced risk of 51\% attacks  & Governance challenges                 \\ \hline
	\end{tabular}
\end{table}

\subsection{Proof of Activity (PoA)}

Proof of Activity (PoA) is a hybrid consensus method that combines the best parts of Proof of Work (PoW) and Proof of Stake (PoS) to make a blockchain network that is more balanced and uses less energy. Iddo Bentov, Charles Lee, Alex Mizrahi, and Meni Rosenfeld came up with the idea in 2014 as an option to PoW and PoS \cite{PoAct}.

The algorithm operations of PoA are as follows:
\begin{enumerate}
	
	\item Mining: Like PoW, miners try solve a cryptographic puzzle. The first miner to figure out the puzzle makes a template block with the block header and the address of where to send the miner's prize.
	
	\item Selection: Validators are picked based on how much they have invested in the network, according to the PoS protocol. The chance of being chosen as a validator goes greater when the stake higher.
	
	\item Block signing: The validators who were chosen sign the sample block and add transactions to it. The deals are verified when each validator adds their signature to the block.
	
	\item Validation threshold: A block is only considered acceptable if it has been signed by a certain number of validators, which is set by the network's consensus rules.
	
	\item Block addition: The block is added to the blockchain when the minimum number of signatures are collected. Validators and miners both get a share of the block payment for what they do for the network.
	
\end{enumerate}

Table \ref{tab:poact} lists advantages and disadvantages of the algorithm.

\begin{table}[htbp]
	\centering
	\caption{Advantages and Disadvantages of PoAct.}
\label{tab:poact}
	\begin{tabular}{|p{0.45\linewidth}|p{0.45\linewidth}|}
		\hline
		\textbf{Advantages} & \textbf{Disadvantages} \\
		\hline
		Energy Efficiency & Complexity \\
		\hline
		Security & Limited adoption \\
		\hline
		Decentralization & PoW drawbacks (energy consumption) \\
		\hline
		Incentivization & Governance challenges \\
		\hline
		Reduced risk of 51\% attacks & Potential centralization \\
		\hline
	\end{tabular}
\end{table}

\subsection{Proof of Authentication (PoAh)}

PoAh is a novel consensus algorithm that replaces PoW with a cryptographic authentication mechanism for resource-constrained devices, thereby rendering the blockchain application-specific. PoAh provides a decentralized security solution that is suitable for both private and permissioned blockchains, unlike other consensus algorithms \cite{PoAh1} \cite{PoAh2}. Table \ref{TAB:PoAhComparison} lists advantages and disadvantages of the algorithm. 

\begin{table}[htbp]
	\caption{Advantages and Disadvantages of PoAh.}
\label{TAB:PoAhComparison}
	\centering
	\begin{tabular}{|p{0.45\linewidth}|p{0.45\linewidth}|}
		\hline
		 \textbf{Advantages} & \textbf{Disadvantages} \\
		\hline
		 More energy-efficient than PoW & Increased complexity compared to PoA\\
		\hline
		 Handles large-scale IoT frameworks & Limited applicability to other blockchain networks \\
		\hline
		 Faster block creation and transaction process & Only one factor (trust Score) \\
		\hline
		 Suitable for resource-constrained IoT devices & Requires further research, development, and testing \\
		\hline
		 Secure consensus through distributed trust & Centralization risk \& potential for Sybil attacks \\
		\hline
	\end{tabular}
\end{table}

\subsection{Proof of PUF-Chain (PoP)}

The focus of ``PUFchain 2.0: Hardware-Assisted Robust Blockchain for Sustaining Simultaneous Device and Data Security in Smart Healthcare'' is the development of a blockchain-based solution for enhancing device and data security in smart healthcare systems. PUFchain 2.0 is a hardware-assisted, secure blockchain that employs Physically Unclonable Functions (PUFs) to improve security \cite{PUFchain1} \cite{PUFchain2}. Table \ref{TAB:pop} lists advantages and disadvantages of the algorithm.

\begin{table}[htbp]
	\caption{Advantages and Disadvantages of PoP.}
\label{TAB:pop}
	\centering
	\begin{tabular}{|p{0.45\linewidth}|p{0.45\linewidth}|}
		\hline
		\textbf{Advantages} & \textbf{Disadvantages} \\
		\hline
		Enhanced security & Complexity \\
		\hline
		Device authentication & Cost \\
		\hline
		Data integrity and privacy & PUF limitations \\
		\hline
		Decentralization & Adoption challenges \\
		\hline
		Scalability & Potential for new attack vectors \\
		\hline
	\end{tabular}
\end{table}

\subsection{Proof of Block \& Trade (PoBT)}

IoT blockchain-specific lightweight proof of block and trade (PoBT) consensus algorithm. It is optimized for IoT systems, which are typically large-scale and heterogeneous networks of devices, in contrast to other consensus algorithms used in enterprise blockchains. PoBT attains scalability by simplifying the complex consensus-based security employed by conventional business blockchain algorithms \cite{PoBT}. It is a research based consensus algorithm. Table \ref{TAB:pobt} lists advantages and disadvantages of the algorithm. 

\begin{table}[htbp]
	\centering
	\caption{Advantages and Disadvantages of PoBT.}
\label{TAB:pobt}
	\begin{tabular}{|p{0.45\linewidth}|p{0.45\linewidth}|}
		\hline
		\textbf{Advantages} & \textbf{Disadvantages} \\
		\hline
		Lightweight and scalable & Relies on business trust \\
		\hline
		Low energy consumption & Potential for trust manipulation \\
		\hline
		Low latency & Less suitable for non-business IoT applications \\
		\hline
	\end{tabular}
\end{table}


\section{Our Proposed Methodology for Analysis of Consensus Algorithms}
\label{sec:Methodology}

In this section, we demonstrate the approach followed to analyze application based multiple consensus algorithms from data collected in section \ref{sec:GeneralOverviewforConsensusAlgorithms} to generally evaluate the suitability of a particular consensus algorithm to a certain application.

\subsection{Multiple Attributes and Their Definitions}

This section presents the attributes as shown in Table \ref{TBL:Attributes} and their levels to evaluate a certain consensus algorithm compared to the required consensus based on application conditions. 

\begin{table}[htbp]
	\caption{Attributes Associated With Consensus Algorithms Analysis.}
\label{TBL:Attributes}
	\centering
		\begin{tabular}{|p{0.15\linewidth}|p{0.75\linewidth}|}
		\hline
			
			\textbf{Attribute}     & \textbf{Definition} \\
			\hline
			Hardware Requirements  &   Refer to node's computational resources, components, and infrastructure for network participation.\\
			\hline
			Pre-Trust Level        &  Refers to that each consensus mechanism requires a certain level of trust.\\
			\hline
			Tolerance Level        &   Refers to consensus mechanism's robustness and adaptability under adversary or system faults.
			\\
			\hline
			Overhead Computation   &   Refers to the need for additional resources to participate in the consensus and sustain the network.\\
			\hline
			Centralization Level   &   Refers to Control, decision-making, and resource concentration within a network.\\
			\hline
			Scalability Level      &   Refers to network's ability to handle more transactions and users while maintaining performance.\\
			\hline
			Latency Level          &  Refers to transaction's processing, confirmation, and ledger entry time.\\
			\hline
			Cost Level             &  Refers to the various expenses associated with operating, maintaining, and participating in a network.\\
			\hline
			Security Level         &  Refers to network's ability to withstand cyber-attacks, fraud, and manipulation.\\
			\hline
			Interoperability        &  Refers to adaptability to diverse communication protocols, data formats, and system architectures.\\
			\hline
			Complexity Level       &  Refers to the complexity associated with the implementation.\\
			\hline
		\end{tabular}
\end{table}

\subsubsection{Hardware Requirements}

Hardware requirements vary depending on the consensus algorithm being used and can include aspects such as Computational Power, Storage Capacity, Energy Consumption, Network Connectivity, Specialized Hardware, and Stake or Investment. In this paper, it is divided to 5 levels: High, Moderate - High, Moderate, Low - Moderate, Low.

\textbf{Definitions:}
High (H): Requires specialized hardware (ASICs/GPUs) for mining, high energy consumption.
Moderate to High (M-H): Requires a computer or server with large storage capacity for plotting, lower energy consumption compared to PoW.
Moderate (M): Requires a computer or server to run a validator node, staking an amount of cryptocurrency. Lower energy consumption. 
Low to moderate (L2M): Requires a computer, server, or IoT device to run a node, lower energy consumption compared to PoW.
Low (L): CPS or IoT device to run a node.

\subsubsection{Pre-Trust Level}

The level of trust required for each consensus algorithm varies depending on the network type, participants, and underlying assumptions. Generally, permissionless DLTs like PoW and PoS require no pre-established trust between participants, while permissioned DLTs using PBFT and its variants require higher levels of trust in the network's structure and participants.

\textbf{Definitions:}
H: Trust is based on the identity and reputation of pre-selected validators.
M2H: Trust is based on a list of trusted validator nodes chosen by each participant.
M: Trust is based on the amount of cryptocurrency staked by validators; higher stake indicates higher trust.  
L2M: Trust is based on pre-existing trust in at least in early stages.
L: Trust is based on computational power and solving mathematical puzzles; no pre-existing trust required.
\subsubsection{Tolerance Level}

Tolerance level requirements can be categorized into the following aspects: Byzantine Fault Tolerance (BFT), Sybil Attack Resistance, Double Spending Resistance, Network Partition Tolerance, Censorship Resistance, Long-Range Attack Resistance, and Selfish Mining Resistance.

\textbf{Definitions:}
H: Capable of tolerating up to a third of faulty or malicious nodes, highly resistant to attacks, and offering strong consistency and dependability.
M2H: Tolerate a moderate number of malfunctioning nodes, achieve a balance between security and performance, and are less susceptible to attacks than algorithms with low tolerance.
M: Manage a limited number of defective nodes, achieve a balance between security and performance, and rely on network configuration and node truthfulness for security.
L2M: Accept a small number of malfunctioning nodes, rely on trusted authorities or centralizing mechanisms, network may be compromised in certain attack scenarios.
L: Handles very few or no defective nodes, is susceptible to a variety of attacks, prioritizes performance over security, and is not advised for high-value or mission-critical applications.
\subsubsection{Overhead Computation}

Overhead computation can have a significant impact on the efficiency, scalability, and resource requirements of a DLT system. Different consensus algorithms have varying levels of overhead computation, which can influence the choice of consensus mechanism for a particular application. Overhead computation can be attributed to various aspects of the consensus process, such as Block validation, Block mining or forging, Network communication, Voting and agreement.

\textbf{Definitions:}
H: require a significant amount of computational resources, power, or effort from nodes to participate in the consensus process.
M2H: A slightly lower resource requirement than high overhead computation algorithms but still demand considerable computational resources from nodes. 
M:Require a fair amount of computational resources but are less resource-intensive compared to high overhead computation algorithms.
L2M: Lower computational resource requirements compared to moderate overhead computation algorithms, but they still demand some level of computational effort from nodes.
L: Requires minimal computational resources from nodes to participate in the consensus process.
\subsubsection{Centralization Level}

A high level of centralization indicates that a small number of nodes or entities exert significant influence over the network, while a low level of centralization suggests a more evenly distributed control among participants. Centralization level is an essential factor in the design and evaluation of consensus algorithms, as it impacts various aspects of a DLT system, such as, Security, Trust, Resilience, Governance.

\textbf{Definitions:} 
H: Single entity or small group controls the network, resources, and decision-making. Relies on central authority.
M2H: Limited number of influential entities; some distribution of control and decision-making.
M: Centralized and decentralized control balance; key functions may be controlled centrally.
L2M: Control distributed among more participants; few entities may have notable influence.
L: Control and resources widely distributed; network resistant to censorship and single points of failure.
\subsubsection{Scalability Level}

Scalability is a critical aspect of DLT systems, as it impacts the capacity, speed, and overall efficiency of the network. Scalability level influences various aspects of a DLT system, such as Transaction throughput, Latency, Network size, and Resource consumption.

\textbf{Definitions:} 
H: Can support a massive number of transactions per second (tps) with minimal latency, suitable for global applications.
M2H: Can handle a significant number of tps with reasonable latency, suitable for large-scale applications.
M: Satisfactory transaction throughput and latency; may require improvements for larger applications.
L2M: Supports moderate tps with increased latency; suitable for small to medium applications.
L: Struggles with low transaction throughput and high latency; not ideal for applications requiring high scalability.
\subsubsection{Latency Level}

Latency is an essential factor in evaluating the performance of a DLT system, as it impacts the responsiveness and overall user experience of the network. Latency level influences various aspects of a DLT system, such as Transaction confirmation time, Block propagation time, Network efficiency, and Consensus convergence time.

\textbf{Definitions:}
H: Significantly delayed transaction confirmation and block validation, not suitable for real-time or time-sensitive applications.
M2H: Slower transaction confirmation and block validation, may not be ideal for time-critical applications.
M: Reasonable transaction confirmation and block validation times, adequate for most general applications.
L2M: Fast transaction confirmation and block validation, suitable for time-sensitive applications.
L: Near-instant transaction confirmation and block validation, ideal for real-time applications and high-frequency trading.
\subsubsection{Cost Level}

Cost is an essential factor to consider when evaluating a DLT system, as it impacts the affordability, accessibility, and sustainability of the network. Cost level influences various aspects of a DLT system, such as Energy consumption, Hardware requirements, Network maintenance and Transaction fees.

\textbf{Definitions:}
H: Significantly high transaction fees and resource consumption, likely to deter frequent transactions or low-value transfers, and may be prohibitive for some applications or users.
M2H: Above-average transaction fees and resource consumption, potentially limiting the frequency of transactions or discouraging low-value transfers.
M: Average transaction fees and resource consumption, acceptable for various use cases but may not be ideal for high-frequency or low-value transactions.
L2M: Low transaction fees and resource consumption, suitable for most applications and users.
L: Minimal transaction fees and resource consumption, ideal for frequent transactions and low-value transfers.
\subsubsection{Security Level}

Security is a crucial factor to consider when evaluating a DLT system, as it impacts the trustworthiness, stability, and overall integrity of the network. Security level influences various aspects of a DLT system, such as Resistance to attacks, Fault tolerance, Data integrity, and Privacy.

\textbf{Definitions:}
H: Extremely resistant to attacks and faults, providing the highest level of security and trustworthiness for critical applications.
M2H: Strong resistance to attacks and faults, suitable for applications that require a high degree of security and trust.
M: Provides a reasonable level of security and attack resistance, appropriate for various use cases but may not be ideal for critical applications.
L2M: Limited resistance to attacks and faults, may not be suitable for applications requiring high security or trustworthiness.
L: Minimal resistance to attacks and faults, not recommended for applications that require a high level of security or trust.

\subsubsection{Interoperability}

Blockchain networks and systems that can communicate, interact, and share data are interoperable. This allows blockchains to efficiently share resources, services, and features. Polkadot, Cosmos, and ICON emphasize compatibility. In the context of DLT and BC, it means providing a layer0 that could serve as a common area between multiple DLTs \cite{alkhodair2023ArXiv}.

\textbf{Definitions:}
H: A consensus algorithm or blockchain platform with high interoperability can easily communicate with many different blockchain networks and platforms. It supports cross-chain communication and collaboration, enabling data and resource sharing between blockchains.
M2H: A consensus algorithm or blockchain platform with moderate to high interoperability can interact with many blockchain networks and platforms, but seamless cross-chain communication may require additional development, configuration, or third-party solutions. Its interoperability features may not function with all blockchain networks.
M: A consensus algorithm or blockchain platform with moderate interoperability can communicate with a few blockchain networks and platforms. Cross-chain communication may need extensive development, configuration, or third-party solutions and support just certain blockchain networks.
L2M: A consensus algorithm or blockchain platform with low to moderate interoperability can interact with a small number of other blockchain networks and platforms, but cross-chain communication may require extensive development, configuration, or third-party solutions. It may support few blockchain networks and have few interoperability features.
L: A consensus method or blockchain platform with low interoperability does not permit cross-chain communication and collaboration. It may not function with other blockchain networks without major modification or third-party solutions.

\subsubsection{Complexity}

The level of difficulty in comprehending, implementing, and maintaining a system's consensus algorithms, cryptographic methods, protocol layers, and general architecture.

\textbf{Definitions:}
H: Complex consensus algorithms and blockchain platforms have extensive designs, many components, and sophisticated architecture. Due to its complexity and technological requirements, it may be difficult to understand, implement, and maintain.
M2H: A consensus algorithm or blockchain platform with moderate to high complexity has a moderately sophisticated design and architecture. It may require several components or processes for implementation and maintenance and demands a strong understanding of the underlying principles.
M: A consensus algorithm or blockchain platform with moderate complexity has a balanced architecture with an average amount of components and operations. It is easy to learn and implement, although some technical knowledge is needed.
L2M: A consensus algorithm or blockchain platform with low to moderate complexity has a simpler design and architecture than those with moderate complexity. Its fewer components and processes make it simpler to understand, implement, and maintain.
L: A simple consensus algorithm or blockchain platform has a minimalist design and architecture. Few components and processes make it simple to understand, implement, and maintain. Since the concepts and needs are simple, it's excellent for beginners.

\subsection{Consensus Based Application (IoT and CPS)}

The perfect consensus algorithm would be the best in terms of performance, security, and efficiency across all key measures, and it would be made just for the needs of Cyber-Physical Systems \cite{Alkhodair2020} \cite{Alkhodair2021}. However, it's important to note that it can be hard to find a consensus method that meets all of these high-level needs. Trade-offs or change on current consensus algorithm to make it work better for your CPS use case \cite{Alkhodair2022} \cite{Alkhodair2023}.

\begin{figure}[htbp]
	\includegraphics[width=16.5cm]{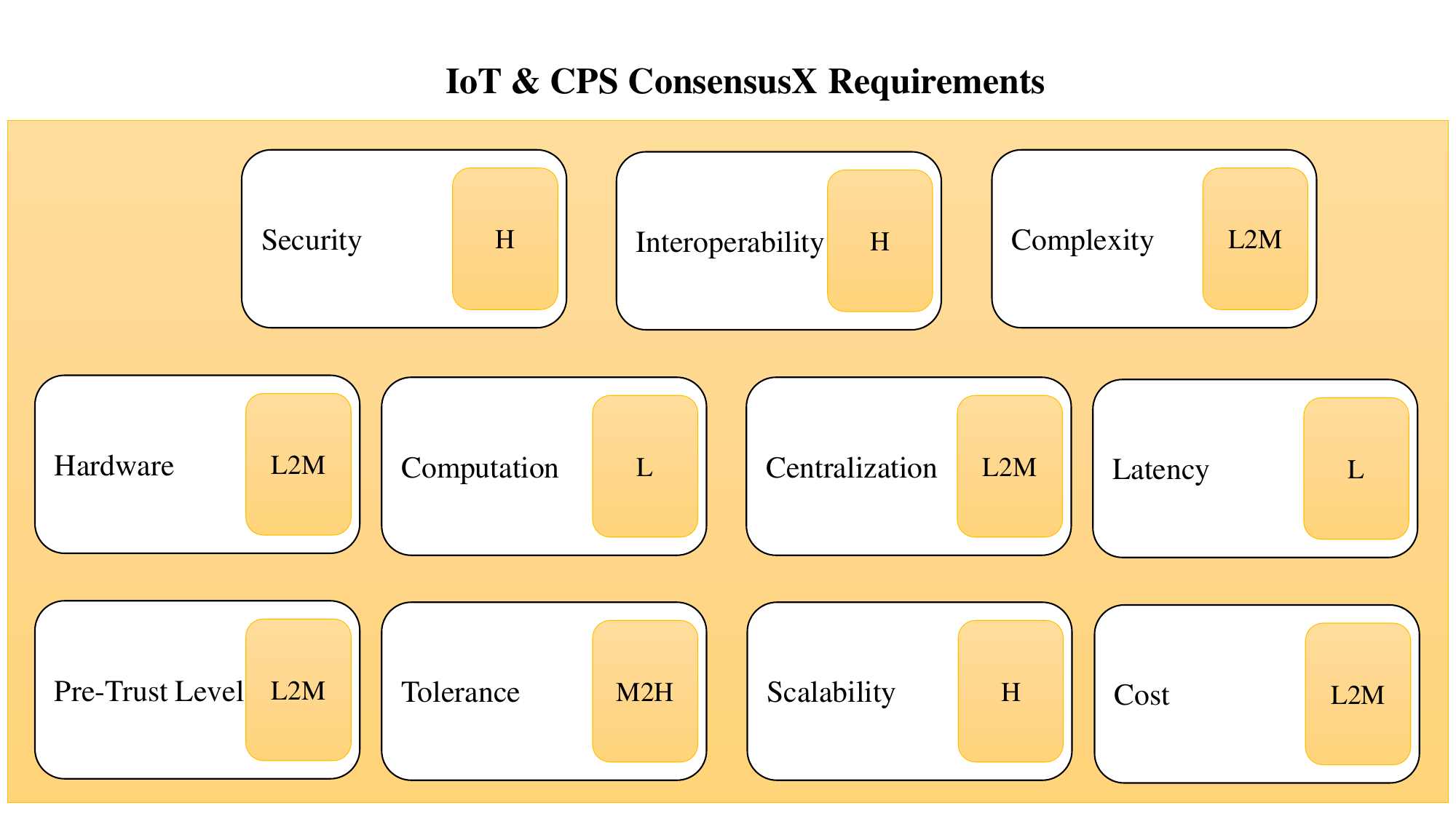}
	\caption{Internet of Things and Cyber Physical Systems desirable ConsensusX Algorithm Attributes.}
\label{FIG:IoTCPSXconsensus}
\end{figure}

A comparative analysis between all candidates consensus algorithms with the required consensus (ConsensusX) to find the optimal choice for each application among the candidates lists is often required. Figure \ref{FIG:IoTCPSXconsensus} presents what has been assigned to each attribute to analyze the consensus algorithm candidate lists. 

The levels for each attributes for the desirable ConsensusX are based on the reasoning below
For a consensus algorithm to be suitable for Cyber Physical Systems (CPS), it should exhibit the discussed desirable levels for attributes in Table \ref{TAB:reasoning}

\begin{table}[htbp]
	\caption{ConsensusX Algorithm Level Assigning and Why?}
\label{TAB:reasoning}
	\centering
	\begin{tabular}{|p{0.15\linewidth}|p{0.15\linewidth}|p{0.60\linewidth}|}
		\hline
		\textbf{Attributes} & \textbf{Assigned Level} & \textbf{Reason} \\
		\hline
		Hardware Requirements & Low to Moderate (L2M) & CPS systems should be cost-effective and capable of running on various hardware without being resource-intensive.\\
		\hline
		Pre-Trust Level & Low to Moderate (L2M) & Trust between network at least initial users should be established to ensure secure and efficient operations, However, some level of trustlessness is still desirable.\\
		\hline
		Tolerance Level & Moderate to High (M2H) & The consensus algorithm should be able to tolerate a number of faulty or malicious nodes without compromising the network's integrity and operations.\\
		\hline
		Overhead Computation & Low (L) &  Computation overhead should be minimized to ensure efficient transaction processing and scalability a CPS.\\
		\hline
		Centralization Level & Low to Moderate (L2M) & Decentralization is essential to prevent single points of failure and enhance network resilience.\\
		\hline
		Scalability Level & High (H) & The consensus algorithm should support high volume of transaction, as CPS systems need to handle a large number of transactions in an efficient way.\\
		\hline
		Latency Level & Low (L) & Fast transaction confirmations are crucial in CPS systems to enable real-time decision-making.\\
		\hline
		Cost Level & Low to Moderate (L2M) & The consensus algorithm should be cost-efficient to implement and maintain, without having significant operational expenses.\\ 
		\hline
		Security Level & High (H) & A secure consensus algorithm is essential to protect, and due to the secure nature of DLTs, security required at its highest to protect data from unauthorized access or tampering.\\
		\hline
		Interoperability & High (H) & Integration and data exchange with other DLT systems and CPS infrastructure is crucial. The consensus algorithm should support a high level of interoperability. \\
		\hline
		Complexity Level & Low to Moderate (L2M) & The consensus algorithm should relatively be simple to implement, understand, and manage for efficient operations in CPS.\\
		\hline
	\end{tabular}
\end{table}

\section{Evaluation of the Algorithms based on the Methodology}
\label{sec:Evaluation}

In this section, evaluation of all candidate consensus algorithms is made using the methodology in section \ref{sec:Methodology} to minimize the list and reach to a list of consensus algorithms that are relatively suitable to CPS. Tables \ref{TAB:CandidateList1}, \ref{TAB:CandidateList2}, \ref{TAB:CandidateList3}, and \ref{TAB:CandidateList4} represent the comparison for candidates' attributes and ConsensusX assigned attributes.

\subsection{Consensus Algorithms Candidates Lists Versus ConsensusX}

%
%
%
%
%
%
%
%
%
%

\begin{table}[htbp]
	\caption{Candidates List 1.}
\label{TAB:CandidateList1}
	\centering
	\begin{tabular}{|p{0.10\linewidth}|p{0.05\linewidth}|p{0.05\linewidth}|p{0.05\linewidth}|p{0.05\linewidth}|p{0.05\linewidth}|p{0.05\linewidth}|p{0.05\linewidth}|p{0.05\linewidth}|p{0.05\linewidth}|p{0.08\linewidth}|p{0.05\linewidth}|}
		\hline
		\textbf{Attributes} & \textbf{Hard.} & \textbf{Trust } & \textbf{Toler.} & \textbf{Compu.} & \textbf{Centra.} & \textbf{Scala.} & \textbf{Late.} & \textbf{Cost} & \textbf{Secu.} & \textbf{Interoper.} & \textbf{Compl.}\\
		\hline
		\textbf{PoS}        & M & M   & M   & M   & M   & M & M & M & H & M & M \\ \hline
		\textbf{DPoS}       & L & M2H & M   & L2M & M2H & H & L & L & M2H & M & M \\ \hline
		\textbf{LPoS}       & M & M   & M   & M   & M   & M & M & M & H & M & M \\ \hline
		\textbf{PPoS}       & M & M   & M   & M   & L   & H & L & M & H & M & M \\ \hline
		\textbf{NPoS}       & M & M   & M   & M   & M   & H & L & M & H & M & M \\ \hline
		\textbf{BPoS}       & M & M   & M   & M   & M   & M & M & M & H & M & M \\ \hline
		\textbf{Ouroboros}  & M & M   & M   & M   & L   & H & L & M & H & M & M \\ \hline
		\textbf{ConsensusX} & L2M & L2M & M2H & L & L2M & H & L & L & H & H & L2M \\ \hline

	\end{tabular}
\end{table}

\begin{figure}[htp]
	\includegraphics[width=16.5cm]{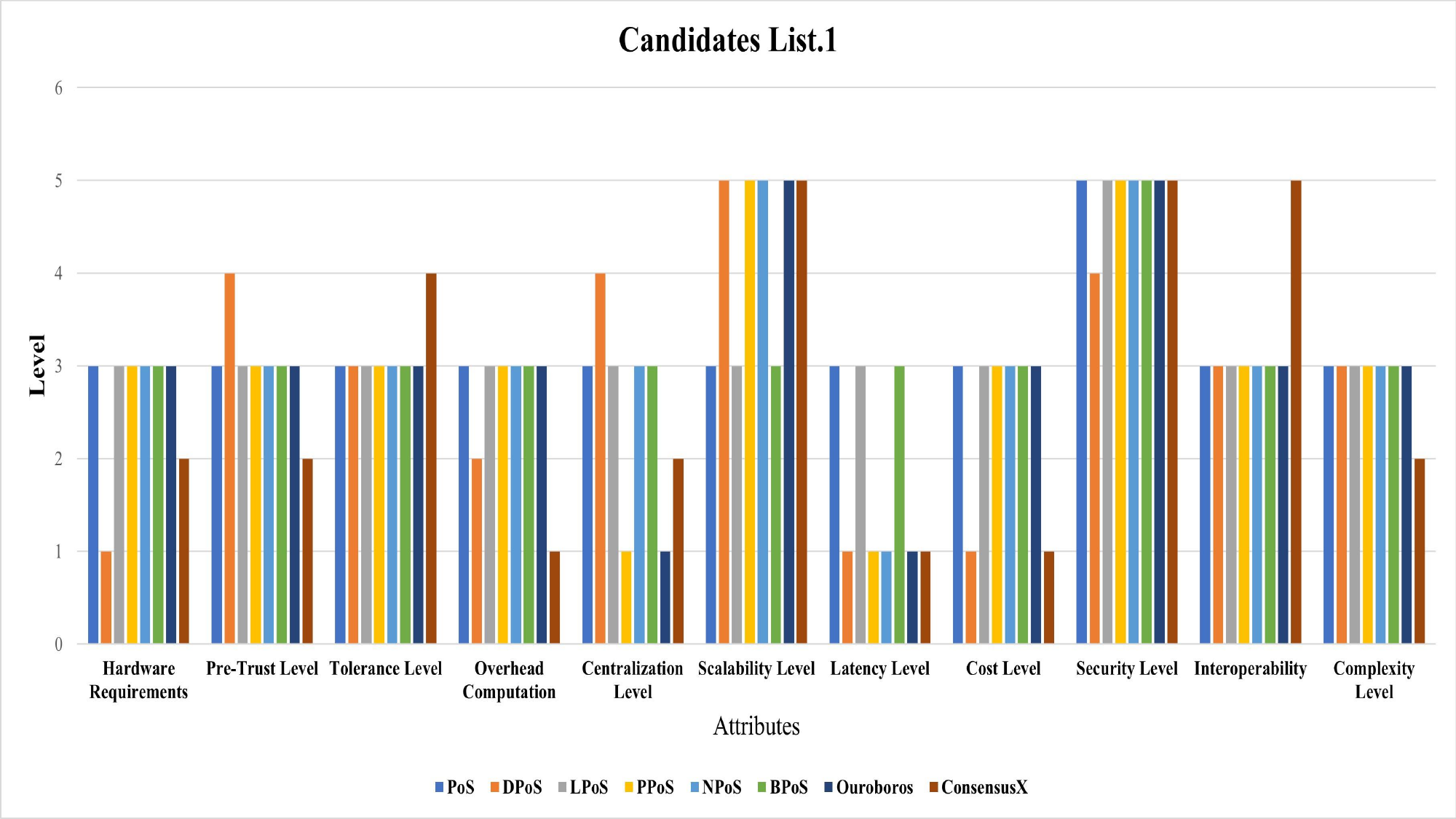}
	\caption{Candidates List 1 Versus ConsensusX.}
\label{FIG:CPSConsensusXChartC1}
\end{figure}

%
%
%
%
%
%
%
%
%
%
%

\begin{table}[htbp]
	\caption{Candidates List 2.}
\label{TAB:CandidateList2}
	\centering
	\begin{tabular}{|p{0.10\linewidth}|p{0.05\linewidth}|p{0.05\linewidth}|p{0.05\linewidth}|p{0.05\linewidth}|p{0.05\linewidth}|p{0.05\linewidth}|p{0.05\linewidth}|p{0.05\linewidth}|p{0.05\linewidth}|p{0.08\linewidth}|p{0.05\linewidth}|}
		\hline
		\textbf{Attributes} & \textbf{Hard.} & \textbf{Trust } & \textbf{Toler.} & \textbf{Compu.} & \textbf{Centra.} & \textbf{Scala.} & \textbf{Late.} & \textbf{Cost} & \textbf{Secu.} & \textbf{Interoper.} & \textbf{Compl.}\\
		\hline
		\textbf{BFT}       & M  & M2H & H & M2H & M & M & M & M & H & M & M2H \\ \hline
		\textbf{PBFT}      & M2H & M2H & H & H & M2H & M & M & M2H & H & M & H \\ \hline
		\textbf{BFA}       & M & M & H & M2H & M & M & M & M & H & M & M2H \\ \hline
		\textbf{IBFT}      & M2H & M2H & H & H & M2H & M & M & M2H & H & M & H \\ \hline
		\textbf{HBBFT}     & M2H & M2H & H & H & M2H & M & M & M2H & H & M & H \\ \hline
		\textbf{RCPA}      & M & M2H & H & M2H & M & M2H & M2H & M & H & M2H & M2H \\ \hline
		\textbf{Tendermint} & M & M & H & M & M & H & M2H & M & H & M2H & M \\ \hline
		\textbf{Hashgraph} & M & M & H & M & L & H & H & M & H & M2H & M \\ \hline
		\textbf{Avalanche} & M & M & H & M & L & H & H & M & H & M2H & M \\ \hline
		\textbf{ConsensusX} & L2M & L2M & M2H & L & L2M & H & L & L & H & H & L2M \\ \hline

	\end{tabular}
\end{table}

\begin{figure}[htp]
	\includegraphics[width=16.5cm]{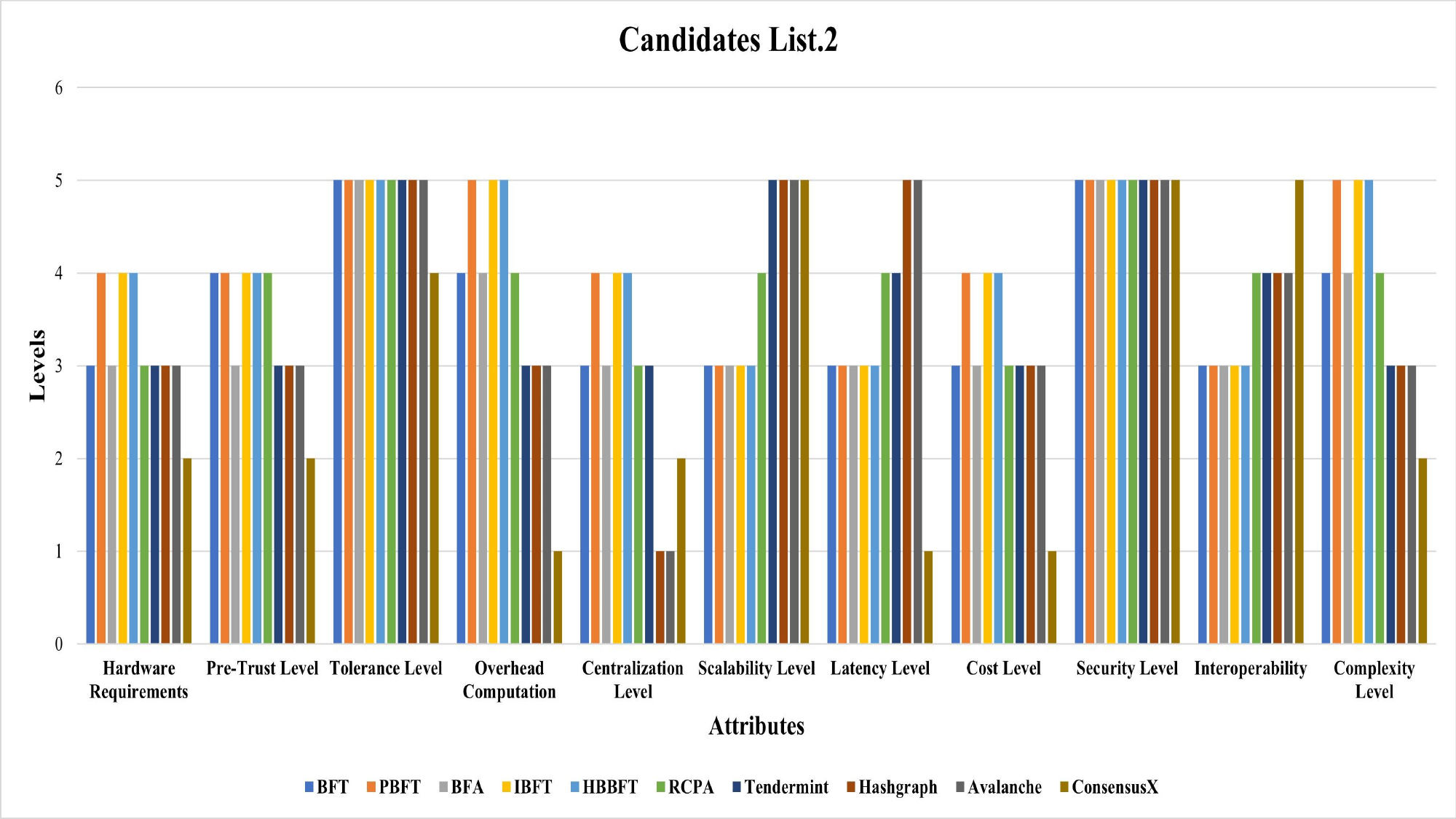}
	\caption{Candidates List 2 Versus ConsensusX.}
\label{FIG:CPSConsensusXChartC2}
\end{figure}

\begin{table}[htbp]
	\caption{Candidates List 3.}
\label{TAB:CandidateList3}
	\centering
	\begin{tabular}{|p{0.10\linewidth}|p{0.05\linewidth}|p{0.05\linewidth}|p{0.05\linewidth}|p{0.05\linewidth}|p{0.05\linewidth}|p{0.05\linewidth}|p{0.05\linewidth}|p{0.05\linewidth}|p{0.05\linewidth}|p{0.08\linewidth}|p{0.05\linewidth}|}
		\hline
		\textbf{Attributes} & \textbf{Hard.} & \textbf{Trust } & \textbf{Toler.} & \textbf{Compu.} & \textbf{Centra.} & \textbf{Scala.} & \textbf{Late.} & \textbf{Cost} & \textbf{Secu.} & \textbf{Interoper.} & \textbf{Compl.}\\
		\hline
		\textbf{PoC}             & M & M & M2H  & M & M & M  & M & M & M2H  & M & M \\ \hline
		\textbf{PoI}             & M & M2H & M2H  & M & M2H & M2H & M2H & M  & M2H & M & M \\ \hline
		\textbf{PoCon}           & M & M & M2H & M & M & M2H & M2H & M & M2H & M & M \\ \hline
		\textbf{PoR}             & M & M2H & M2H & M & M2H & M2H & M2H & M & M2H & M & M \\ \hline
		\textbf{PoW}             & H & L   & M2H & H & H  & L & M & H & H & M & H \\ \hline
		\textbf{Tangle}          & L & L   & H   & L & L  & H & M & L & H & M2H & M \\ \hline
		\textbf{PoA}             & M & H   & H   & L & M2H & H & M2H  & L & H  & M2H & M \\ \hline
		\textbf{PoAct}           & M & M   & M2H & M & M   & M & M    & M & M2H & M  & M \\ \hline
		\textbf{ConsensusX}      & L2M     & L2M & M2H    & L  & L2M & H & L & L  & H & H & L2M \\ \hline

	\end{tabular}
\end{table}

\begin{figure}[htp]
	\includegraphics[width=16.5cm]{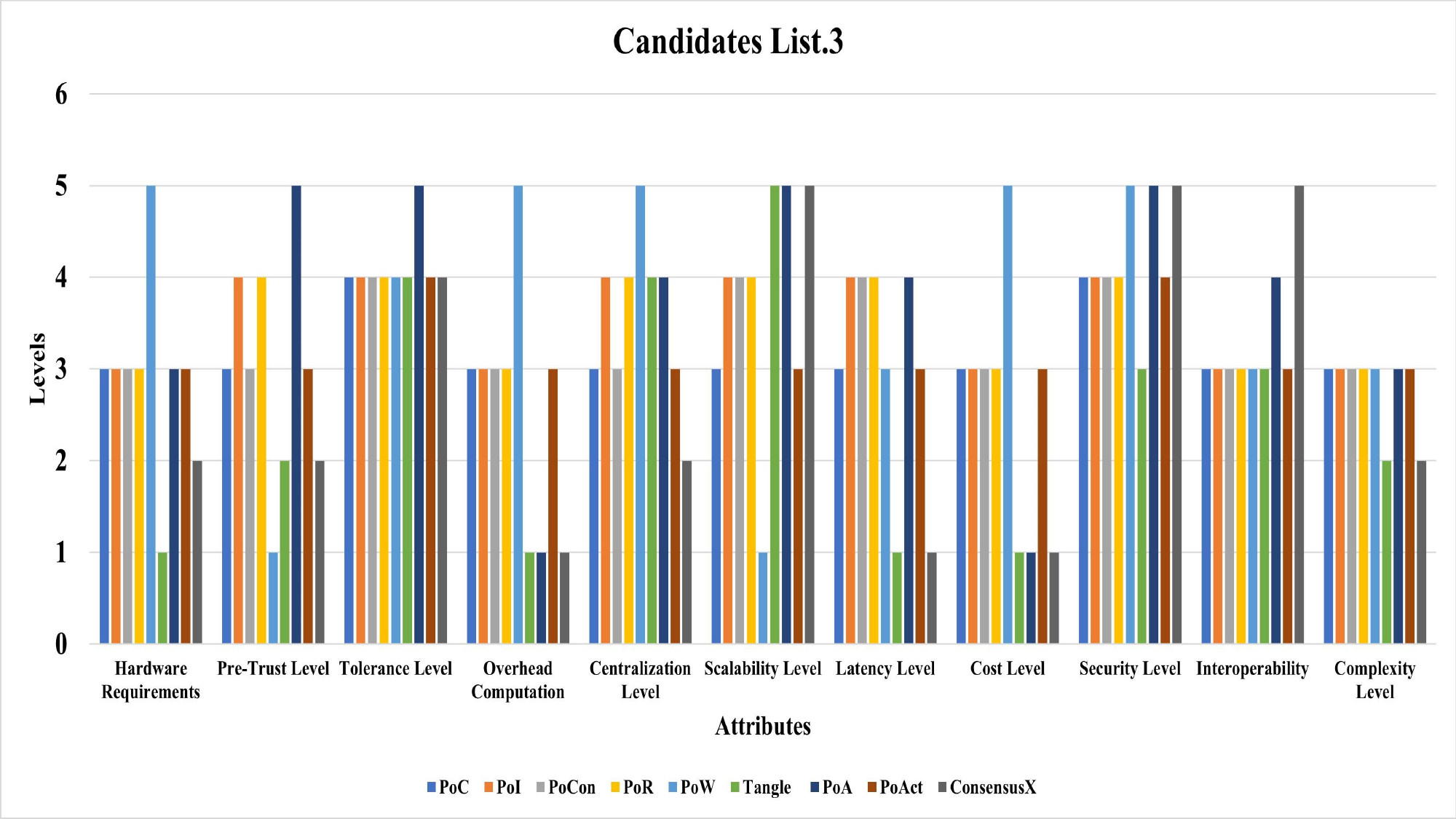}
	\caption{Candidates List 3 Versus ConsensusX.}
\label{FIG:CPSConsensusXChartC3}
\end{figure}

%
%
%
%
%
%
%
%
%
%
%

\begin{table}[htbp]
	\caption{Candidates List 4.}
\label{TAB:CandidateList4}
	\centering
	\begin{tabular}{|p{0.10\linewidth}|p{0.05\linewidth}|p{0.05\linewidth}|p{0.05\linewidth}|p{0.05\linewidth}|p{0.05\linewidth}|p{0.05\linewidth}|p{0.05\linewidth}|p{0.05\linewidth}|p{0.05\linewidth}|p{0.08\linewidth}|p{0.05\linewidth}|}
		\hline
		\textbf{Attributes} & \textbf{Hard.} & \textbf{Trust } & \textbf{Toler.} & \textbf{Compu.} & \textbf{Centra.} & \textbf{Scala.} & \textbf{Late.} & \textbf{Cost} & \textbf{Secu.} & \textbf{Interoper.} & \textbf{Compl.}\\
		\hline
		\textbf{CFT}   & M & M2H & M2H & M & M2H & M & M & M & M2H & M & M \\ \hline
		\textbf{Paxos} & M & M2H & M2H & M & M2H & M & M & M & M2H & M & M \\ \hline
		\textbf{Raft}  & M & M2H & M2H & M & M2H & M & M & M & M2H & M & M \\ \hline
		\textbf{PoET}  & L & M  & M2H & L & M  & M & M & L & M  & M & L \\ \hline
		\textbf{PoB}   & M & M  & M2H & M & M2H & M & M & M & M2H & M & M \\ \hline
		\textbf{PoCov} & M & M  & M2H & M & M  & M & M & M & M2H & M & M \\ \hline
		\textbf{ConsensusX} & L2M & L2M & M2H & L & L2M & H & L & L & H & H & L2M \\ \hline
	\end{tabular}
\end{table}

\begin{figure}[htp]
	\includegraphics[width=16.5cm]{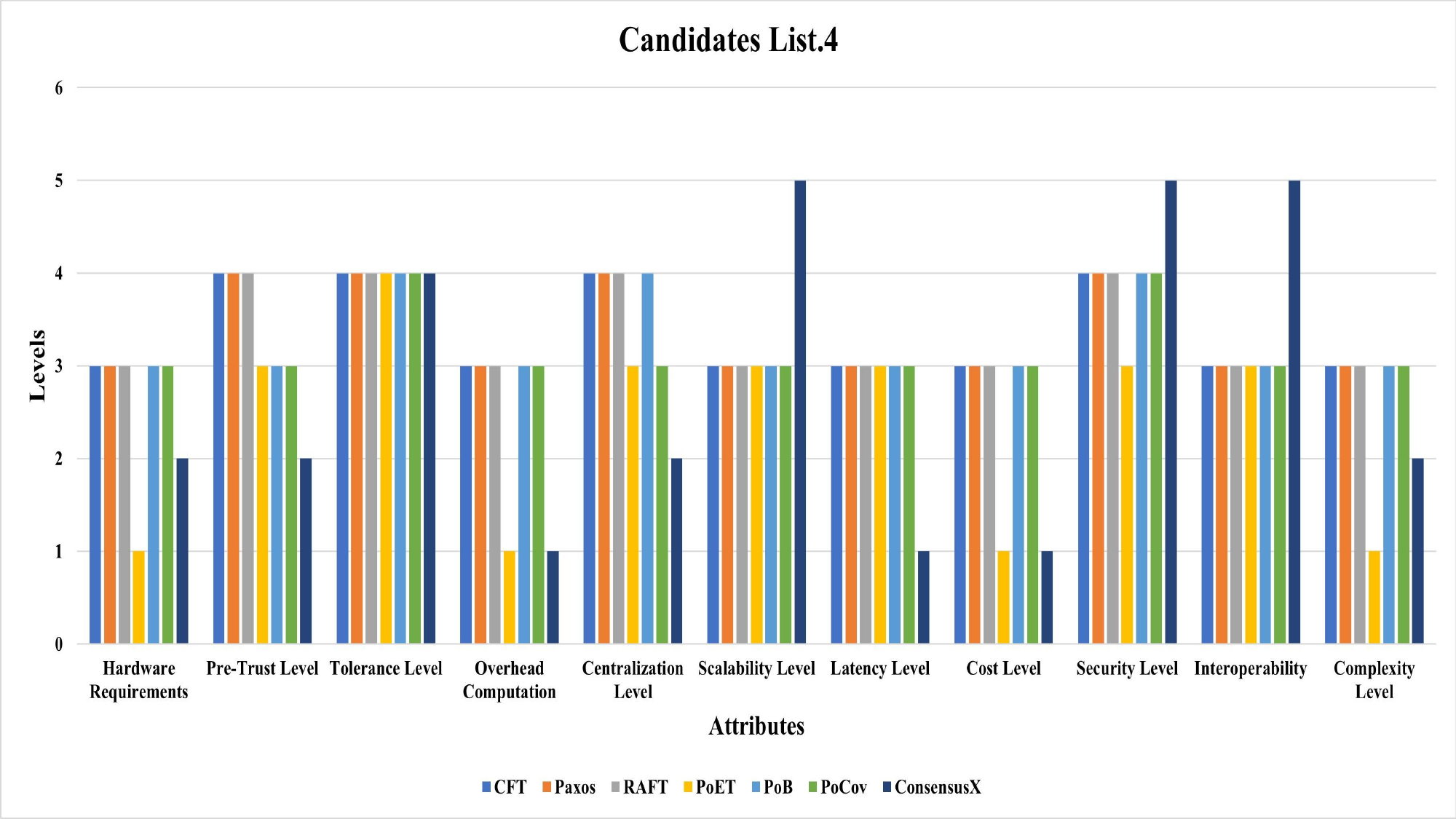}
	\caption{Candidates List 4 Versus ConsensusX.}
\label{FIG:CPSConsensusXChartC4}
\end{figure}

Figures \ref{FIG:CPSConsensusXChartC1}, \ref{FIG:CPSConsensusXChartC2}, \ref{FIG:CPSConsensusXChartC3}, and \ref{FIG:CPSConsensusXChartC4} illustrate the match level between the target ConsensusX and the candidates. The results indicate that PPoS (Pure Proof of Stake), NPoS (Nominated Proof of Stake), Ouroboros, DPoS (Delegated Proof of Stake), Tendermint, Hashgraph, Avalanche, Tangle, PoA (Proof of Authority), and PoET (Proof of Elapsed Time) have at least 8 attributes matching with ConsensusX which is based on our evaluation are suitable consensus algorithms but not the exact required ConsensusX. 
  
\subsection{Discussion}

Cyber-Physical Systems (CPS) necessitate consensus algorithms that offer high security, scalability, and performance while consuming little energy and exhibiting minimal latency. PPoS, NPoS, Ouroboros, DPoS, Tendermint, Hashgraph, Avalanche, Tangle, PoA, and PoET are all acceptable for CPS to varying degrees. DPoS and NPoS, for instance, provide energy efficiency and high throughput, whereas Tendermint and Ouroboros guarantee comprehensive security and decentralization. Hashgraph and Avalanche are promising options due to their scalability and speed of completion, whereas Tangle excels at providing lightweight consensus with minimal resource requirements, making it appropriate for IoT devices. PoA and PoET both take distinctive approaches, with PoA emphasizing trusted authorities for decision-making and PoET utilizing hardware-based random leader selection for impartiality and energy efficiency.
All the listed CAs have similar or close levels for more than 7 attributes out of 11. It is obviously very hard to find the perfect match to certain requirement for a specific field from a pre-established CAs. From the analysis above, PPoS, NPoS, Ouroboros, DPoS, Tendermint, Hashgraph, Avalanche, Tangle, PoA, and PoET are the closest to be used in a CPS environment. Each CA has been used in a certain ledger structure that could impact the overall performance for each one. 

\begin{itemize}
\item Algorithm for PPoS (Pure Proof of Stake): To select validators, this consensus algorithm employs a blockchain-based ledger structure and a novel cryptographic sorting procedure.

\item NPoS (Nominated Proof of Stake) - Polkadot: NPoS employs a blockchain-based ledger structure featuring a Relay Chain, Parachains, and Bridges to connect multiple blockchains.

\item Ouroboros - Cardano: Ouroboros also utilizes a blockchain-based ledger structure, with epochs divided into slots and slot leaders selected via a secure multiparty computation protocol.

\item DPoS (Delegated Proof of Stake) - EOS, Bitshares: DPoS operates on a blockchain-based ledger structure, with a limited number of elected delegates validating transactions and producing blocks.

\item Tendermint - Cosmos: Tendermint is a BFT-based consensus algorithm employing a blockchain-based ledger structure, with a modular architecture that supports multiple application-specific blockchains connected via the Cosmos Hub.

\item Hedera is the hashgraph for the Hedera tree Hashgraph uses a directed acyclic graph (DAG) structure, known as the hashgraph, to record transactions and reach consensus via a virtual voting mechanism.

\item Avalanche - Avalanche Protocol: Avalanche consensus utilizes a directed acyclic graph (DAG) structure with multiple subnets and chains, as well as a novel probabilistic sampling technique for validator selection.

\item Tangle - IOTA: Tangle also employs a directed acyclic graph (DAG) structure, in which transactions are connected via a web-like structure as opposed to a linear blockchain.

\item PoA (Proof of Authority) - VeChain, POA Network: PoA typically uses a blockchain-based ledger structure with a limited number of trusted validators who have the authority to create and validate blocks based on their reputation.

\item Hyperledger Sawtooth's PoET (Proof of Elapsed Time) implementation: PoET utilizes a blockchain-based ledger structure and a novel method that utilizes secure enclaves in hardware to randomly select block producers.
\end{itemize}

\begin{table}[htbp]
	\caption{Existed Implementations for The Final Candidates List with Higher Suitability to CPS.}
\label{TAB:Implementations}
	\centering
	\begin{tabular}{|p{0.15\linewidth}|p{0.35\linewidth}|p{0.40\linewidth}|}
		\hline
		\textbf{Implementations}   &\textbf{Pros} & \textbf{Cons} \\
		\hline
		PPoS (Algorand) & High security, decentralization, energy efficiency & Complex cryptographic processes, scalability challenges \\
		\hline
		NPoS (Polkadot) & Scalability, interoperability, efficient consensus & Complex architecture, centralization risk \\
		\hline
		Ouroboros (Cardano) & Energy efficiency, strong security, adaptability & Complexity, potential latency, evolving protocol \\
		\hline
		DPoS (EOS, Bitshares) & High performance, scalability, reduced energy consumption & Centralization risk, voter apathy, reliance on delegates \\
		\hline
		Tendermint (Cosmos) & High throughput, modularity, interoperability & Centralization risk, complex architecture, reliance on validators \\
		\hline
		Hashgraph (Hedera Hashgraph) & Fast transaction processing, high throughput, asynchronous BFT & Limited decentralization, patented technology, complex data structure \\
		\hline
		Avalanche & High scalability, robust security, adaptable architecture & New protocol, complex validator selection, unproven long-term stability \\
		\hline
		Tangle (IOTA) & Scalability, zero transaction fees, IoT suitability & Lack of robust security, reliance on coordinator nodes, complex data structure \\
		\hline
PoA (VeChain, POA Network) & High performance, reduced energy consumption, predictable validator selection & Centralization risk, reliance on trusted authorities, limited decentralization \\
		\hline
PoET (Hyperledger Sawtooth) & Fairness, energy efficiency, hardware-based security & Dependence on specific hardware, scalability challenges, limited applicability outside enterprise use cases \\
		\hline
	\end{tabular}
\end{table}

\section{Conclusion}
\label{sec:Conclusion}

In this study, 30 consensus algorithms, including PPoS, NPoS, Ouroboros, DPoS, Tendermint, Hashgraph, Avalanche, Tangle, PoA, and PoET, were investigated. Eleven attributes, including security, scalability, decentralization, energy efficiency, and performance, were used to evaluate each algorithm based on the requirements and trade-offs of cyber-physical systems. In addition, a detailed comparison of PoS family algorithms, BFT family algorithms, weight family, and others, as well as the benefits and drawbacks of each consensus algorithm based on their ledger structures, were provided.
While each consensus algorithm has its own strengths and limitations, it is crucial to consider the use case's specific requirements and constraints when selecting an appropriate consensus mechanism. In addition, it is essential to investigate potential modifications and adaptations of these algorithms to address their limitations and difficulties. In the end, the selection of a consensus algorithm will hinge on the optimal balance between the desired level of security, performance, decentralization, and other application-specific factors. The work resulted in a list of candidates with greater compatibility with CPS and their existing implementations, as listed in Table \ref{TAB:Implementations}.

\bibliographystyle{IEEEtran}


\begin{IEEEbiography}
	[{\includegraphics[height=1.25in, keepaspectratio]{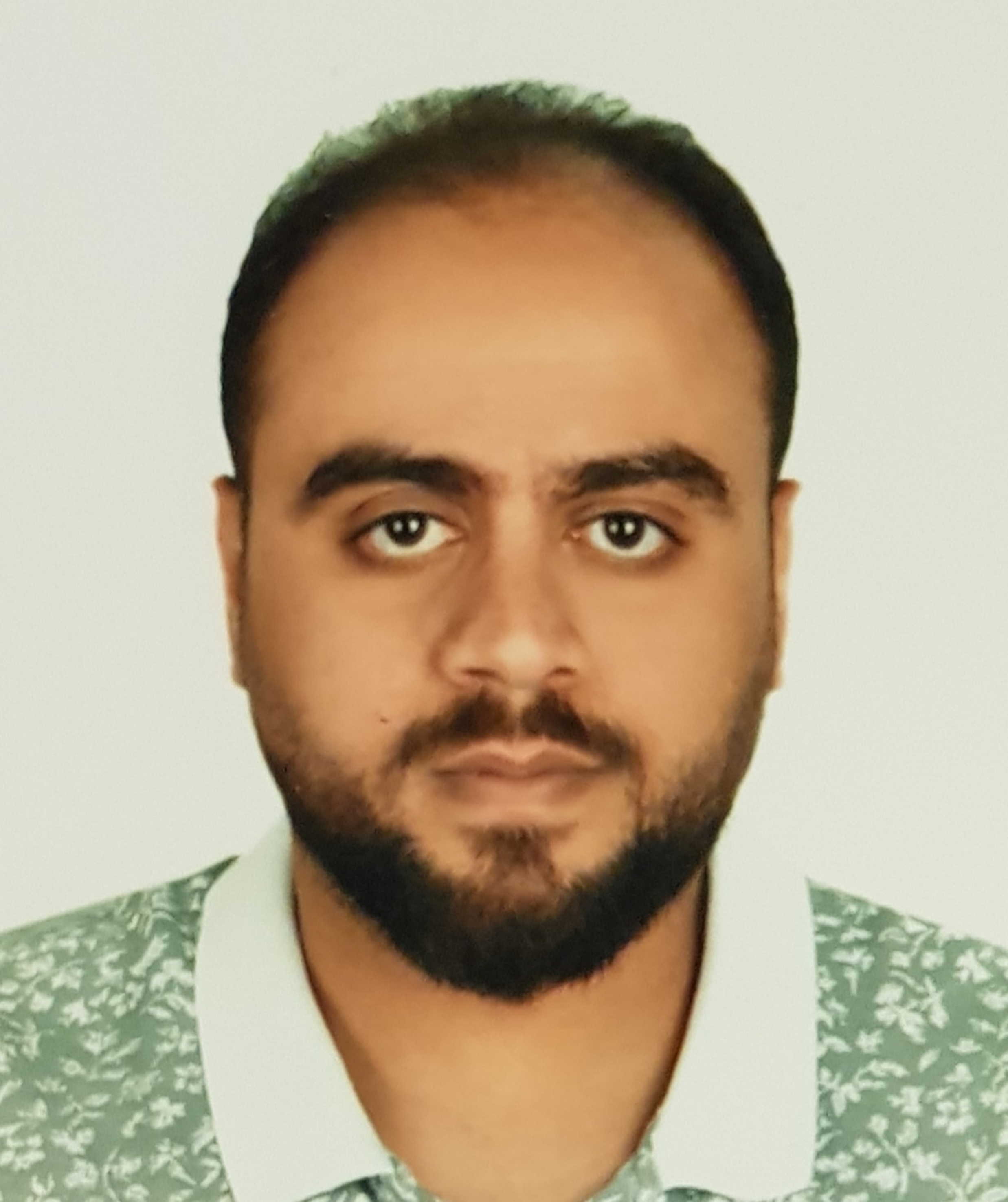}}]
	{Ahmad J. Alkhodair} received a bachelor's degree (Honors) in computer engineering from Fahad bin Sultan University (FBSU), Saudi Arabia-Tabuk, in 2012 and master's in computer engineering in 2017 from Denver University (DU), Colorado-Denver, USA. A Faculty Member in University of Tabuk, Saudi Arabia-Tabuk at the department of Computer Engineering. Currently a Ph.D. candidate in the research group at Smart Electronics Systems Laboratory (SESL) at Computer Science and Engineering at the University of North Texas, Denton, TX. Research interests include Distributed Ledger Technology (DLT), Cyber Physical Systems (CPS), Artificial Intelligence (AI) in Smart Cities. 
\end{IEEEbiography}

\begin{IEEEbiography}
[{\includegraphics[height=1.25in, keepaspectratio]{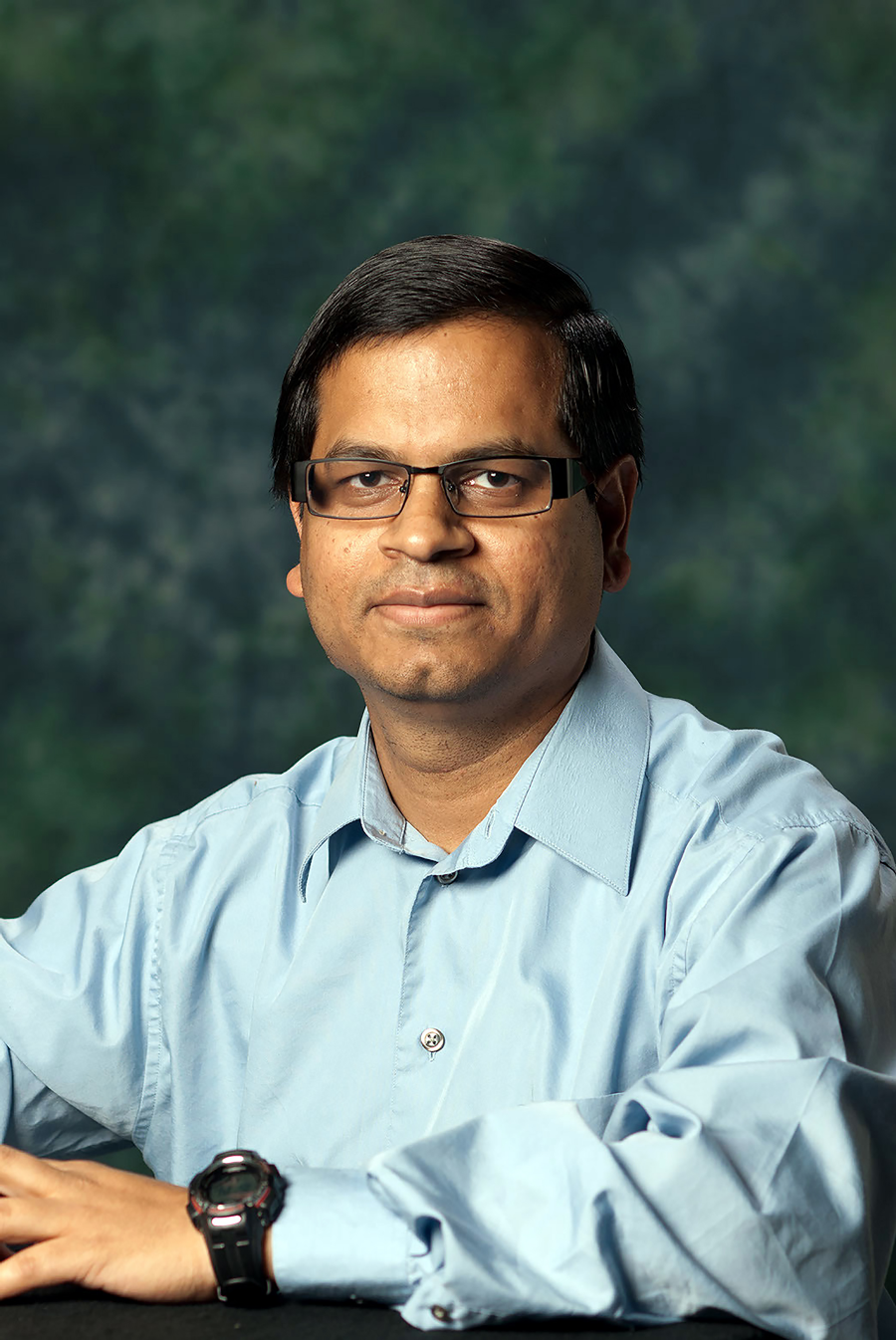}}]
{Saraju P. Mohanty} received the bachelor's degree (Honors) in electrical engineering from the Orissa University of Agriculture and Technology, Bhubaneswar, in 1995, the master's degree in Systems Science and Automation from the Indian Institute of Science, Bengaluru, in 1999, and the Ph.D. degree in Computer Science and Engineering from the University of South Florida, Tampa, in 2003. He is a Professor with the University of North Texas. His research is in ``Smart Electronic Systems'' which has been funded by National Science Foundations (NSF), Semiconductor Research Corporation (SRC), U.S. Air Force, IUSSTF, and Mission Innovation. He has authored 450 research articles, 5 books, and 9 granted and pending patents. His Google Scholar h-index is 54 and i10-index is 227 with 12,000 citations. He is regarded as a visionary researcher on Smart Cities technology in which his research deals with security and energy aware, and AI/ML-integrated smart components. He introduced the Secure Digital Camera (SDC) in 2004 with built-in security features designed using Hardware Assisted Security (HAS) or Security by Design (SbD) principle. He is widely credited as the designer for the first digital watermarking chip in 2004 and first the low-power digital watermarking chip in 2006. He is a recipient of 16 best paper awards, Fulbright Specialist Award in 2020, IEEE Consumer Electronics Society Outstanding Service Award in 2020, the IEEE-CS-TCVLSI Distinguished Leadership Award in 2018, and the PROSE Award for Best Textbook in Physical Sciences and Mathematics category in 2016. He has delivered 22 keynotes and served on 14 panels at various International Conferences. He has been serving on the editorial board of several peer-reviewed international transactions/journals, including IEEE Transactions on Big Data (TBD), IEEE Transactions on Computer-Aided Design of Integrated Circuits and Systems (TCAD), IEEE Transactions on Consumer Electronics (TCE), and ACM Journal on Emerging Technologies in Computing Systems (JETC). He has been the Editor-in-Chief (EiC) of the IEEE Consumer Electronics Magazine (MCE) during 2016-2021. He served as the Chair of Technical Committee on Very Large Scale Integration (TCVLSI), IEEE Computer Society (IEEE-CS) during 2014-2018 and on the Board of Governors of the IEEE Consumer Electronics Society during 2019-2021. He serves on the steering, organizing, and program committees of several international conferences. He is the steering committee chair/vice-chair for the IEEE International Symposium on Smart Electronic Systems (IEEE-iSES), the IEEE-CS Symposium on VLSI (ISVLSI), and the OITS International Conference on Information Technology (OCIT). He has mentored 2 post-doctoral researchers, and supervised 15 Ph.D. dissertations, 26 M.S. theses, and 20 undergraduate projects.
\end{IEEEbiography}

\begin{IEEEbiography}
	[{\includegraphics[height=1.25in, keepaspectratio]{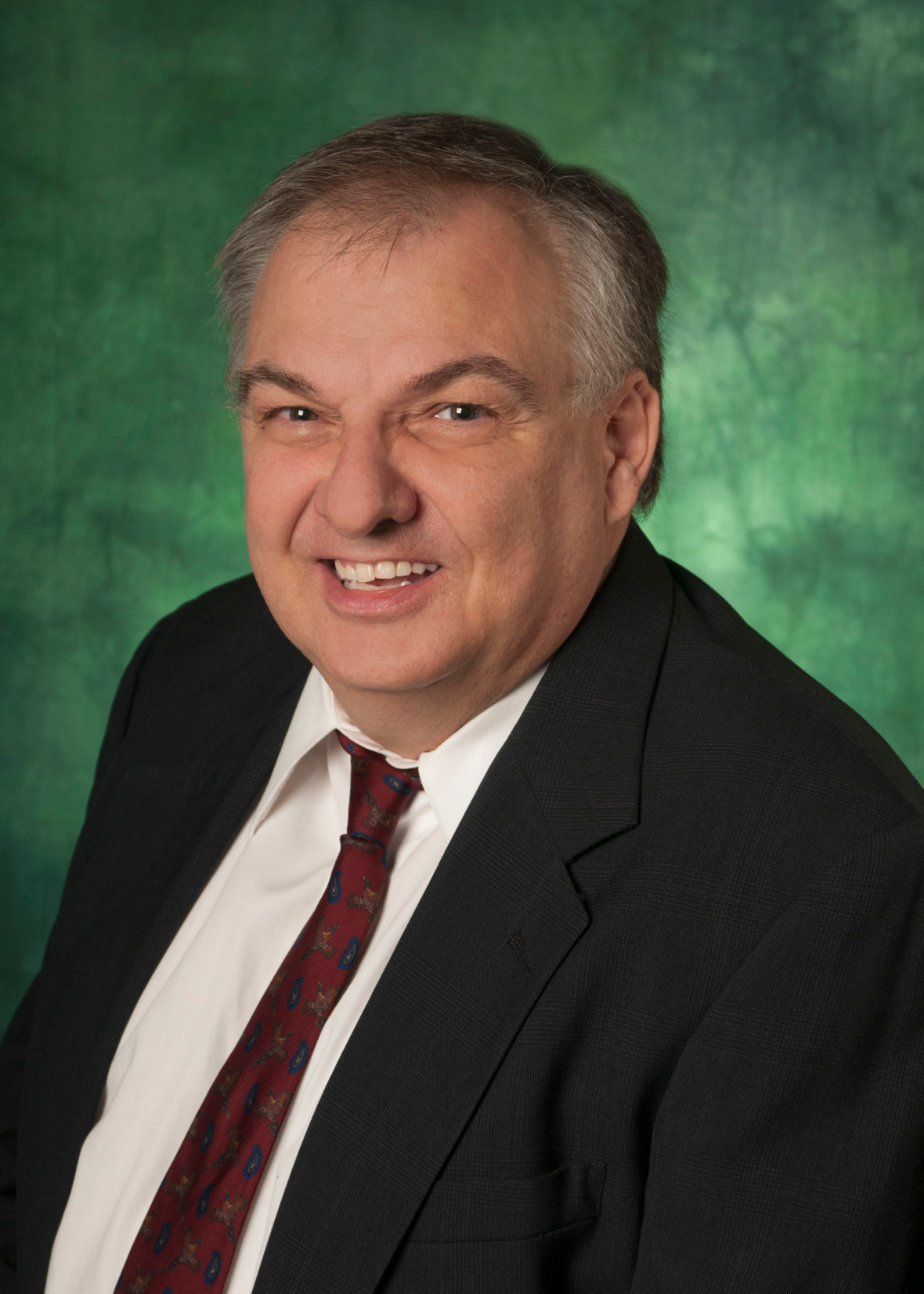}}]
	{Elias Kougianos} received a BSEE from the University of Patras, Greece in 1985 and an MSEE in 1987, an MS in Physics in 1988 and a Ph.D. in EE in 1997, all from Louisiana State University. From 1988 through 1998 he was with Texas Instruments, Inc., in Houston and Dallas, TX. In 1998 he joined Avant! Corp. (now Synopsys) in Phoenix, AZ as a Senior Applications engineer and in 2000 he joined Cadence Design Systems, Inc., in Dallas, TX as a Senior Architect in Analog/Mixed-Signal Custom IC design. He has been at UNT since 2004. He is a Professor in the Department of Electrical Engineering, at the University of North Texas (UNT), Denton, TX. His research interests are in the area of Analog/Mixed-Signal/RF IC design and simulation and in the development of VLSI architectures for multimedia applications. He is an author of over 200 peer-reviewed journal and conference publications.
\end{IEEEbiography}







\end{document}